\documentclass{LMCS}

\def\dOi{11(1:5)2015}
\lmcsheading%
{\dOi}
{1--50}
{}
{}
{Apr.~17, 2013}
{Mar.~\phantom06, 2015}
{}

\ACMCCS{[{\bf Software and its engineering}]: Software organization
  and properties---Software functional properties---Formal
  methods---Software verification; Software notations and
  tools---General programming languages---Language features; [{\bf
      Theory of computation}]: Semantics and reasoning---Program
  semantics---Axiomatic semantics; Logic}


\usepackage{ifthen}
\usepackage{amssymb}
\newboolean{showcomments}
\setboolean{showcomments}{true}
\ifthenelse{\boolean{showcomments}}
  {\newcommand{\mynote}[2]{
  \marginpar{{\Huge !}}
    \fbox{\bfseries\sffamily\scriptsize#1}
    {\small$\blacktriangleright$\textsf{\emph{#2}}$\blacktriangleleft$}
   }
  }
  {\newcommand{\mynote}[2]{}
  }

\newcommand{\EEE}[1]{\CAL{E}^{#1}[\cdot]}
\newcommand{\PROVESTCAP}{\vdash^{\mathsf{tcap}}}
\newcommand{\MODALFORALL}[2]{\forall #1^{\mathsf{\square #2}}}
\newcommand{\MODALEXISTS}[2]{\exists #1^{\mathsf{\square #2}}}
\newcommand{\LIFT}[1]{\mathsf{lift}_{#1}}
\newcommand{\EVAL}{\mathsf{eval}}
\newcommand{\POWER}{\mathsf{power}}
\newcommand{\EXT}[1]{\mathsf{Ext}{(#1)}}
\newcommand{\EXTQ}[1]{\mathsf{Ext_{q}}{(#1)}}
\newcommand{\PCF}{\textsc{Pcf}}
\newcommand{\PCFDP}{\textsc{Pcf$_{\text{\textsc{dp}}}$}}
\newcommand{\FORMULATYPES}[2]{#1 \vdash {#2}}
\newcommand{\JUDGEMENTTYPES}[2]{\FORMULATYPES{#1}{#2}}
\newcommand{\FV}[1]{\mathsf{fv}(#1)}
\newcommand{\QQ}[1]{\langle #1 \rangle}
\newcommand{\QQEVAL}[3]{#1 = \QQ{#2}\{#3\}}
\newcommand{\ONEEVAL}[4]{#1 \bullet #2=#3 \{#4\}}
\newcommand{\SIMPLEEVAL}[3]{#1 \bullet #2=#3}
\newcommand{\VERYSIMPLEEVAL}[2]{#1 \bullet #2}
\newcommand{\EVALCONV}[2]{#1 \bullet #2 \CONV}
\newcommand{\DEFEQ}{\stackrel{\text{\emph{def}}}{=}}
\newcommand{\CAL}[1]{\mathcal{#1}}
\newcommand{\RRR}{\CAL{R}}
\newcommand{\INT}{\mathsf{Int}}
\newcommand{\BOOL}{\mathsf{Bool}}
\newcommand{\UNIT}{\mathsf{Unit}}
\newcommand{\IFTHEN}[2]{\mathtt{if}\; #1\;\mathtt{then}\; #2}
\newcommand{\IFTHENELSE}[3]{\IFTHEN{#1}{#2}\;\mathtt{else}\;#3}
\newcommand{\AST}[1]{\QQ{#1}}
\newcommand{\LET}[3]{\mathtt{let}\ #1 = #2\ \mathtt{in}\ #3}
\newcommand{\LETQQ}[3]{\LET{\QQ{#1}}{#2}{#3}}
\newcommand{\EXPRESSIONTYPES}[3]{#1 \vdash #2 : #3}
\newcommand{\DOM}[1]{\mathsf{dom}(#1)}
\newcommand{\TYPES}[3]{#1 \vdash #2 : #3}
\newcommand{\NI}{\noindent}
\newcommand{\CONV}{\Downarrow}
\newcommand{\LEQ}{\lesssim}
\newcommand{\CONG}{\simeq}
\newcommand{\DIV}{\Uparrow}
\newcommand{\infer}[2]{\frac{\displaystyle{ #1 }}{\displaystyle{ #2 }}}
\newcommand{\ZEROPREMISERULE}[1]{\infer{}{#1}}
\newcommand{\ONEPREMISERULE}[2]{\infer{#1}{#2}}
\newcommand{\TWOPREMISERULE}[3]{\infer{#1 \quad #2}{#3}}
\newcommand{\THREEPREMISERULE}[4]{\infer{#1 \quad #2 \quad #3}{#4}}
\newcommand{\FOURPREMISERULE}[5]{\infer{#1 \quad #2 \quad #3 \quad #4}{#5}}
\newcommand{\FIVEPREMISERULE}[6]{\infer{#1 \quad #2 \quad #3 \quad #4 \quad #5}{#6}}

\newcommand{\RULENAME}[1]{\textsc{#1}}
\newcommand{\SMALLRULENAME}[1]{\textsc{\tiny #1}}
\newcommand{\SUBST}[2]{[#1/#2]}
\newcommand{\MINUS}{{\mbox{\bf\small -}}}
\newcommand{\LOGIC}[1]{\mathsf{#1}}
\newcommand{\PROGRAM}[1]{\mathtt{#1}}
\newcommand{\SEMB}[1]{\lbrack\!\lbrack #1 \rbrack\!\rbrack}
\newcommand{\ZEROPREMISERULENAMEDRIGHT}[2]{\ZEROPREMISERULE{#1}\,\SMALLRULENAME{#2}}
\newcommand{\ONEPREMISERULENAMEDRIGHT}[3]{\ONEPREMISERULE{#1}{#2}\,\SMALLRULENAME{#3}}
\newcommand{\TWOPREMISERULENAMEDRIGHT}[4]{\TWOPREMISERULE{#1}{#2}{#3}\,\SMALLRULENAME{#4}}
\newcommand{\THREEPREMISERULENAMEDRIGHT}[5]{\THREEPREMISERULE{#1}{#2}{#3}{#4}\,\SMALLRULENAME{#5}}
\newcommand{\FOURPREMISERULENAMEDRIGHT}[6]{\FOURPREMISERULE{#1}{#2}{#3}{#4}{#5}\,\SMALLRULENAME{#6}}
\newcommand{\FIVEPREMISERULENAMEDRIGHT}[7]{\FIVEPREMISERULE{#1}{#2}{#3}{#4}{#5}{#6}\,\SMALLRULENAME{#7}}

 \newenvironment{FIGURE}{\begin{figure}\rule{\linewidth}{0.5pt}
 }{\rule{\linewidth}{0.5pt}\end{figure}}
\newcommand{\NOVSPACEPARAGRAPH}[1]{\NI\textbf{\emph{#1}.}}
\newcommand{\PARAGRAPH}[1]{\vspace{2mm}\NOVSPACEPARAGRAPH{#1}}
\theoremstyle{plain}
\theoremstyle{plain}
\theoremstyle{plain}
\theoremstyle{plain}
\theoremstyle{plain}
\newcommand{\TRUTH}{\LOGIC{T}}
\newcommand{\TRUE}{\LOGIC{t}}
\newcommand{\FALSITY}{\LOGIC{F}}
\newcommand{\FALSE}{\LOGIC{f}}
\newcommand{\FS}{\rightarrow}
\newcommand{\IMPLIES}{\supset}
\newcommand{\VEC}[1]{\tilde{#1}}
\newenvironment{GRAMMAR}{\[\begin{array}{rcl c rcl c rcl}}{\end{array}\]}
\newenvironment{RULES}{\[\begin{array}{c}}{\end{array}\]}
\newcommand{\VERTICAL}{\;  \mid\hspace{-3.0pt}\mid \; }
\newcommand{\AND}{\mathbin{\wedge}}
\newcommand{\BIGAND}{\bigwedge}
\newcommand{\BIGOR}{\bigvee}
\newcommand{\OR}{\vee}
\newcommand{\RED}{\rightarrow}
\newcommand{\NRED}{\rightarrow\hspace{-3.0mm}\rightarrow}
 \newcommand{\nextLine}{\\[1mm] \hline \\[-3mm]}

\newenvironment{NDERIVATION}[1]{\setcounter{line}{#1}\[\begin{array}{ll}}{\end{array}\]}
\newcommand{\NLINESKELETON}[2]{\theline &\quad  #1\ \quad\hfill \text{\emph{#2}}\addtocounter{line}{1}}
\newcommand{\NLINE}[2]{\NLINESKELETON{#1}{#2}\nextLine}
\newcommand{\NLASTLINE}[2]{\NLINESKELETON{#1}{#2}}
\newcommand{\ASSERT}[4]{\{#1\}\; #2 :_{#3} \{#4\}}
\newcommand{\QUOTATION}[1]{
\hfill
\hspace{92mm}
\begin{minipage}{70mm}\tiny #1\end{minipage}
\vspace{-7mm}}

\begin{document}

\title[Program Logics for Homogeneous Generative Run-Time Meta-Programming]{Program Logics for Homogeneous Generative Run-Time Meta-Programming}

\author[M.~Berger]{Martin Berger\rsuper a}
\address{{\lsuper a}Department of Informatics, University of Sussex, Falmer, Brighton BN1 9QJ, United Kingdom}
\email{M.F.Berger@sussex.ac.uk}
\author[L.~Tratt]{Laurence Tratt\rsuper b}
\address{{\lsuper b}Software Development Team, Department of Informatics, King's College London, Strand, London WC2R 2LS, United Kingdom}
\email{laurie@tratt.net}


\newcounter{line}
\setcounter{line}{1}

\thispagestyle{plain}

\begin{abstract}

\NI This paper provides the first program logic for homogeneous
generative run-time meta-programming---using a variant of
MiniML$_e^{\square}$ by Davies and Pfenning as its underlying
meta-programming language.  We show the applicability of our approach
by reasoning about example meta-programs from the literature. We also
demonstrate that our logics are relatively complete in the sense of
Cook, enable the inductive derivation of characteristic formulae, and
exactly capture the observational properties induced by the
operational semantics.

\end{abstract}

\maketitle

\QUOTATION{Dedicated to the memory of Kohei Honda.}

\section{Introduction}

\NI Meta-programming (MP) is the generation or manipulation of programs, or
parts of programs, by other programs, i.e.~in an algorithmic way.
Many programming languages, going back at least as far as Lisp, have
explicit MP features. These can be classified in various ways such as:
generative (program creation), intensional (program analysis),
compile-time (happening while programs are compiled), run-time (taking
place as part of program execution), heterogeneous (where the system
generating or analysing the program is different from the system being
generated or analysed), homogeneous (where the systems involved are
the same), and lexical (working on simple strings) or syntactical
(working on abstract syntax trees). Compilers use MP to compile
programs; web system languages such as PHP use MP to produce web pages
containing Javascript; Javascript (in common with some other
languages) performs MP by dynamically generating strings and then
executing them using its $\PROGRAM{eval}$ function. In short, MP is a
mainstream activity.

One of the most important types of MP is homogeneous generative
meta-programming. The first language to support this was Lisp with its
S-expression based macros; Scheme's macros improve upon Lisp's by
being fully hygienic, but are conceptually similar. Perhaps
unfortunately, the power of Lisp-based macros was long seen to rest
largely on Lisp's minimalistic syntax and subsequent work on HGMP struggled
to transfer Lisp's power to languages with modern, large syntaxes. 
MetaML~\cite{TahaW:mulstapitaa} was the first syntactically rich
language capable of homogeneous generative meta-programming in a manner
convenient enough to rival Lisp's, albeit it could only generate code at
run-time rather than at compile-time. Since then, MetaOCaml has taken
MetaML's~\cite{TahaW:envClas} ideas further; while Template
Haskell~\cite{SheardT:temmetpfh} and
Converge~\cite{tratt__compile_time_meta_programming_in_a_dynamically_typed_oo_language}
have developed compile-time generative meta-programming. These
languages have clearly shown that a wide variety of modern programming
languages can house homogeneous generative meta-programming, and that
this allows powerful, safe programming of a type previously
impractical or impossible.

This paper develops program logics for generative MP. An important question
is which flavour? From this paper's perspective, the most obvious division is
whether generative MP occurs solely at run-time (\`a la MetaML) or also at
compile-time (\`a la Lisp). Since the latter case includes the former, a
reasonable first step is to tackle run-time generative MP. In
other words, this paper develops logics for languages in the MetaML vein, and we
hope it provides a basis for extending that work to languages that can support
compile-time generative MP. So that we are clear about what precisely form of MP
we are tackling, we use the term homogeneous generative run-time
meta-programming (HGRTMP).  We appreciate that HGRTMP is not a snappy acronym,
but, in the process of developing this work, we have found that MP's many
flavours are too easily confused with one another.

\PARAGRAPH{Meta-programming \& verification} 
There are currently no logics for MP capable languages, HGRTMP or otherwise.  We
believe that the following reasons might be partly responsible:
\begin{itemize}

\item Reasoning about MP languages is a strict superset of reasoning
      about non-MP languages. Developing logics for non-MP programming
      languages is a hard problem on its own, and satisfactory
      solutions for reasoning about programs with higher-order
      functions, state, pointers, continuations, or concurrency have
      only recently been
      discovered~\cite{BergerM:prologfshoc,REYNOLDS,YHB07:local:full}.

\item MP correctness can sometimes be side-stepped by ignoring the MP itself
      and looking only at its output. Compilation is an example where
      the MP machinery is more complex than the program's
      output. However, verifying only the output of MP is limiting,
      because knowledge gathered from the program's input, and during
      the MP process, is lost.

\item Static typing of generative MP still lacks a satisfactory 
      solution. Consequently, most generative MP languages are at
      least partly dynamically typed (including MetaOCaml which checks
      for certain forms of code extrusion at run-time); Template
      Haskell on the other hand intertwines code generation with
      type-checking in complicated ways.  Logics for such languages
      are not well understood in the absence of other MP features;
      moreover, many MP languages have additional features such as
      capturing substitution, pattern matching of code, and splicing
      of types, which are largely unexplored
      theoretically. Heterogeneous MP adds the complication of
      multi-language verification.

\end{itemize}

\PARAGRAPH{Contributions}
The present paper is an extended version of~\cite{BergerM:prologfhml}
with proofs, simplifications, and other improvements. It is the first
to investigate the use of program logics for the specification and
verification of HGRTMP\footnote{Since the original
publication \cite{BergerM:prologfhml}, Charlton has developed a logic
for a simple, first-order HGRTMP language with a
Javascript-like \texttt{eval} feature \cite{CharltonN:reaabusbrcg}.}.
The aim of the paper is to explore the axiomatic foundations of
HGRTMP.  The specific technical contributions of this paper are as
follows:
\begin{itemize}

\item We provide the first program logic for an HGRTMP language---\PCFDP, 
      a variant of Davies and Pfenning's
      MiniML$_e^{\square}$~\cite{DaviesR:modanaosc}, itself an
      extension of \PCF~\cite{GunterCA:semprol}. The logic is for
      total correctness and smoothly generalises previous work on
      axiomatic semantics for the ML family of
      languages~\cite{BergerM:prologfshoc,ALIASfull,HondaK:froproltop,HY04PPDP,GLOBAL,YHB07:local:full}.
      A key feature of our logic is that \PCFDP\ programs in
      the \PCF\ fragment (i.e.~those that do not perform HGRTMP)
      can be reasoned about in the simpler \PCF\ 
      logic~\cite{HondaK:froproltop,HY04PPDP}. Reasoning
      about HGRTMP therefore imposes no additional burden over reasoning about
      non-MP programs.

\item We show that our logic is relatively complete in the sense of
      Cook~\cite{CookSA:soucomaspv}.

\item We demonstrate that the axiomatic semantics induced by our logic
      coincides precisely with the contextual semantics given by the
      reduction rules of \PCFDP.

\item We present an additional inference system for characteristic
      formulae which enables, for each program $M$, the inductive
      derivation of a pair $A, B$ of formulae which describe
      completely $M$'s behaviour (descriptive
      completeness~\cite{HondaK:descriptive}).

\end{itemize}

\NI As the first work in this area, we do not pretend to tackle all the intricacies
involved in a modern programming language. Instead, we work on a simplified
language which allows us to focus on the fundamental issues.

\section{\texorpdfstring{\PCFDP}{Pcf sub dp}}\label{language}

\NI This section introduces \PCFDP\footnote{The name is our tip of the
  hat to Davies and Pfenning's work.}, the MP language that is the
basis of our study.  \PCFDP\ is a variant of call-by-value
\PCF\ \cite{GunterCA:semprol}, extended with the HGRTMP features of Davies
and Pfenning's Mini-ML$^{\square}_e$~\cite[Section
  3]{DaviesR:modanaosc}.  Mini-ML$^{\square}_e$ was the first typed MP
language to provide facilities for executing generated code. Typing
the execution of generated code is a difficult problem.
Mini-ML$^{\square}_e$ achieves type-safety with two substantial
restrictions on meta-programming:
\begin{itemize}

\item Only code without free variables can be run (i.e.~generated code
  which is not closed cannot be run).

\item Variables free in code cannot be $\lambda$-abstracted or be
  recursion variables.

\end{itemize}
Mini-ML$^{\square}_e$ was one of the first MP languages with a
Curry-Howard correspondence, although the present paper does not
investigate the connection between our program logic and the
Curry-Howard correspondence. \PCFDP\ is essentially
Mini-ML$^{\square}_e$, but with a slightly different form of recursion
that can be given a moderately simpler logical characterisation.

\PCFDP\ is an ideal vehicle for our investigation for two reasons.
First, \PCFDP\ is designed to be a simple language, yet it has all the
key features of HGRTMP; \PCFDP's operational semantics
is substantially simpler than that of MetaML~\cite{TahaW:mulstapitaa}
and its descendants, for example.  Second, \PCFDP\ is built atop
\PCF, a well-understood idealised programming language with
existing program logics~\cite{HondaK:froproltop,HY04PPDP}. This
allows us to compare reasoning in the \PCF-fragment with reasoning in
full \PCFDP.

\subsection{Language basics}

\NI \PCF\ is a traditional $\lambda$-calculus and we assume readers
are familiar with such languages. \PCFDP\ extends \PCF\ with two new
constructs, $\QQ{M}$ and $\LETQQ{x}{M}{N}$, as well as a new type
$\QQ{\alpha}$.

\emph{Quasi-quotes} $\QQ{M}$ were invented in the context of logic
\cite{QuineWVO:froalogpov,QuineWVO:mathlog}, and introduced to
programming languages in Lisp~\cite{BawdenA:quailisp}. A quasi-quote
$\QQ{M}$ represents the code of $M$, and allows code fragments to be
expressed using concrete syntax.  If $M$ has type $\alpha$, then
$\QQ{M}$ is typed $\QQ{\alpha}$. For example, $\QQ{1 + 7}$ is the code
of the program $1 + 7$, and $\QQ{1 + 7}$ has type $\QQ{\INT}$.
$\QQ{M}$ is a value for all $M$ and hence $\QQ{1 + 7}$ does not reduce
to $\QQ{8}$.  Note that \PCFDP's quasi-quotes are subtly different
from the abstract syntax trees (ASTs) used in some languages
(e.g.~Template Haskell and Converge). In such languages, quasi-quotes are a
`front end' for ASTs, but ASTs can be manually instantiated
to represent any program. In \PCFDP, in contrast, quasi-quotes are the only
term constructors for meta-programs. This makes our formalism more tractable but
prevents some seemingly reasonable meta-programs from being
expressed (e.g.~those that generate an \texttt{if}
with an arbitrary number of \texttt{else if} clauses).

The \emph{unquote} construct $\LETQQ{x}{M}{N}$ extracts code from a
quasi-quote. It evaluates $M$ to code $\QQ{M'}$, extracts $M'$ from the
quasi-quote, names it $x$ and makes $M'$ available in $N$ without reducing
$M'$. The fact that $M'$ is not evaluated after extraction from a
quasi-quote is the essence of generative MP as it enables the construction of
code \emph{other than values} under $\lambda$-abstractions.

\PCFDP's unquote unifies MetaML's separate notions of splicing
(inserting quasi-quoted code into another quasi-quoted fragment) and
executing quasi-quotes. The following example shows it being used for
splicing:
\[
	\LETQQ{x}{(\lambda z.z)\QQ{1 + 7}}{\QQ{\lambda n.x^n}}
\]
This first reduces the application to $\QQ{1 + 7}$, then extracts the code
from $\QQ{1 + 7}$, names it $x$ and makes it available unevaluated to
the code $\QQ{\lambda n.x^n}$:
\begin{eqnarray*}
	\LETQQ{x}{(\lambda z.z)\QQ{1 + 7}}{\QQ{\lambda n.x^n}}
			&\ \quad \RED \quad\ &
	\LETQQ{x}{\QQ{1 + 7}}{\QQ{\lambda n.x^n}}
			\\
			&\RED&
	\QQ{\lambda n.x^n}\SUBST{1 + 7}{x}
                        \\
			&=&
	\QQ{\lambda n.(1 + 7)^n}
\end{eqnarray*}
\NI The program $\LETQQ{x}{\QQ{N}}{x}$ is an example of unquote executing a
program, since $N$ will be extracted from the quasi-quote and bound to $x$,
which is then run.

\label{language:page:1}
\subsection{Syntax and types} 
\label{syntax}
We now formalise \PCFDP's syntax and semantics, assuming a set of
variables, ranged over by $x, y, g, u, m, ...$ (for more details see
\cite{DaviesR:modanaosc,GunterCA:semprol}).
\begin{GRAMMAR}
	\alpha
		&\ ::=\ &
	\UNIT
		\VERTICAL
	\BOOL
		\VERTICAL
	\INT
		\VERTICAL
	\alpha \FS \beta
		\VERTICAL
	\AST{\alpha}
		\\[1mm]
	V
		&::=&
	\PROGRAM{c}
		\VERTICAL
	x
		\VERTICAL
	\lambda x^{\alpha}.M
		\VERTICAL
	\mu g^{\alpha \FS \beta}.\lambda x^{\alpha}.M
		\VERTICAL
	\QQ{M}
		\\[1mm]
	M
		&::=&
        V
		\VERTICAL
        \PROGRAM{op}(\VEC{M})
		\VERTICAL
	MN
		\VERTICAL
	\IFTHENELSE{M}{N}{N'}
		\VERTICAL
        \LETQQ{x}{M}{N}
\end{GRAMMAR}

\NI Here, $\alpha$ ranges over \emph{types}, $V$ over \emph{values},
and $M$ over \emph{programs}.  Constants $\PROGRAM{c}$ range over the
integers $0, 1, 2, -1, ...$, booleans $\TRUE, \FALSE$, and $()$ of
type $\UNIT$, $\PROGRAM{op}$ ranges over the usual first-order
operators like addition, multiplication, equality, conjunction,
negation, comparison, etc., with the restriction that equality is
\emph{not} defined on expressions of function type or of type
$\QQ{\alpha}$.  The abbreviation $\VEC{M}$ means a (possibly empty)
tuple $(M_1, ..., M_n)$.  The recursion operator is $\mu g.\lambda
x.M$.  The \emph{free variables} $\FV{M}$ of $M$ are defined as usual
with two new clauses: $\FV{\QQ{M}} \DEFEQ \FV{M}$ and
$\FV{\LETQQ{x}{M}{N}} \DEFEQ \FV{M} \cup (\FV{N} \setminus \{x\})$. We
write $\lambda ().M$ for $\lambda x^{\UNIT}.M$ and $\LET{x}{M}{N}$ for
$(\lambda x.N)M$, assuming that $x \notin \FV{M}$ in both cases. We
assume Barendregt's variable condition, and tacitly rename bound
variables where necessary.

The \emph{reduction relation} $\RED$ is unchanged from \PCF\ for the
\PCF-fragment of \PCFDP, and adapted to \PCFDP\ as follows. First we
define \emph{reduction contexts}, by extending those for \PCF\ with a
construct for unquoting.
\begin{GRAMMAR}
	\EEE{}
		&\ ::=\ &
        [.]
		\VERTICAL
        \EEE{}M
		\VERTICAL
        V\EEE{}
		\VERTICAL
        \PROGRAM{op}(\VEC{V}\EEE{}\VEC{M})
		\VERTICAL
        \IFTHENELSE{\EEE{}}{M}{N}
                \\[1mm]
		&\VERTICAL&
        \LETQQ{x}{\EEE{}}{M}
\end{GRAMMAR}
Now $\RED$ is defined on \emph{closed} programs by the 
clauses given next:
\begin{itemize}

\item $(\lambda x.M)V \RED M\SUBST{V}{x}$.
\item $(\mu g.\lambda x.M)V \RED M\SUBST{\mu g.\lambda x.M}{g}\SUBST{V}{x}$.
\item $\IFTHENELSE{\TRUE}{M}{N} \RED M$.
\item $\LETQQ{x}{\QQ{M}}{N} \RED N\SUBST{M}{x}$.
\item $M \RED N$ implies $\CAL{E}[M] \RED \CAL{E}[N]$.

\end{itemize}
We write $\NRED$ for $\RED^*$. $M \CONV V$ means that $M \NRED V$ for
some value $V$. We write $M\CONV$ if $M \CONV V$ for some appropriate $V$,
and $M\DIV$ if not $M\CONV$.

\begin{FIGURE}
\begin{RULES}
	\ONEPREMISERULE
        {
		(x, \alpha) \in \Gamma \cup \Delta
        }
	{
		\TYPES{\Gamma; \Delta}{x}{\alpha}
	}
		\quad
	\ONEPREMISERULE
	{
		\TYPES{\Gamma, x : \alpha; \Delta}{M}{\beta}
	}
	{
		\TYPES{\Gamma; \Delta}{\lambda x^{\alpha}.M}{\alpha \FS \beta}
	}
		\quad
        \TWOPREMISERULE
        {
		\TYPES{\Gamma; \Delta}{M}{\alpha \FS \beta}
        }
        {
		\TYPES{\Gamma; \Delta}{N}{\alpha}
        }
	{
		\TYPES{\Gamma; \Delta}{MN}{\beta}
	}
                \\\\
	\ONEPREMISERULE
	{
		\TYPES{\Gamma, f : (\alpha \FS \beta); \Delta}{\lambda x^{\alpha}.M}{\alpha \FS \beta}
	}
        {
		\TYPES{\Gamma; \Delta}{\mu f^{\alpha \FS \beta}.\lambda x^{\alpha}.M}{\alpha \FS \beta}
        }
                \quad
	\THREEPREMISERULE
	{
		\TYPES{\Gamma; \Delta}{M}{\BOOL}
	}
	{
		\TYPES{\Gamma; \Delta}{N}{\alpha}
	}
	{
		\TYPES{\Gamma; \Delta}{N'}{\alpha}
	}
	{
		\TYPES{\Gamma; \Delta}{\IFTHENELSE{M}{N}{N'}}{\alpha}
	}
		\\\\
	\TWOPREMISERULE
	{
		\TYPES{\Gamma; \Delta}{M}{\INT}
	}
	{
		\TYPES{\Gamma; \Delta}{N}{\INT}
	}
	{
		\TYPES{\Gamma; \Delta}{M + N}{\INT}
	}
        \quad
	\ONEPREMISERULE
	{
		\TYPES{\epsilon; \Delta}{M}{\alpha}
	}
	{
		\TYPES{\Gamma; \Delta}{\QQ{M}}{\AST{\alpha}}
	}
		\quad
	\TWOPREMISERULE
	{
		\TYPES{\Gamma; \Delta}{M}{\AST{\alpha}}
	}
        {
		\TYPES{\Gamma; \Delta, x : \alpha}{N}{\beta}
        }
	{
		\TYPES{\Gamma; \Delta}{\LETQQ{x}{M}{N}}{\beta}
	}

\end{RULES}
\caption{Key typing rules for \PCFDP.}\label{figure:language:typing}
\end{FIGURE}

A \emph{typing environment} ($\Gamma, \Delta, ...$) is a finite map
$x_1 : \alpha_1, ..., x_k : \alpha_k$ from variables to types.  The
\emph{domain} $\DOM{\Gamma}$ of $\Gamma$ is the set $\{x_1, ...,
x_n\}$, assuming that $\Gamma$ is $x_1 : \alpha_1, ..., x_n :
\alpha_n$. We write $\epsilon$ for the empty environment.  The
\emph{typing judgement} is written $\TYPES{\Gamma; \Delta}{M}{\alpha}$
where we assume that $\DOM{\Gamma} \cap \DOM{\Delta} = \emptyset$. We
write $\TYPES{}{M}{\alpha}$ for $\TYPES{\epsilon;
  \epsilon}{M}{\alpha}$.  We say a program $M$ is \emph{closed} if
$\TYPES{}{M}{\alpha}$.  We call $\Delta$ a \emph{modal context} in
$\TYPES{\Gamma; \Delta}{M}{\alpha}$. We say a variable $x$ is
\emph{modal} or \emph{modally typed} in $\TYPES{\Gamma; \Delta}{M}{\alpha}$ if $x \in
\DOM{\Delta}$. Modal variables represent code inside other code, and
code to be run.  The key type-checking rules are given in Figure
\ref{figure:language:typing}. Typing for constants and first-order
operations is standard. 

Noteworthy features of the typing system are that modal variables
cannot be $\lambda$- or $\mu$-abstracted, that all free variables in
quasi-quotes must be modal, and that modal variables can only be
generated by unquotes.~\cite{DaviesR:modanaosc} gives detailed
explanations of this typing system and its relationship to modal
logics.

\PARAGRAPH{Contextual congruence}
By $\LEQ_{\Gamma; \Delta; \alpha}$ (often abbreviated to just
$\LEQ$) we denote the usual \emph{typed contextual precongruence}: if
$\TYPES{\Gamma; \Delta}{M_i}{\alpha}$ for $i = 1, 2$ then: $M_1
\LEQ_{\Gamma; \Delta; \alpha} M_2$ iff for all closing context
$C[\cdot]$ such that $\TYPES{}{C[M_i]}{\UNIT}$ $(i = 1, 2)$ we have
\[
    C[M_1] \CONV\  \text{implies}\  C[M_2] \CONV.
\]
We write $\CONG$ for $\LEQ \cap \LEQ^{-1}$ and call $\CONG$
\emph{contextual congruence}.  Other forms of congruence are
possible, but we will use $\CONG$ in the rest of this paper. 
Our choice means that code can only be observed
contextually, i.e.~by running it in a context. Hence for example
$\QQ{M}$ and $\QQ{\lambda x.Mx}$ are contextually indistinguishable if
$x \notin \FV{M}$, as are $\QQ{1 + 2}$ and $\QQ{3}$. This facilitates
a smooth integration of the logics for \PCFDP\ with the logics for
\PCF.\footnote{Some MP languages are more discriminating, allowing,
  e.g.~printing of code, which can distinguish $\alpha$-equivalent
  programs. It is unclear how to design logics for such languages.  A
  detailed discussion of program equalities in meta-programming
  languages can be found in~\cite{InoueJ:reaabomsp}.}

\subsection{Basic lemmas}
We now present a collection of simple facts that we use later.

\begin{prop}\label{app:completenessProofs:lemma:1}\label{lamguage:theorem:1}
\
\begin{enumerate}[label={\cW0}(\arabic*)]

\item \label{lamguage:theorem:1:5} If $M$ is closed and $M \CONG V$ then $M \CONV W$ for
  some value $W$ with $V \CONG W$.

\item \label{lamguage:theorem:1:2} If $N$ is closed and $\QQ{M} \CONG N$ then $N \CONV \QQ{M'}$
  and $M \CONG M'$ for some $M'$.

\item \label{lamguage:theorem:1:3} $M\SUBST{\mu g.M}{g} \CONG \mu g.M$.

\item \label{lamguage:theorem:1:4} Let $M_1$ and $M_2$ be closed. If
  $\TYPES{\Gamma; \Delta}{M_i}{\alpha}$ for $i = 1, 2$ then $M_1
  \Uparrow$ and $M_2 \Uparrow$ implies $M_1 \CONG M_2$

\item\label{app:completenessProofs:lemma:1:5} If $M \LEQ N$ then
  $L\SUBST{M}{x} \LEQ L\SUBST{N}{x}$.

\item \label{lamguage:theorem:1:1} $\NRED \ \subseteq\ \CONG \ \subseteq\ \LEQ$.

\item\label{app:completenessProofs:lemma:1:1} 
If $M \NRED N$ and $L \LEQ N$ then also $L \LEQ M$.

\item\label{app:completenessProofs:lemma:1:2} $M \LEQ N$ if and only if
$\QQ{M} \LEQ \QQ{N}$.

\item\label{app:completenessProofs:lemma:1:6} If $MN \LEQ M'N$ for all
  $N$ then $M \LEQ M'$.

\item\label{app:completenessProofs:lemma:1:3} If $M\SUBST{N}{x} \CONV$
  but $N \DIV$ then for all closed $N'$: $M\SUBST{N'}{x} \CONV$. 

\item\label{app:completenessProofs:lemma:1:4} If $M\SUBST{N}{x} \CONV$
  and $N \LEQ N'$ (with $N'$ closed) then also $M\SUBST{N'}{x} \CONV$.

\item \label{lamguage:theorem:1:6} If for all $n$ we have $W_n \LEQ
  V$, then also $\mu g.\lambda x.M \LEQ V$ where $W_0 \DEFEQ \Omega$
  and $W_{n+1} \DEFEQ \lambda x.M\SUBST{W_n}{g}$,
  cf.~\cite{PittsAM:opebtp}. 

\end{enumerate}
\end{prop}

\begin{proof}
All are straightforward yet laborious when carried out in detail,
and can be tackled with standard techniques of operational semantics
\cite{GunterCA:semprol,PittsAM:opebtp}.  As just one example, take
(\ref{lamguage:theorem:1:6}): if for all $n$ we have $W_n \LEQ V$, but
at the same time $\mu g.\lambda x.M \not \LEQ V$, we could find a
closing context $C[\cdot]$ such that $C[\mu g.\lambda x.M]\CONV$ but
$C[V]\DIV$. But $C[\mu g.\lambda x.M]\CONV$ means that the computation
towards a value is of finite length, hence only a finite number of recursive
calls were made, so some $n$ must exist, such that $C[W_n]\CONV$. This
in turn means $C[V]\CONV$ by our assumptions, contradicting
$C[V]\DIV$.
\end{proof}

\subsection{Some example programs}

Lifting is an important construct in generative MP, taking a run-time
value and converting it into its quasi-quoted equivalent. For example
$\QQ{3}$ is the lifting of 3, and $\QQ{\lambda x^{\alpha}.x}$ is the
lifting of the identity function of type $\alpha$.

We call a type $\alpha$ \emph{basic} if it does not contain the function
space constructor, i.e.~if it has no sub-expressions of the form
$\beta \FS \beta'$.  In \PCFDP, lifting takes an arbitrary value $V$
of basic type $\alpha$, and converts it to code $\QQ{V}$ of type
$\QQ{\alpha}$.  Note that we cannot simply write $\lambda x.\QQ{x}$
because modal variables (i.e.~variables free in code) cannot be
$\lambda$-abstracted.  For $\alpha = \INT$ the function is defined as
follows:
\[
	\LIFT{\INT} 
              \quad \DEFEQ\quad 
	\mu g.\lambda n^{\INT}.\IFTHENELSE{n \leq 0}{\QQ{0}}{\LETQQ{x}{g(n-1)}{\QQ{x+1}}}.
\]
Note that $\LIFT{\INT}$ works properly only on non-negative integers. 
Note also that $\LIFT{\INT}\ 3$ evaluates to $\QQ{0 + 1 + 1 + 1}$, not $\QQ{3}$.  
In more expressive meta-programming languages such as
Converge the corresponding program would evaluate to $\QQ{3}$, which
is more efficient, although $\QQ{0 + 1 + 1 + 1}$ and $\QQ{3}$ are
observationally indistinguishable in \PCFDP.

Lifting is easily extended to $\UNIT$ and $\BOOL$, but not to function
types, because of \PCFDP's inability to abstract modal variables.  For
basic types $\QQ{\alpha}$ we can define lifting as follows.
\begin{eqnarray*}
	\LIFT{\QQ{\alpha}}
		      &\DEFEQ&
	\lambda x^{\QQ{\alpha}}.\LETQQ{a}{x}{\QQ{\QQ{a}}}
\end{eqnarray*}
We reason about $\LIFT{\INT}$ in Section \ref{examples}.

Another example is the function $\EVAL$, a function of type
$\QQ{\alpha} \FS \alpha$ for running code~\cite{DaviesR:modanaosc}.
This function is essentially a wrapper around unquoting:
\[
	\EVAL
		\ \DEFEQ\
	\lambda x^{\QQ{\alpha}}.\LETQQ{y}{x}{y}.
\]

\NI Clearly, $\EVAL\ \QQ{17 + 3}$ converges to 20.

The last example in this section is the well-known $\POWER$ generative
MP program which creates a function that raises a number to a given
power~\cite{TahaW:mulstapitaa}. Although somewhat contrived, this
function shows how generative MP can be used for efficiency purposes:
rather than using run-time recursion on every call, HGRTMP turns this
into a fixed expression. In essence, if a program contains many applications
$(\lambda na.a^n)\ 3$, it makes sense to specialise such applications
to $\lambda a. a \times a \times a$. A simple encoding of $\POWER$ in
\PCFDP\ is the following:
\[
	\POWER
		\ \DEFEQ\ 
	\mu p. \lambda n.
	\IFTHENELSE{n \leq 0}{\QQ{\lambda x. 1}}{\LETQQ{q}{p (n-1)}{\QQ{\lambda x. x \times (q\ x)}}}
\]
This function has type $\TYPES{}{\POWER}{\INT \FS \QQ{\INT \FS
    \INT}}$. This type says that $\POWER$ takes an integer and returns
code. That code, when run, is a function from integers to integers.
$\POWER$ can can be used as follows:
\[
	\POWER\ 2
		\quad \NRED \quad
	\QQ{\lambda  a. a \times ((\lambda  b.  b \times ((\lambda  c.  1) b)) a)}
\]

\section{A logic for total correctness}\label{logic}
 
\NI Our logic is a Hoare logic with pre- and post-conditions in the
tradition of logics for ML-like
languages~\cite{BergerM:prologfshoc,ALIASfull,HY04PPDP,GLOBAL}. In
this section we define its syntax and semantics.

\subsection{Syntax and types}

\emph{Expressions}, ranged over by $e, e', ...$ and \emph{formulae} $A, B, ...$ of the
logic are given by the grammar below, using the types and variables of
\PCF:
\begin{GRAMMAR}
	e
		&\ \quad::=\quad\ &
	\LOGIC{c}
		\VERTICAL
	x
		\VERTICAL
        \LOGIC{op}(\VEC{e})
		\\[1mm]
	A
		&::=&
	e = e'
		\VERTICAL
	\neg A
		\VERTICAL
	A \AND B
		\VERTICAL
	\forall x^{\alpha}.A
		\VERTICAL
	\ONEEVAL{u}{e}{m}{A}              
		\VERTICAL
	\QQEVAL{u}{m}{A}
\end{GRAMMAR}
 
\NI Our logical language is an extension of first-order logic with
equality (and axioms for arithmetic e.g.~Peano arithmetic or some set
theory). Other quantifiers, logical constants like $\TRUTH, \FALSITY$ and
propositional connectives like $\IMPLIES$ (implication) are defined by
de Morgan duality. Quantifiers range over values of appropriate type.
Constants $\LOGIC{c}$ and operations $\LOGIC{op}$ are those of
Section~\ref{syntax}.

Our logic extends that of
\PCF~\cite{HondaK:froproltop,HondaK:descriptive,HY04PPDP} with a new
\emph{code evaluation} predicate \mbox{$\QQEVAL{u}{m}{A}$}. It says
that $u$, which must be of type $\QQ{\alpha}$, denotes (up to
contextual congruence) a quasi-quoted program $\QQ{M}$, such that
whenever $M$ is unquoted and executed, it converges to a value; if
that value is denoted by $m$ then $A$ makes a true statement about
that value.  We recall
from~\cite{HondaK:froproltop,HondaK:descriptive,HY04PPDP} that
$\ONEEVAL{u}{e}{m}{A}$ says that (assuming $u$ is of the function
type) $u$ denotes a function, which, when fed with the value denoted
by $e$, terminates and yields another value. If we name this latter
value $m$, $A$ holds.  The variable $m$ is an \emph{anchor} in both
$\ONEEVAL{u}{e}{m}{A}$ and $\QQEVAL{u}{m}{A}$, bound within scope $A$.
The \emph{free variables} of $e$ and $A$, written $\FV{e}$ and
$\FV{A}$, respectively, are defined by the following clauses:
\begin{itemize}

\item $\FV{\LOGIC{c}} \DEFEQ \emptyset$.
\item $\FV{x} \DEFEQ \{x\}$.
\item $\FV{\LOGIC{op}(\VEC{e})} \DEFEQ \bigcup_{i}\FV{e_i}$.
\item $\FV{e = e'} \DEFEQ \FV{e} \cup \FV{e'}$.
\item $\FV{\neg A} \DEFEQ \FV{A}$.
\item $\FV{A \AND B} \DEFEQ \FV{A} \cup \FV{B}$.
\item $\FV{\forall x^{\alpha}.A} \DEFEQ \FV{A} \setminus \{x\}$.
\item $\FV{\ONEEVAL{u}{e}{m}{A}} \DEFEQ (\FV{A} \setminus \{m\}) \cup \{u\} \cup \FV{e}$.
\item $\FV{\QQEVAL{u}{m}{A}} \DEFEQ (\FV{A} \setminus \{m\}) \cup \{u\}$.

\end{itemize}

\NI In the  presentation below we often use the following abbreviations and
conventions:

\begin{itemize}

\item $A^{\MINUS x}$ means that $x \notin \FV{A}$.

\item $x\CONV$ means $\exists y^{\alpha}.x = y$, assuming that $x$ has
  type $\alpha$, $y$ is fresh and not modally typed. This
  abbreviation is interesting primarily when $x$ is modally typed.

\item $\EVALCONV{x}{e}$ for $\ONEEVAL{x}{e}{m}{\TRUTH}$.

\item $m = \QQ{\cdot}$ is a shorthand for $\QQEVAL{m}{x}{\TRUTH}$
  where $x$ is fresh.

\item $m = \QQ{e}$ is short for $\QQEVAL{m}{x}{x = e}$ where $x$ is
  fresh, e.g.~$m = \QQ{x}$ is short for $\QQEVAL{m}{y}{x = y}$. Note that
  e.g.~$x$ is free in $m = \QQ{x}$, unlike in $\QQEVAL{m}{x}{A}$.

\item $m \bullet e = e'$ abbreviates $\ONEEVAL{m}{e}{x}{x = e'}$ where
  $x$ is fresh.

\item We often omit typing annotations in expressions and formulae.

\end{itemize}
We have the usual capture avoiding substitutions of expressions for
variables in expressions $e\SUBST{e'}{x}$ and formulae
$A\SUBST{e}{x}$. They are defined by the following straightforward
clauses.
\begin{itemize}

\item $y\SUBST{e}{x} \DEFEQ 
\begin{cases}
   y & x \neq y\\
   e & x = y
\end{cases}$.

\item $\LOGIC{c}\SUBST{e}{x} \DEFEQ \LOGIC{c}$.

\item $\LOGIC{op}(\VEC{e})\SUBST{e'}{x} \DEFEQ \LOGIC{op}(\VEC{e}\SUBST{e'}{x})$.

\item $(e_1 = e_2)\SUBST{e}{x} \DEFEQ (e_1\SUBST{e}{x}) = (e_2\SUBST{e}{x})$.

\item $(\neg A)\SUBST{e}{x} \DEFEQ \neg (A \SUBST{e}{x})$.

\item $(A \AND B)\SUBST{e}{x} \DEFEQ (A\SUBST{e}{x}) \AND (B\SUBST{e}{x})$.

\item $(\forall y.A)\SUBST{e}{x} \DEFEQ \forall y.(A\SUBST{e}{x})$ assuming
  $x \neq y$ and $y \notin \FV{e}$.

\item $(\ONEEVAL{u}{e}{m}{A})\SUBST{e'}{x} \DEFEQ
  \ONEEVAL{u\SUBST{e'}{x}}{e\SUBST{e'}{x}}{m}{A\SUBST{e'}{x}}$
  assuming $m \neq y$ and $m \notin \FV{e'}$.

\item $(\QQEVAL{u}{m}{A})\SUBST{e}{x} \DEFEQ \QQEVAL{u\SUBST{e}{x}}{m}{A\SUBST{e}{x}}$,
 assuming $m \neq x$ and $m \notin \FV{e}$.

\end{itemize}

\begin{FIGURE}
\begin{RULES}
	\ONEPREMISERULE
        {
		(x, \alpha) \in \Gamma \cup \Delta          
        }
	{
		\EXPRESSIONTYPES{\Gamma; \Delta}{x}{\alpha}
	}
		\quad
	\THREEPREMISERULE
	{
		\EXPRESSIONTYPES{\Gamma; \Delta}{u}{\alpha \FS \beta}
	}
	{
		\EXPRESSIONTYPES{\Gamma; \Delta}{e}{\alpha}
	}
	{
		\FORMULATYPES{\Gamma, m : \beta; \Delta}{A}
	}
	{
		\FORMULATYPES{\Gamma; \Delta}{\ONEEVAL{u}{e}{m}{A}}
	}
		\\\\
 	\TWOPREMISERULE
	{
		\EXPRESSIONTYPES{\Gamma; \Delta}{e}{\alpha}
	}
	{
		\EXPRESSIONTYPES{\Gamma; \Delta}{e'}{\alpha}
	}
	{
		\FORMULATYPES{\Gamma; \Delta}{e = e'}
	}
		\quad
	\TWOPREMISERULE
	{
		\FORMULATYPES{\Gamma; \Delta}{A}
	}
	{
		\FORMULATYPES{\Gamma; \Delta}{B}
	}
	{
		\FORMULATYPES{\Gamma; \Delta}{A \AND B}
	}
		\quad
	\ONEPREMISERULE
	{
		\FORMULATYPES{\Gamma, x : \alpha; \Delta}{A}
	}
	{
		\FORMULATYPES{\Gamma; \Delta}{\forall x^{\alpha}. A}
	}
		\\\\
	\TWOPREMISERULE
	{
		\EXPRESSIONTYPES{\Gamma; \Delta}{u}{\QQ{\alpha}}
	}
	{
		\FORMULATYPES{\Gamma; \Delta, m : \alpha}{A}
	}
	{
		\FORMULATYPES{\Gamma; \Delta}{\QQEVAL{u}{m}{A}}
	}
 		\quad
	\ONEPREMISERULE
	{
		\FORMULATYPES{\Gamma; \Delta}{A}
	}
	{
		\FORMULATYPES{\Gamma; \Delta}{\neg A}
	}
		\\\\
	\FOURPREMISERULE
	{
		\FORMULATYPES{\Gamma; \Delta}{A}
	}
	{
		m \notin \DOM{\Gamma} \cup \DOM{\Delta}
	}
	{
		\TYPES{\Gamma; \Delta}{M}{\alpha}
	}
	{
		\FORMULATYPES{\Gamma, m : \alpha; \Delta}{B}
	}
	{
		\JUDGEMENTTYPES{\Gamma; \Delta; \alpha}{\ASSERT{A}{M}{m}{B}}
	}
\end{RULES}
\caption{Typing rules for expressions, formulae and judgements. Rules for constants and 
first-order operations omitted.}\label{figure:logic:typing}
\end{FIGURE}

\NI In the last two cases we assume that if $x = u$ then $e'$ must be a
variable.

The \emph{judgements} for total correctness are of the form
\[
          \ASSERT{A}{M}{m}{B}.
\]  
The variable $m$ is the \emph{anchor} of the judgement, is a bound
variable with scope $B$, and cannot be modal.  The judgement is to be
understood as follows: if $A$ holds, then $M$ terminates to a value
(more precisely, the closure of $M$ with arbitrary values meeting the
precondition\footnote{In the remainder, we will sometimes be informal
  and say that a program $M$ reduces or terminates, even when $M$ may
  not be closed.  What we mean is that the closure of $M$ in the
  ambient model reduces or terminates.}), and if we denote that value
by $m$, then $B$ holds. In other words, our judgements are entirely
conventional for total correctness program logics. If a variable $x$
occurs freely in $A$ or in $B$, but not in $M$, then $x$ is an
\emph{auxiliary variable} of the judgement $\ASSERT{A}{M}{m}{B}$.

\PARAGRAPH{Typing expressions, formulae and judgements}
Program logics are typed (although for simple programming languages,
types can be implicit), and ours is no exception. We use the
following typing judgements.
\begin{itemize}

\item For expressions, the typing judgement is $\EXPRESSIONTYPES{\Gamma; \Delta}{e}{\alpha}$.
\item For formulae, the typing judgement is $\FORMULATYPES{\Gamma; \Delta}{A}$.
\item For judgements, the typing judgement is $\JUDGEMENTTYPES{\Gamma; \Delta; \alpha}{\ASSERT{A}{M}{m}{B}}$.

\end{itemize}
The typing rules for all three judgements are given in Figure
\ref{figure:logic:typing}. Several points are worth noting.

\begin{itemize}

\item The anchor in $\QQEVAL{u}{m}{A}$ is modal, while it is not
modal in $\ONEEVAL{u}{e}{m}{A}$ and in judgements. 

\item Normal quantification $\forall x.A$ quantifies only
  non-modal variables $x$.

\end{itemize}
 \emph{From now on,
  we assume all occurring programs, expressions, formulae and
  judgements to be well-typed.}

\PARAGRAPH{Examples of assertions \& judgements} 
We continue with a few simple examples to help explain the use of our
logic.
\begin{itemize}

\item The assertion $m = \QQ{3}$, which is short for $\QQEVAL{m}{x}{x
  = 3}$ says that $m$ denotes code which, when executed, will evaluate
  to 3. It can be used to make the following assertion on the program $\QQ{1 + 2}$:
  \[
     \ASSERT{\TRUTH}{\QQ{1 + 2}}{m}{m = \QQ{3}}.
  \]

\item Let $\Omega_{\alpha}$ be a non-terminating program of type
  $\alpha$ (we usually drop the type subscript). When we quasi-quote
  $\Omega$, the judgement $\ASSERT{\TRUTH}{\QQ{\Omega}}{m}{\TRUTH}$
  says (\emph{qua} precondition) that $\QQ{\Omega}$ is a terminating
  program. Indeed, that is the strongest statement we can make about
  $\QQ{\Omega}$ in a logic for total correctness, cf.~Section
  \ref{completeness2}.

\item The assertion  $\forall x^{\INT}.  \ONEEVAL{m}{x}{y}{y = \QQ{x}}$ says 
that $m$ denotes a terminating function  which receives an integer and
returns code which evaluates to that integer. Later, we use this assertion
when reasoning about $\LIFT{\INT}$ which has the following specification:
\[
	\ASSERT{\TRUTH}{\LIFT{\INT}}{u}{\forall n. n \geq 0 \IMPLIES \ONEEVAL{u}{n}{m}{m = \QQ{n}}}
\]

\item The formula 
  \[
     A_u \DEFEQ \forall n^{\INT} \geq 0. \exists f^{\INT \FS \INT}. (u \bullet n = \QQ{f}
     \AND \forall x^{\INT}. f \bullet x = x^n)
  \] 
  says that $u$ denotes a function which receives an integer $n$ as
  argument, to return code which when evaluated and fed another
  integer $x$, computes the power $x^n$, provided $n \geq 0$. We can
  then show that 
  \[
  \ASSERT{\TRUTH}{\POWER}{u}{A_u} 
  \]
  and
  \[
     \ASSERT{A_u}{u\ 7}{r}{\QQEVAL{r}{f}{\forall x.f \bullet x = x^7}}.
  \]

\item The formula $\forall x^{\QQ{\alpha}}y^{\alpha}.(x = \QQ{y}
  \IMPLIES u \bullet x = y)$ can be used to specify the evaluation
  function from Section \ref{language}:
  \[
  \ASSERT{\TRUTH}{\EVAL}{u}{\forall x^{\QQ{\alpha}}y^{\alpha}.(x = \QQ{y}
  \IMPLIES u \bullet x = y)}.
  \]

\end{itemize}

\subsection{Models and the satisfaction relation}

This subsection formally presents the semantics of our logic. We begin
with the notion of model. Our models are conventional, with the key
difference from the models of \PCF-logics ~\cite{HY04PPDP} being that modal
variables denote possibly non-terminating programs. 

Let $\Gamma, \Delta$ be two contexts with disjoint domains (the idea
is that $\Delta$ is modal while $\Gamma$ is not). A \emph{model} of
type $\Gamma; \Delta$ is a pair $(\xi, \sigma)$ such that:
\begin{itemize}

\item $\xi$ is a map from $\DOM{\Gamma}$ to closed \emph{values} such
  that $\TYPES{}{\xi(x)}{\Gamma(x)}$;

\item $\sigma$ is a map from $\DOM{\Delta}$ to closed \emph{programs}
  $\TYPES{}{\sigma(x)}{\Delta(x)}$. 

\end{itemize}

\NI We use the following conventions in our subsequent presentation:
\begin{itemize}

\item We write $(\xi, \sigma)^{\Gamma; \Delta}$ to indicate that $(\xi,
\sigma)$ is a model of type $\Gamma; \Delta$. 

\item We write $\xi \cdot x : V$ for $\xi \cup \{(x, V)\}$ assuming
  that $x \notin \DOM{\xi}$.

\item Likewise for $\sigma \cdot x : M$. 

\item Let $\eta = (\xi, \sigma)$ be a
model of type $\Gamma; \Delta$.
\begin{itemize}

\item We write $\DOM{\eta}$ for $\DOM{\Gamma} \cup \DOM{\Delta}$.

\item We write $\eta(x) = V$ to indicate that $x \in \DOM{\eta}$,
  and $(x, V) \in (\xi \cup \sigma)$.

\end{itemize}
\end{itemize}
We can now present the semantics of expressions.  Let
$\EXPRESSIONTYPES{\Gamma; \Delta}{e}{\alpha}$ and assume that $(\xi,
\sigma)$ is a $\Gamma; \Delta$-model, we define $\SEMB{e}_{(\xi,
\sigma)}$ by the following inductive clauses:
\begin{itemize}

\item  $\SEMB{\LOGIC{c}}_{(\xi, \sigma)} \DEFEQ \PROGRAM{c}$,

\item $\SEMB{\LOGIC{op}(\VEC{e})}_{(\xi, \sigma)} \DEFEQ
\PROGRAM{op}(\SEMB{\VEC{e}}_{(\xi, \sigma)})$, 

\item $\SEMB{x}_{(\xi, \sigma)} \DEFEQ (\xi \cup \sigma)(x)$. 

\end{itemize}
The satisfaction relation for formulae has the following shape.  Let
$\FORMULATYPES{\Gamma; \Delta}{A}$ and assume that $(\xi, \sigma)$ is
a $\Gamma; \Delta$-model. 
\begin{itemize}

\item $(\xi, \sigma) \models e = e'$ iff $\SEMB{e}_{(\xi, \sigma)}
 \CONG \SEMB{e'}_{(\xi, \sigma)}$.

\item $(\xi, \sigma) \models \neg A$ iff $(\xi, \sigma) \not \models
 A$.

\item $(\xi, \sigma) \models A \AND B$ iff $(\xi, \sigma) \models A$
and $(\xi, \sigma) \models B$.

\item $(\xi, \sigma) \models \forall x^{\alpha}.A$ iff for all closed
values $V$ of type $\alpha$: $(\xi \cdot x : V, \sigma) \models A$.

\item $(\xi, \sigma) \models \ONEEVAL{u}{e}{x}{A}$ iff
  $(\SEMB{u}_{(\xi, \sigma)} \SEMB{e}_{(\xi, \sigma)}) \CONV V$ and
  $(\xi \cdot x : V, \sigma) \models A$.

\item $(\xi, \sigma) \models \QQEVAL{u}{m}{A}$ iff $\SEMB{u}_{(\xi, \sigma)} \CONV
\QQ{M}$, $M \CONV V$ and $(\xi, \sigma \cdot m : V) \models A$.

\end{itemize}

\NI The concept of upwards-closedness is important in the context of
completeness and defined as follows. Let $A$ be a formula typeable
under $\FORMULATYPES{\Gamma, u : \alpha; \Delta}{A}$.  We say $A$  is
\emph{upwards closed at $u$} if whenever $V \LEQ W$ then also
\[
    (\xi \cdot u : V, \sigma) \models A 
        \qquad\text{implies}\qquad 
    (\xi \cdot u : W, \sigma) \models A
\]
for all suitable $\xi$ and $\sigma$.

For defining the semantics of judgements, we need to explain what it means
to apply a model $\eta \DEFEQ (\xi, \sigma)$ to a program $M$, written
$M\eta$. We also refer to $M\eta$ as the \emph{closure} of $M$ with
$\eta$.  That is defined as usual, using the following inductive
clauses, where we assume that free variables are not caught when a
model is moved under a binder:
\begin{itemize} 
\item $x\eta \DEFEQ \eta(x)$.

\item $(MN)\eta \DEFEQ (M\eta)(N\eta)$. 

\item $\PROGRAM{c}\eta \DEFEQ \PROGRAM{c}$.

\item $(\lambda x.M)\eta \DEFEQ \lambda x.(M\eta)$.

\item $(\mu g.\lambda x.M)\eta \DEFEQ \mu g.\lambda x.(M\eta)$.

\item $\QQ{M}\eta \DEFEQ \QQ{M\eta}$.

\item $(\PROGRAM{op}(\VEC{M}))\eta \DEFEQ \PROGRAM{op}(\VEC{M}\eta)$.

\item $(\IFTHENELSE{M}{N}{N'})\eta \DEFEQ \IFTHENELSE{M\eta}{N\eta}{N'\eta}$.

\item $(\LETQQ{x}{M}{N})\eta \DEFEQ \LETQQ{x}{M\eta}{N\eta}$.

\end{itemize}

\NI We record the following simple fact for subsequent use.

\begin{obs}\label{completeness:lemma:15} Let $\RRR$ be one of $\LEQ,
  \cong$, then: $M\ \RRR\ N$ if and only iff for all appropriately
  typed models $\eta$: $M\eta\ \RRR\ N\eta$.
\end{obs}

\NI The \emph{satisfaction relation} $\models \ASSERT{A}{M}{m}{B}$ is
given next.  Let $\JUDGEMENTTYPES{\Gamma; \Delta;
  \alpha}{\ASSERT{A}{M}{m}{B}}$. Then $\models \ASSERT{A}{M}{m}{B}$
holds if and only if for all models $(\xi, \sigma)^{\Gamma; \Delta}$:
\[
        (\xi, \sigma) \models A\qquad \text{implies}\qquad \exists V.(M(\xi, \sigma) \CONV V\ \text{and}\  (\xi \cdot m : V, \sigma) \models B)).
\]
This is the standard notion for total correctness, adapted to the
present logic.

\PARAGRAPH{A note on models} The reader might wonder why our notion of
model uses \emph{values} (which always terminate) as denotations
for non-modal variables, but general programs (which may not
terminate) for modal variables. The answer is a combination of two
factors:

\begin{itemize}

\item Our logic is part of a tradition of constructing
  Hoare logics, where models provide denotations for the free
  variables of the program that a judgement is about. Moreover, the
  type of the denotation should be the same as the type of the
  corresponding free variable.  This simple model-building heuristic
  has proven to be robust for a wide variety of programming
  languages~\cite{BergerM:comlogfamltmp}, and we decided to build our logic for
  \PCFDP\ in the same way.

\item Although our logic is for total correctness, we can still
  make assertions about non-terminating programs, and programs that
  contain non-terminating sub-programs, for example:
  \begin{itemize}

    \item $\ASSERT{\FALSITY}{\Omega}{u}{A}$.
    \item $\ASSERT{\TRUTH}{\lambda x.\Omega}{u}{\TRUTH}$.
    \item $\ASSERT{\TRUTH}{\QQ{\Omega}}{u}{\TRUTH}$.

  \end{itemize}
  Since judgements like $\ASSERT{\TRUTH}{\lambda x.\Omega}{u}{\TRUTH}$
  are already derivable in the logic for total correctness for \PCF,
  the question arises as to how models used in logics for \PCF\ need
  only values as denotations?  The answer is that there is a
  substantial difference between quasi-quotes and
  $\lambda$-abstractions in how the (non-terminating) sub-programs
  they harbour are accessed. The only way $\Omega$ can be executed in
  $\lambda x.\Omega$ is by application. This does not involve creating
  a new free variable bound to $\Omega$ (which would need a denotation
  in a corresponding model).

  In contrast, when unquoting a \PCFDP\ quasi-quote, e.g.
  \[
     \LETQQ{x}{\QQ{\Omega}}{M},
  \]
  then $x$ is free (as well as modal) in $M$, and will be bound to
  $\Omega$, but without attempting to evaluate $\Omega$.  This is
  quite different from evaluating e.g.~$\LET{x}{\lambda x.\Omega}{M}$,
  as we can see when comparing the evaluation of both terms
  side-by-side.
  \begin{align*}
    \LETQQ{x}{\Omega}{M} &\quad \RED\quad M\SUBST{\Omega}{x} \\
    \LET{x}{\lambda ().\Omega}{M} &\quad \RED\quad M\SUBST{\lambda ().\Omega}{x} 
  \end{align*}
Models must accommodate this behaviour, and allowing modal
variables to denote non-value programs does just this.

\end{itemize}

\subsection{Axioms and rules} We have now ready to present the rules and 
axioms of our logic.

\PARAGRAPH{Axioms} The  axioms come in two
forms: those that are germane to \PCFDP's meta-programming extensions,
and those that are not.  All axioms for the \PCF\ logics
of~\cite{HondaK:froproltop,HondaK:descriptive,HY04PPDP} remain valid,
and are listed in Appendix \ref{app:axioms} for completeness. Here we
present only the axioms for the logical constructs not already
available in the logics for \PCF, i.e.~for the code evaluation
predicate $\QQEVAL{x}{m}{A}$.

Tacitly, we assume typability of all axioms. That means not only that
all axioms must be typable, but conversely also that whenever an axiom
is typable, it is a valid axiom.  The axioms are given in Figure
\ref{figure:logic:axioms}.  The presentation uses the
following abbreviations:
\begin{center}
 $\EXTQ{xy}$ stands for $\forall a.(\QQEVAL{x}{z}{z = a} \equiv
  \QQEVAL{y}{z}{z = a})$. 
\end{center}

\begin{FIGURE}
\begin{RULES}
\begin{array}{lrcll}
	(q1) & \QQEVAL{x}{m}{A} \AND \QQEVAL{x}{m}{B} 
	     &\equiv&
	\QQEVAL{x}{m}{A \AND B}
  			    \\
	(q2) & \QQEVAL{x}{m}{\neg A} 
	     &\IMPLIES& 
	\neg\QQEVAL{x}{m}{A}
  			    \\
	(q3) & \QQEVAL{x}{m}{A} \AND \neg \QQEVAL{x}{m}{B} 
	     &\equiv &
	\QQEVAL{x}{m}{A \AND \neg B}
  			    \\
	(q4) & \QQEVAL{x}{m}{A \AND B} 
	     &\equiv& 
	A \AND \QQEVAL{x}{m}{B} 
          	   &
	m \notin \FV{A}
  			    \\
	(q5) & \QQEVAL{x}{m}{\forall a^{\alpha}. A} 
	     &\equiv& 
	\forall a^{\alpha}.\QQEVAL{x}{m}{A}
                      & a \neq x, m
  			    \\
	(q6) & (A \IMPLIES B) \AND \QQEVAL{x}{m}{A}
	     &\IMPLIES& 
	\QQEVAL{x}{m}{B}
                        \\
        (term) & 
        x\CONV
                        &&&
        x\ \text{non modal}
			\\
        (term_{q}) & \QQEVAL{x}{m}{A}
	     &\equiv& 
        \QQEVAL{x}{m}{A \AND m\CONV}
			\\
        (div) & \neg \forall m^{\QQ{\alpha}}.m = \QQ{\cdot}
                        \\
	(ext_{q}) & x = y
	      &\equiv&
	\EXTQ{xy}
           &
        x, y \ \text{of type}\ \QQ{\alpha} \\
          &&&& \text{both non-modal} 
                      \\
        (q_{\alpha}) &\QQEVAL{x}{m}{A}
                        &\equiv&
                      \QQEVAL{x}{n}{\QQEVAL{x}{m}{A \AND m = n}}\hspace{-13mm}\\
                      &&&& \hspace{-22mm}n \neq x, n\in \FV{A}\ \text{implies}\ n = m
\end{array}
\end{RULES}
\caption{Key total correctness axioms for \PCFDP. The remaining axioms
  are as for \PCF. Except where noted otherwise, free variables can be
  modal or non-modal.}\label{figure:logic:axioms}
\end{FIGURE}

\NI Axiom $(q1)$ says that if the quasi-quote denoted by $x$ makes $A$
true (assuming the program in that quasi-quote is denoted by $y$), and
in the same way makes $B$ true, then it also makes $A \AND B$ true,
and vice versa. Axiom $(q2)$ says that if the quasi-quote denoted by
$x$ contains a program, denoted by $y$, and makes $\neg A$ true, then
it cannot be the case that under the same conditions $A$ holds.  The
reverse implication is false, because $\neg \QQEVAL{x}{m}{A}$ is also
true when $x$ denotes a quasi-quote whose contained program is
diverging. But in this case, $\QQEVAL{x}{m}{\neg A}$ is still false
due to lacking termination.  Next is $(q3)$: $\QQEVAL{x}{m}{A}$ says
in particular that $x$ denotes a quasi-quote containing a terminating
program, so $\neg \QQEVAL{x}{m}{B}$ can only be true because $B$ is
false.  Axioms $(q4, q5)$ let us move formulae and quantifiers in and
out of code-evaluation formulae, as long as free variables do not
become bound in the process nor bound variables become free.  Axiom
(q6) allows us to weaken the assertion inside the code evaluation
predicate. The reverse implication is trivially false.  The axiom
$(term)$ formalises that denotations of non-modal variables always
terminate. The axiom $(term_q)$ enables us explicitly to express as a
logical formula the fact that $\QQEVAL{x}{m}{A}$ guarantees that the
code denoted by $x$ terminates.  The axiom $(q_{\alpha})$ may appear
confusing on first sight, but it states something simple: namely that
we can easily nest code evaluation predicates.  The equality $m = n$
relates the two anchors.  The axiom $(div)$ simply states that not
every quasi-quote holds code that terminates when executed.  The
code-extensionality axiom $(ext_q)$ formalises what it means for two
quasi-quotes to be equal: they must contain observationally
indistinguishable code. The corresponding axiom $(ext)$ for functions
can be found in Appendix \ref{app:axioms} together with other axioms
for the \PCF-part of the language.  Note that it is vital for $x$ and
$y$ to be non-modal. The direction $\EXTQ{xy} \IMPLIES x = y$ is
unsound otherwise, because $\EXTQ{xy}$ cannot distinguish between
e.g.~appropriately typed $\Omega$ and $\QQ{\Omega}$.

\begin{FIGURE}
\begin{RULES}
	\ZEROPREMISERULENAMEDRIGHT
        {
		\ASSERT{A\SUBST{x}{m} \AND x\CONV}{x}{m}{A}
        }{Var}
		\quad
	\ZEROPREMISERULENAMEDRIGHT
        {
		\ASSERT{A\SUBST{\LOGIC{c}}{m}}{\PROGRAM{c}}{m}{A}
        }{Const}
		\quad
	\ONEPREMISERULENAMEDRIGHT
	{
		\ASSERT{A^{\MINUS g}}{M}{u}{B}
	}
	{
		\ASSERT{A}{\mu g.M}{u}{B\SUBST{u}{g}}
	}{Rec}
		\\\\		
	\ONEPREMISERULENAMEDRIGHT
	{
		\ASSERT{A^{\MINUS x} \AND B}{M}{m}{C}
	}
	{
		\ASSERT
                {A}
		{\lambda x^{\alpha}.M}{u}
		{\forall x.(B \IMPLIES \ONEEVAL{u}{x}{m}{C})}
	}{Abs}
		\quad
	\TWOPREMISERULENAMEDRIGHT
	{
		\ASSERT{A}{M}{m}{B}
	}
	{
		\ASSERT{B}{N}{n}{C\SUBST{m + n}{u}}
	}
	{
		\ASSERT{A}{M+N}{u}{C}
	}{Add}
		\\\\
	\FIVEPREMISERULENAMEDRIGHT
	{
		\ASSERT{A}{M}{m}{B}
	}
	{
		\ASSERT{B\SUBST{b_i}{m}}{N_i}{u}{C}
	}
	{
		b_1 = \TRUE
        }
        {
                b_2 = \FALSE
	}
	{
		i = 1, 2
	}
	{
		\ASSERT{A}{\IFTHENELSE{M}{N_1}{N_2}}{u}{C}
	}{If}
		\\\\
	\TWOPREMISERULENAMEDRIGHT
	{
		\ASSERT{A}{M}{m}{B}
	}
	{
		\ASSERT{B}{N}{n}{\ONEEVAL{m}{n}{u}{C}}
	}
	{
		\ASSERT{A}{MN}{u}{C}
	}{App}
		\quad
	\ONEPREMISERULENAMEDRIGHT
	{
		\ASSERT{A}{M}{m}{B}
	}
	{
		\ASSERT{\TRUTH}{\QQ{M}}{u}{A \IMPLIES \QQEVAL{u}{m}{B}}
	}{Quote}
		\\\\
	\ONEPREMISERULENAMEDRIGHT
	{
	  \begin{array}{l}
	       \ASSERT{A}{M}{m}{E^{\MINUS x} \AND (B^{\MINUS x} \IMPLIES \QQEVAL{m}{x}{C^{\MINUS m}})}
               \\
		\ASSERT{E \AND (B \IMPLIES C) \AND (m = \QQ{\cdot} \IMPLIES m = \QQ{x})}{N}{u}{D^{\MINUS xm}}
          \end{array}
        }
	{
		\ASSERT{A}{\LETQQ{x}{M}{N}}{u}{D}
	}{Unquote$^+$}
\end{RULES}
\caption{\PCFDP\ inference rules for total correctness.}\label{figure:logic:rules}
\end{FIGURE}

\PARAGRAPH{Rules} The rules of inference can be found in Figures
\ref{figure:logic:rules} and \ref{figure:logic:structuralRules}.  We
write $\vdash \ASSERT{A}{M}{m}{B}$ to indicate that
$\ASSERT{A}{M}{m}{B}$ is derivable using these rules.  Structural
rules like Hoare's rule of consequence, are standard (see
e.g.~\cite{HondaK:froproltop,HondaK:descriptive,HY04PPDP}) and used
without further comment. All rules are typed. The typing of rules
follows the corresponding typing of the programs occurring in the
judgements, but with additions to account for auxiliary
variables. Rather than detailing the typing for all rules, we exhibit
an example. The typing rule for the unquote-construct is this:
\[
	\TWOPREMISERULE
	{
		\TYPES{\Gamma; \Delta}{M}{\AST{\alpha}}
	}
        {
		\TYPES{\Gamma; \Delta, x : \alpha}{N}{\beta}
        }
	{
		\TYPES{\Gamma; \Delta}{\LETQQ{x}{M}{N}}{\beta}
	}
\]
The corresponding typing for [\RULENAME{Unquote}$^{+}$] is rather similar:
\[
	\ONEPREMISERULE
	{
          \begin{array}{c}
		\FORMULATYPES{\Gamma; \Delta; \QQ{\alpha}}{\ASSERT{A}{M}{m}{E^{\MINUS x} \AND (B^{\MINUS x} \IMPLIES \QQEVAL{m}{x}{C^{\MINUS m}})}} \\
		\FORMULATYPES{\Gamma, m : \QQ{\alpha}; \Delta, x : \alpha; \beta}{\ASSERT{E \AND (B \IMPLIES C) \AND (m = \QQ{\cdot} \IMPLIES m = \QQ{x})}{N}{u}{D^{\MINUS xm}}}
          \end{array}
        }
	{
		\FORMULATYPES{\Gamma; \Delta; \beta}{\ASSERT{A}{\LETQQ{x}{M}{N}}{u}{D}}
	}
\]

\NI All rules in Figure \ref{figure:logic:structuralRules} and most
rules in Figure \ref{figure:logic:rules} are standard and unchanged
from \cite{HondaK:froproltop,HondaK:descriptive,HY04PPDP} with three
significant exceptions, explained next.

[\RULENAME{Var}] adds $x \CONV$, i.e.~$\exists a.x = a$ in the
precondition.  By construction of our models, $x\CONV$ is trivially
true if $x$ is non-modal. If $x$ is modal, the situation is different
because $x$ may denote a non-terminating program. In this case
$x\CONV$ constrains $x$ so that it really denotes a value, as is
required in a logic for total correctness.

[\RULENAME{Quote}] says that $\QQ{M}$ always terminates (because the
conclusion's precondition is simply $\TRUTH$). Moreover, if $u$
denotes the result of evaluating $\QQ{M}$, i.e.~$\QQ{M}$ itself, then,
assuming $A$ holds (i.e., given the premise, if $M$ terminates), $u$
contains a terminating program, denoted $m$, making $B$ true. Clearly,
in a logic for total correctness, if $M$ is not a terminating program,
$A$ will be equivalent to $\FALSITY$, in which case,
[\RULENAME{Quote}] does not make a non-trivial assertion about
$\QQ{M}$ beyond stating that it terminates.

[\RULENAME{Unquote}$^+$] is similar to the usual rule for
$\LET{x}{M}{N}$ which is easily derivable using [\RULENAME{Abs},
  \RULENAME{App}]:
\[
	\TWOPREMISERULENAMEDRIGHT
	{
		\ASSERT{A}{M}{x}{B}
	}
	{
		\ASSERT{B}{N}{u}{C}
	}
	{
		\ASSERT{A}{\LET{x}{M}{N}}{u}{C}
	}{Let}
\]
The rule for $\LETQQ{x}{M}{N}$ is more difficult because a quasi-quote
always terminates, but the code it contains may not. Moreover, even if
$M$ evaluates to a quasi-quote containing a divergent program, the
overall expression may still terminate, because $N$ uses the
destructed quasi-quote in a way that cannot detect divergence.  An
example is as follows:
\[
	\LETQQ{x}{\QQ{\Omega}}{\lambda y.x}.
\]

\NI Our rule [\RULENAME{Unquote}$^+$] deals with this complication in the
following way.  Assume 
\[
   \ASSERT{A}{M}{m}{B \IMPLIES \QQEVAL{m}{x}{C}}
\]
holds. If $M$ evaluates to a quasi-quote containing a divergent
program, $B$ would be equivalent to $\FALSITY$.  This is because in a
logic for total correctness, $\QQEVAL{m}{x}{C}$ means that the
quasi-quote denoted by $m$ must contain a converging program. Hence
the only way that
\[
   B \IMPLIES \QQEVAL{m}{x}{C} 
\]
can be true if it doesn't is if $B$ is equivalent to $\FALSITY$.  In
this case $B \IMPLIES \QQEVAL{m}{x}{C}$ as a whole is equivalent to
$\TRUTH$, i.e.~conveys no information Hence, since $x$ does not occur
freely in $E$, the denotation of $x$ is not constrained by the left
premise, hence the termination behaviour of $N$ cannot depend on $x$.
In other words $N$ uses whatever $x$ denotes in a way that makes the
termination or otherwise of N independent of $x$.  The additional
formula $E$ enables us easily to carry information from the conclusion
of the assertion for $M$ to the premise of the assertion about $N$.

\begin{FIGURE}
\begin{RULES}
	\TWOPREMISERULENAMEDRIGHT
	{
		\ASSERT{A'}{M}{m}{B'}
	}
	{
		A \IMPLIES (A' \AND (B' \IMPLIES B))
	}
	{
		\ASSERT{A}{M}{m}{B}
	}{Conseq-Kl}
		\quad
	\ONEPREMISERULENAMEDRIGHT
	{
		\ASSERT{A \AND B}{V}{m}{C}
	}
	{
		\ASSERT{A}{V}{m}{B \IMPLIES C}
	}{$\AND$-$\IMPLIES$}
		\\\\
        \ONEPREMISERULENAMEDRIGHT
        {
          \ASSERT{\FALSITY \AND A}{M}{m}{B}
        }
        {
          \ASSERT{\FALSITY}{M}{m}{A \AND B}
        }{$\AND$-$\FALSITY$}
                 \quad
	\ONEPREMISERULENAMEDRIGHT
	{
		\ASSERT{A}{M}{m}{B \IMPLIES C}
	}
	{
		\ASSERT{A \AND B}{M}{m}{C}
	}{$\IMPLIES$-$\AND$}
		\quad
	\TWOPREMISERULENAMEDRIGHT
	{
		\ASSERT{A}{M}{m}{B}
	}
	{
		\ASSERT{A'}{M}{m}{B}
	}
	{
		\ASSERT{A \OR A'}{M}{m}{B}
	}{$\OR$-Pre}
		\\\\
	\TWOPREMISERULENAMEDRIGHT
	{
		\ASSERT{A}{M}{m}{B}
	}
	{
		\ASSERT{A}{M}{m}{B'}
	}
	{
		\ASSERT{A}{M}{m}{B \AND B'}
	}{$\AND$-Post}
		\quad
	\TWOPREMISERULENAMEDRIGHT
	{
		\ASSERT{A}{M}{m}{B^{\MINUS i}}
	}
	{
		\text{$i$ auxiliary}
	}
	{
		\ASSERT{\exists i.A}{M}{m}{B}
	}{Aux$_{\exists}$}
		\\\\
	\TWOPREMISERULENAMEDRIGHT
	{
		\ASSERT{A^{\MINUS i}}{M}{m}{B}
	}
	{
		\text{$i$ auxiliary}
	}
	{
		\ASSERT{A}{M}{m}{\forall i.B}
	}{Aux$_{\forall}$}
		\quad
	\ONEPREMISERULENAMEDRIGHT
	{
		\ASSERT{A}{M}{m}{B}
	}
	{
		\ASSERT{A \AND C}{M}{m}{B \AND C}
	}{Invar}

\end{RULES}
\caption{Structural rules for total correctness.}\label{figure:logic:structuralRules}
\end{FIGURE}

Finally, the requirement 
\[
  m = \QQ{\cdot} \IMPLIES m = \QQ{x}
\]
in the precondition of the assertion for $N$ makes the following fact
available for reasoning about $N$: whenever $M$ evaluates to a
quasi-quote $\QQ{M'}$, then $M'$ is bound to $x$. This fact is not
used in the reasoning about example programs in this paper. However,
it appears to be vital for proving completeness, see Proposition
\ref{completeness:proposition:tcapSoundness} in Section
\ref{completeness2}.\footnote{The previous, short version of this
  paper \cite{BergerM:comlogfamltmp} used only [\RULENAME{Unquote}],
  not [\RULENAME{Unquote}$^+$]. It is unclear if
  Prop.~\ref{completeness:proposition:tcapSoundness} can be
  established with [\RULENAME{Unquote}] alone.} In reasoning about
programs we typically use the following simpler rule.
\[
	\TWOPREMISERULENAMEDRIGHT
	{
		\ASSERT{A}{M}{m}{E^{\MINUS x} \AND (B^{\MINUS x} \IMPLIES \QQEVAL{m}{x}{C^{\MINUS m}})}
	}
        {
		\ASSERT{E \AND (B \IMPLIES C)}{N}{u}{D^{\MINUS xm}}
        }
	{
		\ASSERT{A}{\LETQQ{x}{M}{N}}{u}{D}
	}{Unquote}
\]

\NI In many derivations, $E$ is simply $\TRUTH$ and omitted. Clearly,
    [\RULENAME{Unquote}] is easily derivable from
    [\RULENAME{Unquote}$^+$].

The rule [\RULENAME{Conseq-Kl}] is slightly more elaborate than
Hoare's original rule of consequence, present already in~\cite{HOARE}, and
repeated below for comparison:

\[
	\THREEPREMISERULENAMEDRIGHT
	{
		A \IMPLIES A'
	}
	{
		\ASSERT{A'}{M}{m}{B'}
	}
	{
		B' \IMPLIES B
	}
	{
		\ASSERT{A}{M}{m}{B}
	}{Conseq}
\]

\NI
[\RULENAME{Conseq}] is usually sufficient in practise. But for proving
  relative completeness in Section \ref{completeness2},
  [\RULENAME{Conseq-Kl}], going back at least as far
    Kleymann~\cite{KleymannT:hoalogaav}, is more convenient.  Using
    [\RULENAME{Conseq-Kl}], Hoare's [\RULENAME{Conseq}] is easily
    derivable.

We note that the rules for programs in the \PCF-fragment of
\PCFDP\ are the same as those in the logic for \PCF~\cite{HY04PPDP},
apart from a slightly different presentation. The only apparent
difference is in the respective rules for variables (with \PCFDP\ on
the left, \PCF\ on the right):
\[
   \ASSERT{A\SUBST{x}{m} \AND x\CONV}{x}{m}{A}
      \qquad\qquad
   \ASSERT{A\SUBST{x}{m}}{x}{m}{A}
\]
However, this is misleading for two reasons:
\begin{itemize}

\item For non-modal variables $x$, by axiom $(term)$, $x\CONV$ always
  holds, so $A\SUBST{x}{m}$ can be inferred trivially from
  $A\SUBST{x}{m} \AND x\CONV$ and vice versa.

\item We could have split the \PCFDP\ rule for variables into two as
  follows:
  \[
        \ONEPREMISERULENAMEDRIGHT
        {
                x\ \text{non-modal}
        }
        {
                \ASSERT{\TRUTH}{x}{m}{x = m}
        }{Var}
                \quad
        \ONEPREMISERULENAMEDRIGHT
        {
                x\ \text{modal}
        }
        {
                \ASSERT{x \CONV}{x}{m}{x = m}
        }{Var$_m$}
  \]
  Indeed that is what we will do later in Section
  \ref{completeness2}. The reason for using a combined rule in this
  Section is economy of presentation.
\end{itemize}

\NI The ability to reason about the \PCF-fragment in our logic for
\PCFDP\ is significant for two reasons. First, on the theoretical side, it
shows that adding MP features is a modular extension
of our base language and base logic, raising the intriguing question
if modularity can be retained in situations where the base language
has rich effects like state or exceptions, or where the
MP features are more extensive (e.g.~compile-time
meta-programming), or allow MP on open code (that is, code with
free variables). Secondly, on the pragmatic side, it makes life
easier, because the specification and verification of programs and
program parts that do not use MP features do not
have to pay a price in terms of additional complexity vis-a-vis the
logic for \PCF.

\subsection{Soundness}

We now establish that the axioms and rules introduced in the previous subsection are sound.

\begin{thm}\label{logic:theorem:1}
\ 
\begin{enumerate}

\item\label{logic:theorem:1:1} All axioms are sound.
\item\label{logic:theorem:1:2} All rules are sound.

\end{enumerate}
\end{thm}

\NI Proofs for axioms and rules not relating to \PCFDP's
meta-programming extensions are straightforward extensions of the
corresponding proofs for \PCF-logics like
\cite{ALIASfull,HY04PPDP,HondaK:obscomplfihofTECREP,YHB07:local:full}
and mostly omitted.  Before embarking on proofs, we collect facts that
will be useful later.

\begin{prop}\label{app:soundness:9qw384}
\
\begin{enumerate}

\item\label{app:soundness:9qw384:1}
Assume the formula $A$ is typable under $\Gamma, x : \alpha; \Delta$
and $\Gamma; \Delta, x : \alpha$. Let $(\xi \cdot x : V, \sigma)$ be a
model of type $\Gamma, x : \alpha; \Delta$ and $(\xi, \sigma \cdot x :
V)$ be a model of type $\Gamma; \Delta, x : \alpha$. Then
\[
	(\xi \cdot x : V, \sigma) \models A
	     \qquad\text{iff}\qquad
	(\xi, \sigma \cdot x : V) \models A.
\]

\item\label{app:soundness:9qw384:3}Let $\EXPRESSIONTYPES{\Gamma_i; \Delta_i}{e}{\alpha}$,
  $\FORMULATYPES{\Gamma_i; \Delta_i}{A}$, and assume that $\eta_i$ is a
  $\Gamma_i; \Delta_i$-model for $i = 1, 2$. Then:
  \begin{itemize}

    \item $\eta_1(x) \CONG \eta_2(x)$ for all $x \in \FV{e}$ implies
      $\SEMB{e}_{\eta_1} \CONG \SEMB{e}_{\eta_2}$.

    \item $\eta_1(x) \CONG \eta_2(x)$ for all $x \in \FV{A}$ implies
      $\eta_1 \models A$ iff $\eta_2 \models A$.

  \end{itemize}

\item\label{app:soundness:9qw384:2}
Let $\EXPRESSIONTYPES{\Gamma; \Delta}{e}{\alpha}$,
$\EXPRESSIONTYPES{\Gamma, x : \alpha; \Delta}{e'}{\beta}$,
$\FORMULATYPES{\Gamma, x : \alpha; \Delta}{A}$ and 
 $\eta \DEFEQ (\xi, \sigma)$ be an appropriately typed model such that $\SEMB{e}_{\eta} \CONV$. 
\begin{itemize}

\item $\SEMB{e'\SUBST{e}{x}}_{\eta} \CONG
\SEMB{e'}_{(\xi \cdot x : \SEMB{e}_{\eta}, \sigma)}$.

\item $\eta \models A\SUBST{e}{x}$ iff  $\forall V.(V \CONG
  \SEMB{e}_{\xi, \sigma} \IMPLIES (\xi \cdot x : V, \sigma) \models
  A)$.

\item $\eta \models A\SUBST{e}{x}$ iff  $\eta \models \exists x.(A \AND x = e).$

\end{itemize}

\item\label{app:soundness:9qw384:4}
Let $\TYPES{\Gamma; \Delta, x : \alpha}{M}{\beta}$, assume $N$ is a
closed program of type $\alpha$. Given a $\Gamma; \Delta$-model $\eta \DEFEQ (\xi, \sigma)$
it holds that
\[
	M\eta\SUBST{N}{x} = M(\xi, \sigma \cdot x : N).
\]

\end{enumerate}
\end{prop}
\begin{proof}
The content of this proposition is straightforward, hence proofs are
omitted.
\end{proof}

%\PARAGRAPH{Proof of Theorem \ref{logic:theorem:1}.\ref{logic:theorem:1:1}}
\noindent\emph{Proof} \relax[\hspace{1 pt}of Theorem \ref{logic:theorem:1}.\ref{logic:theorem:1:1}].
The proof of axioms $(q1)$ - $(q5)$ for quasi-quotes are essentially
just trivial instances of first-order logical laws and omitted, except
that we explicitly prove $(q3)$ as a representative example.

To establish $(q3)$, let $\eta \DEFEQ (\xi, \sigma)$ be an
appropriately typed model. Then we reason as follows:
\begin{NDERIVATION}{1}
	\NLINE{\eta \models \QQEVAL{x}{y}{A} \AND \neg \QQEVAL{x}{y}{B}}{}
	\NLINE{\exists M, V. \eta(x) = M, M \CONV V, (\xi, \sigma \cdot y : V) \models A}{1}
	\NLINE{\eta(x) = M, M \CONV V}{2}
	\NLINE{(\xi, \sigma \cdot y : V) \not \models B}{1, 3}
	\NLASTLINE{(\xi, \sigma \cdot y : V) \models A \AND \neg B}{2, 4}
\end{NDERIVATION}
The reverse implication is similar.

The axiom $(term)$ is immediate from the definition of models.

Regarding $(term_q)$, let  $\eta \DEFEQ (\xi, \sigma)$ be an
appropriately typed model.
\begin{NDERIVATION}{1}
	\NLINE{\eta \models \QQEVAL{x}{m}{A}}{}
	\NLINE{\exists M, V. \eta(x) = M, M \CONV V, (\xi, \sigma \cdot m : V) \models A}{1}
	\NLINE{\exists M, V. \eta(x) = M, M \CONV V, (\xi \cdot y : V, \sigma \cdot m : V) \models m = y }{}
	\NLINE{\exists M, V. \eta(x) = M, M \CONV V, (\xi,  \sigma \cdot m : V) \models \exists y. m = y}{3}
	\NLINE{\exists M, V. \eta(x) = M, M \CONV V, (\xi,  \sigma \cdot m : V) \models m\CONV}{4}
	\NLINE{\exists M, V. \eta(x) = M, M \CONV V, (\xi,  \sigma \cdot m : V) \models A \AND m\CONV}{2, 5}
	\NLASTLINE{(\xi, \sigma) \models \QQEVAL{x}{m}{ A \AND m\CONV}}{6}
\end{NDERIVATION}
The reverse implication is immediate.

The soundness of $(div)$ is immediate from the construction of the
model and the satisfaction relation: $(div)$ states that not every
quasi-quote contains a terminating program, e.g.~$\QQ{\Omega}$.

Finally, for $(ext_q)$, let $\eta \DEFEQ (\xi, \sigma)$ be an
appropriately typed model, where $x, y \in \DOM{\eta}$. We assume that
$x, y$ are of type $\QQ{\alpha}$. 
\begin{NDERIVATION}{1}
	\NLINE{\eta \models x = y}{}
        \NLINE{\eta' \DEFEQ (\xi \cdot a : V, \sigma)}{V\ \text{arbitrary value}}
	\NLINE{\eta' \models \QQEVAL{x}{m}{m = a}}{Assumption}
	\NLINE{
          \begin{array}[t]{l}
            \exists M, W. \SEMB{x}_{\eta'} = \QQ{M}, M \CONV W, \eta'' \models m = a, \\
            \text{where}\ \eta'' \DEFEQ (\xi \cdot a : V, \sigma \cdot m : W)
          \end{array}
        }{3}
        \NLINE{W = \SEMB{m}_{\eta''} \cong \SEMB{a}_{\eta''} = V}{2, 4}
        \NLINE{\SEMB{x}_{\eta} \CONG \SEMB{y}_{\eta}}{1}
        \NLINE{\SEMB{x}_{\eta'} \CONG \SEMB{y}_{\eta'}}{6, Prop.~\ref{app:soundness:9qw384}.\ref{app:soundness:9qw384:3}}
	\NLINE{\begin{array}[t]{l}
            \exists M', W'. \SEMB{y}_{\eta'} = \QQ{M'}, M' \CONV W', \eta''' \models m = a,\\
            \text{where}\ \eta''' \DEFEQ (\xi \cdot a : V, \sigma \cdot m : W')
          \end{array}
        }{4, 7, Prop.~\ref{lamguage:theorem:1}}
	\NLASTLINE{\eta' \models \QQEVAL{y}{m}{m = a}}{8}
\end{NDERIVATION}

\NI For the reverse implication we assume that $x$ and $y$ are both of
type $\QQ{\alpha}$, and that $\eta = (\xi, \sigma)$.
\[
   \eta \models \EXTQ{xy}
\]
which means there are two cases for any chosen value $V$ of
appropriate type, where $\eta' = (\xi \cdot a : V, \sigma)$.
\begin{itemize}

\item $\SEMB{x}_{\eta'} \Downarrow \QQ{M}, M \Downarrow W, W \CONG V$, and also
$\SEMB{y}_{\eta'} \Downarrow \QQ{M'}, M' \Downarrow W', W' \CONG V$. By
Proposition  \ref{lamguage:theorem:1}.\ref{lamguage:theorem:1:1} this implies
\[
   \SEMB{x}_{\eta'} \CONG\QQ{M}, M \CONG W, W \CONG V
\]
and
\[
   \SEMB{y}_{\eta'} \CONG \QQ{M'}, M' \CONG W', W' \CONG V.
\]
Hence $W \CONG W'$.  As $\CONG$ is a congruence, the above in turn
implies $\SEMB{x}_{\eta'} \CONG \SEMB{y}_{\eta'}$, whence
\[
   \eta \models x = y
\]
using the fact that $a$ is different from $x, y$ and
Proposition \ref{app:soundness:9qw384}.\ref{app:soundness:9qw384:3}.

\item $\SEMB{x}_{\eta'} \Uparrow$ and $\SEMB{y}_{\eta'} \Uparrow$. In
  this case $\SEMB{x}_{\eta'} \CONG \SEMB{y}_{\eta'}$ by Proposition
  \ref{lamguage:theorem:1}.\ref{lamguage:theorem:1:4}. Hence again
  \[
     \eta \models x = y
  \]
  as in the previous case.\qed

\end{itemize}

%\PARAGRAPH{Proof of Theorem \ref{logic:theorem:1}.\ref{logic:theorem:1:2}}
\noindent\emph{Proof} \relax[\hspace{1 pt}of Theorem
  \ref{logic:theorem:1}.\ref{logic:theorem:1:2}]. 
We proceed by induction on the derivation of the inference.
All rules but [\RULENAME{Var, Quote, Unquote$^+$}] are essentially
unchanged from ~\cite{HY04PPDP} so we concentrate on these three.

For [\RULENAME{Var}] we have two subcases, depending on whether the
variable under assertion is modal or not. The latter case is trivial,
so we deal only with the former (which is also easy). In this case the
rule is typed as follows:
\[
	\ZEROPREMISERULE
        {
		\FORMULATYPES{\Gamma; \Delta, x : \alpha; \alpha}{\ASSERT{A\SUBST{x}{m} \AND x\CONV}{x}{m}{A}}
        }
\]
Let $\eta \DEFEQ (\xi, \sigma \cdot x : M)$ be a $\Gamma; \Delta, x :
\alpha$-model such that
\begin{align}\label{app:soundness:s8sdfg}
	\eta \models A\SUBST{x}{m} \AND x\CONV.
\end{align}
Then in particular
\[
	\eta \models \exists a^{\alpha}. x = a.
\]
By the semantics of quantification, we know that some \emph{value} $V$
must exist with $(\xi \cdot a : V, \sigma \cdot x : M) \models x =
a$. Thus by definition of the interpretation of equality,
\[
	M \CONG V.
\]
Hence, since $x\eta = M$  immediately
\[
	x\eta \CONV V
\]
which means the program under assertion terminates.
From (\ref{app:soundness:s8sdfg}) we also get that
\[
	\eta \models A\SUBST{x}{m}
\]
and since $\SEMB{x}_{\eta} \CONV V$, we can apply Proposition
\ref{app:soundness:9qw384}.\ref{app:soundness:9qw384:2} and obtain.
\[
	 (\xi \cdot m : V, \sigma \cdot x : M)\models A.
\]
This concludes our discussion of [\RULENAME{Var}].

The case of [\RULENAME{Quote}] is straightforward and based on Proposition
\ref{app:soundness:9qw384}.\ref{app:soundness:9qw384:1}. The rule is typed as follows.
\[
	\ONEPREMISERULE
	{
		\FORMULATYPES{\Gamma; \Delta; \alpha}{\ASSERT{A}{M}{m}{B}}
	}
	{
		\FORMULATYPES{\Gamma; \Delta; \QQ{\alpha}}{\ASSERT{\TRUTH}{\QQ{M}}{u}{A \IMPLIES \QQEVAL{u}{m}{B}}}
	}
\]
Let $\eta = (\xi, \sigma)$ be a $\Gamma; \Delta$-model and define
$\eta' \DEFEQ (\xi \cdot u : \QQ{M\eta}, \sigma)$.  Since $\QQ{M}\eta = \QQ{M\eta}$ is
already a value, we need only show that $\eta' \models A \IMPLIES
\QQEVAL{u}{m}{B}$.  To this end, let $\eta' \models A$.  Then we reason as
follows (in this derivation and others, (IH) is used as an abbreviation of
`induction hypothesis'):
\begin{NDERIVATION}{1}
  \NLINE{\eta' \models A}{}
  \NLINE{\eta \models A}{1, $u \notin \FV{A}$, Prop.~\ref{app:soundness:9qw384}.\ref{app:soundness:9qw384:3}}
  \NLINE{M\eta \CONV V\ \text{and}\ (\xi \cdot m : V, \sigma) \models B}{(IH), 2}
  \NLINE{(\xi, \sigma \cdot m : V) \models B}{3, Prop.~\ref{app:soundness:9qw384}.\ref{app:soundness:9qw384:1}}
  \NLINE{M\eta \CONG V}{3, Prop.~\ref{lamguage:theorem:1}.\ref{lamguage:theorem:1:1}}
  \NLINE{(\xi, \sigma \cdot m : M\eta) \models B}{4, 5, Prop.~\ref{app:soundness:9qw384}.\ref{app:soundness:9qw384:3}}
  \NLASTLINE{(\xi \cdot u : \QQ{M\eta}, \sigma \cdot m : M\eta) \models B}{6, $u \notin \FV{B}$, Prop.~\ref{app:soundness:9qw384}.\ref{app:soundness:9qw384:3}}
\end{NDERIVATION}
This concludes the case of [\RULENAME{Quote}].

Finally, we establish the soundness of [\RULENAME{Unquote$^+$}].
Choose a $\Gamma; \Delta$-model $\eta \DEFEQ (\xi, \sigma)$ such that
$\eta \models A$. By the (IH) we know that 
\begin{align}\label{app:soundness:394859}
	M\eta \CONV \QQ{M'}
	      \qquad\qquad
	\underbrace{(\xi \cdot m : \QQ{M'}, \sigma)}_{\eta'} \models E \AND (B \IMPLIES \QQEVAL{m}{x}{C})
\end{align}
Now we have two cases:
\begin{itemize}

\item\label{app:soundness:24358:1} $\eta' \models E$, $\eta' \models
  B$ and $\eta' \models \QQEVAL{m}{x}{C}$.

\item\label{app:soundness:24358:2} $\eta' \models E$ but $\eta'
  \not\models B$.

\end{itemize}

\NI We start with the former:

\begin{NDERIVATION}{1}
  \NLINE{\eta' \models E\ \text{and}\ \eta' \models B\ \text{and}\ \eta' \models \QQEVAL{m}{x}{C}}{}
  \NLINE{\SEMB{m}_{\eta'} = \QQ{M'}, M' \CONV V\ \text{and}\ (\xi \cdot m : \QQ{M'}, \sigma \cdot x : V) \models C}{1}
  \NLINE{\eta'' \DEFEQ (\xi \cdot m : \QQ{M'}, \sigma \cdot x : V)}{}
  \NLINE{\eta'' \models E \AND B}{1, Prop.~\ref{app:soundness:9qw384}.\ref{app:soundness:9qw384:3}}
  \NLINE{\eta'' \models E \AND B \AND C}{2, 4}
  \NLINE{\eta'' \models E \AND (B \IMPLIES C)}{5}
  \NLINE{\eta''' \DEFEQ (\xi \cdot m : \QQ{M'}, \sigma \cdot x : M')}{}
  \NLINE{M' \CONG V}{2, Prop.~\ref{lamguage:theorem:1}.\ref{lamguage:theorem:1:1}}
  \NLINE{\eta''' \models E \AND (B \IMPLIES C)}{6, 8, Prop.~\ref{app:soundness:9qw384}.\ref{app:soundness:9qw384:3}}
  \NLINE{\eta''' \models m = \QQ{\cdot} \AND m = \QQ{x}}{2, 8, Prop.~\ref{app:soundness:9qw384}.\ref{app:soundness:9qw384:3}}
  \NLINE{\eta''' \models m = \QQ{\cdot} \IMPLIES m = \QQ{x}}{10}
  \NLASTLINE{N\eta''' \CONV W\ \text{and}\ (\xi \cdot m : \QQ{M'} \cdot u : W, \sigma \cdot x : M') \models D}{11, (IH)}
\end{NDERIVATION}
Now we can consider the reductions of $(\LETQQ{x}{M}{N})\eta$:
\[
\begin{array}{lclcr}
	(\LETQQ{x}{M}{N})\eta
		&\ \quad=\quad\ &
	\LETQQ{x}{M\eta}{N\eta}
		\\
		&\NRED&
	\LETQQ{x}{\QQ{M'}}{N\eta}
		&\qquad&
		by\ (\ref{app:soundness:394859})
		\\
		&\RED&
	N\eta\SUBST{M'}{x}
		\\
		&=&
	N\eta'''
		&&
	Prop.~~\ref{app:soundness:9qw384}.\ref{app:soundness:9qw384:4}
		\\
		&\CONV&
	W
		&&
	by\ 10
\end{array}
\]
By Line 10 and Proposition
\ref{app:soundness:9qw384}.\ref{app:soundness:9qw384:3} we know that
$(\xi \cdot u : W, \sigma) \models D$ since $m, x
\notin \FV{D}$.

Now we consider the second case $\eta' \not \models B$. 
In this simpler case we reason as follows:
\begin{NDERIVATION}{1}
	\NLINE{\eta' \models E\ \text{and}\ \eta' \not \models B}{}
	\NLINE{\eta'' \DEFEQ (\xi \cdot m : \QQ{M'}, \sigma \cdot x : M')}{}
	\NLINE{\eta'' \models E\ \text{and}\ \eta'' \not \models B}{1, Prop.~\ref{app:soundness:9qw384}.\ref{app:soundness:9qw384:3}}
	\NLINE{\eta'' \models E \AND (B \IMPLIES C)}{3}	
        \NLINE{\eta'' \models m = \QQ{\cdot} \IMPLIES m = \QQ{x}}{by construction}
        \NLINE{\eta'' \models E \AND (B \IMPLIES C) \AND (m = \QQ{\cdot} \IMPLIES m = \QQ{x})}{4, 5}
	\NLASTLINE{N\eta'' \Downarrow W\ \text{and}\ (\xi \cdot m : \QQ{M'} \cdot u : W, \sigma \cdot x : M') \models D}{6, (IH)}
\end{NDERIVATION}
As in the previous case we now consider the reductions of
 $(\LETQQ{x}{M}{N})\eta$:
\[
\begin{array}{lclcr}
	(\LETQQ{x}{M}{N})\eta
		&\ \quad=\quad\ &
	\LETQQ{x}{M\eta}{N\eta}
		\\
		&\NRED&
	\LETQQ{x}{\QQ{M'}}{N\eta}
		&\qquad&
		by\ (\ref{app:soundness:394859})
		\\
		&\RED&
	N\eta\SUBST{M'}{x}
		\\
		&=&
	N\eta''
		&&
	Prop.~\ref{app:soundness:9qw384}.\ref{app:soundness:9qw384:4}
		\\
		&\CONV&
	W
		&&
	by\ 7
\end{array}
\]
By Line 7 and Proposition
\ref{app:soundness:9qw384}.\ref{app:soundness:9qw384:3} we know that
$(\xi \cdot u : W, \sigma) \models D$ since $m, x
\notin \FV{D}$.  This concludes the reasoning for
[\RULENAME{Unquote$^+$}].

We now verify the structural rule [\RULENAME{$\AND$-$\IMPLIES$}] to
demonstrate that the soundness of the structural rules (which have already been
shown sound in the context of \PCF~\cite{HY04PPDP}) is not affected
by our extension of \PCF\ with facilities for meta-programming. The
proofs for the other structural rules are similarly straightforward.

Let $\eta = (\xi, \sigma)$ be an appropriately typed model. Then we
reason as follows:
\begin{NDERIVATION}{1}
  \NLINE{\eta \models A}{} 
  \NLINE{V\eta \Downarrow V\eta}{V value}
  \NLINE{\eta' \DEFEQ (\xi \cdot m : V\eta, \sigma)}{}
  \NLINE{\eta' \models B}{Assumption}
  \NLINE{\eta \models B}{4, $m \notin \FV{B}$, Prop.~\ref{app:soundness:9qw384}.\ref{app:soundness:9qw384:3}}
  \NLINE{\eta \models A \AND B}{1, 5}
  \NLINE{V\eta \Downarrow V\eta, \eta' \models C}{V value, 6, (IH)}
  \NLINE{V\eta \Downarrow V\eta, \eta' \models B \IMPLIES C}{4, 7}
  \NLASTLINE{\eta \models \ASSERT{A}{V}{m}{B \IMPLIES C}}{1, 8}
\end{NDERIVATION}

\NI Note that as with \PCF, the condition that the program be a value
in [\RULENAME{$\AND$-$\IMPLIES$}] cannot be dropped in a logic for
total correctness, because
\[
   	\ONEPREMISERULE
	{
		\ASSERT{\TRUTH \AND \FALSITY}{\Omega}{m}{C}
	}
	{
		\ASSERT{\TRUTH}{\Omega}{m}{\FALSITY \IMPLIES C}
	}
\]
is unsound.

Finally, we prove sound [\RULENAME{Rec}], the rule for recursion in a
total correctness setting. It's compellingly simple form was first
given in~\cite{HondaK:descriptive}. Here we need to show that the
addition of MP facilities does not void soundness. This
is straightforward.

Let $\eta = (\xi, \sigma)$ be an appropriately typed model with $g
\notin \DOM{\eta}$, and $\eta' \DEFEQ (\xi \cdot g : \mu g.M\eta, \sigma)$.
\begin{NDERIVATION}{1}
  \NLINE{\eta \models A}{Assumption}
  \NLINE{\eta' \models A}{1, $g \notin \FV{A}$, Prop.~\ref{app:soundness:9qw384}.\ref{app:soundness:9qw384:3}}
  \NLINE{M\eta' \Downarrow V,  \quad \overbrace{(\xi \cdot g : \mu g.M\eta, u : V\sigma)}^{\eta''} \models B}{2, (IH)}
  \NLASTLINE{V \cong M\eta' \cong \mu g.M\eta}{3, Prop.~\ref{lamguage:theorem:1}.\ref{lamguage:theorem:1:3}}
\end{NDERIVATION}

\begin{NDERIVATION}{5}
  \NLINE{\eta'' \models g = u}{3, 4}
  \NLINE{\eta'' \models B \AND g = u}{5}
  \NLINE{(\xi \cdot u : V, \sigma) \models \exists g.(B \AND g = u)}{6}
  \NLINE{(\xi \cdot u : V, \sigma) \models B\SUBST{u}{g}}{7, Prop.~\ref{app:soundness:9qw384}.\ref{app:soundness:9qw384:2}}
  \NLINE{(\xi \cdot u : \mu g.M\eta, \sigma) \models B\SUBST{u}{g}}{8, Prop.~\ref{app:soundness:9qw384}.\ref{app:soundness:9qw384:3}}
  \NLINE{(\xi \cdot u : \mu g.M\eta, \sigma) \models B}{3, $g \notin \FV{B}$, Prop.~\ref{app:soundness:9qw384}.\ref{app:soundness:9qw384:3}}
  \NLASTLINE{\mu g.M\eta \Downarrow \mu g.M\eta, \quad (\xi \cdot u : \mu g.M\eta, \sigma) \models B}{10}
\end{NDERIVATION}

\NI This last proof is unchanged from the soundness proof for
    [\RULENAME{Rec}] in \PCF, although `under the hood' some used
    propositions need additional work to do with the generalised
    language and notion of model. This is also true for all other
    rules and axioms that involve only \PCF\ syntax.\qed

\section{Reasoning examples}\label{examples}

\NI We now put our logic to use by reasoning about some of the
programs introduced in Section \ref{language}.  The derivations use the
abbreviations of Section \ref{logic} and and often omit steps that are
trivial, or irrelevant from the perspective of meta-programming.

\begin{exa}%\PARAGRAPH{Example 1}
We begin with the simple program $\ASSERT{\TRUTH}{\QQ{1 + 2}}{m}{m = \QQ{3}}$.
The derivation is straightforward.

\begin{NDERIVATION}{1}
  \NLINE{\ASSERT{\TRUTH}{1 + 2}{a}{a = 3}}{}
  \NLINE{\ASSERT{\TRUTH}{\QQ{1 + 2}}{m}{\TRUTH \IMPLIES \QQEVAL{m}{a}{a = 3}}}{\RULENAME{Quote}, 1}
  \NLASTLINE{\ASSERT{\TRUTH}{\QQ{1 + 2}}{m}{m = \QQ{3}}}{\RULENAME{Conseq}, 2}
\end{NDERIVATION}
\end{exa}

\begin{exa}%\PARAGRAPH{Example 2}
This example deals with the code of a non-terminating program. We derive
$\ASSERT{\TRUTH}{\QQ{\Omega}}{m}{\TRUTH}$. This is the strongest total correctness assertion 
about $\QQ{\Omega}$. In the proof, we
assume that $\ASSERT{\FALSITY}{\Omega}{a}{\TRUTH}$ is derivable, which
is easy to show.

\begin{NDERIVATION}{1}
  \NLINE{\ASSERT{\FALSITY}{\Omega}{a}{\TRUTH}}{}
  \NLINE{\ASSERT{\TRUTH}{\QQ{\Omega}}{m}{\FALSITY \IMPLIES \QQEVAL{m}{a}{\TRUTH}}}{\RULENAME{Quote}, 1}
  \NLASTLINE{\ASSERT{\TRUTH}{\QQ{\Omega}}{m}{\TRUTH}}{\RULENAME{Conseq}, 2}
\end{NDERIVATION}
\end{exa}

\begin{exa}%\PARAGRAPH{Example 3}
The third example destructs a quasi-quote and then injects the resulting program into
another quasi-quote. 
\[
	\ASSERT{\TRUTH}{\LETQQ{x}{\QQ{1 + 2}}{\QQ{x + 3}}}{m}{m = \QQ{6}}
\]
We derive the assertion in small steps to demonstrate how to apply our
logical rules.

\begin{NDERIVATION}{1}
  \NLINE{\ASSERT{\TRUTH}{\QQ{1 + 2}}{m}{m = \QQ{3}}}{Ex.~1}
  \NLINE{\ASSERT{(a = 3)\SUBST{x}{a} \AND x\CONV}{x}{a}{a = 3}}{\RULENAME{Var}}
  \NLINE{\ASSERT{x = 3 \AND x\CONV}{x}{a}{a = 3}}{\RULENAME{Conseq}, 2}
  \NLINE{\ASSERT{\TRUTH}{3}{b}{b = 3}}{\RULENAME{Const, Conseq}}
  \NLINE{\ASSERT{a = 3 \AND x\CONV}{3}{b}{a = 3 \AND b = 3}}{\RULENAME{Invar}, 4}
  \NLINE{\ASSERT{a = 3}{3}{b}{(c = 6)\SUBST{a + b}{c}}}{\RULENAME{Conseq}, 5}
  \NLINE{\ASSERT{x = 3 \AND x\CONV}{x + 3}{c}{c = 6}}{\RULENAME{Add}, 3, 6}
  \NLINE{\ASSERT{\TRUTH}{\QQ{x + 3}}{u}{(x = 3 \AND x\CONV) \IMPLIES \QQEVAL{u}{c}{c = 6}}}{\RULENAME{Quote}, 7}
  \NLINE{\ASSERT{x = 3 \AND x\CONV}{\QQ{x + 3}}{u}{\QQEVAL{u}{c}{c = 6}}}{$\IMPLIES$-$\AND$, 8}
  \NLINE{\ASSERT{\TRUTH}{\QQ{1 + 2}}{m}{\TRUTH \IMPLIES \QQEVAL{m}{x}{x = 3 \AND x\CONV}}}{\RULENAME{Conseq}, 1}
  \NLINE{\ASSERT{\TRUTH \IMPLIES (x = 3 \AND x\CONV)}{\QQ{x + 3}}{u}{u = \QQ{6}}}{\RULENAME{Conseq}, 9}
  \NLASTLINE{\ASSERT{\TRUTH}{\LETQQ{x}{\QQ{1 + 2}}{\QQ{x + 3}}}{u}{u = \QQ{6}}}{\RULENAME{Unquote}, 10, 11}
\end{NDERIVATION}

\NI In Line 10, we use the $(term_q)$ axiom among others.
\end{exa}

\begin{exa}%\PARAGRAPH{Example 4}
We now show that when a quasi-quote containing a non-terminating sub-program
is destructed, but the resulting sub-program is not used, the overall program
still terminates.
This reflects the operational semantics in Section \ref{language}.
\[
	\ASSERT{\TRUTH}{\LETQQ{x}{\QQ{\Omega}}{\QQ{1 + 2}}}{m}{m = \QQ{3}}
\]
The derivation follows:

\begin{NDERIVATION}{1}
  \NLINE{\ASSERT{\TRUTH}{\QQ{\Omega}}{m}{\TRUTH}}{Ex.~2}
  \NLINE{\ASSERT{\TRUTH}{\QQ{\Omega}}{m}{\FALSITY \IMPLIES \QQEVAL{m}{a}{\TRUTH}}}{\RULENAME{Conseq}, 1}
  \NLINE{\ASSERT{\TRUTH}{\QQ{1 + 2}}{m}{m = \QQ{3}}}{Ex.~1}
  \NLINE{\ASSERT{\FALSITY \IMPLIES \TRUTH}{\QQ{1 + 2}}{m}{m = \QQ{3}}}{\RULENAME{Conseq}, 3}
  \NLASTLINE{\ASSERT{\TRUTH}{\LETQQ{x}{\QQ{\Omega}}{\QQ{1 + 2}}}{m}{m = \QQ{3}}}{\RULENAME{Unquote}, 2, 4}
\end{NDERIVATION}
\end{exa}

\begin{exa}%\PARAGRAPH{Example 5}
This example extracts a non-terminating program from a quasi-quote, and
injects it into a new quasi-quote. Our total-correctness logic cannot
say anything non-trivial about the resulting quasi-quote (cf.~Example
2):
\[
	\ASSERT{\TRUTH}{\LETQQ{x}{\QQ{\Omega}}{\QQ{x}}}{u}{\TRUTH}
\]
The derivation is straightforward.

\begin{NDERIVATION}{1}
  \NLINE{\ASSERT{\TRUTH}{\QQ{\Omega}}{m}{\TRUTH}}{Ex.~2}
  \NLASTLINE{\ASSERT{\TRUTH}{\QQ{\Omega}}{m}{\FALSITY \IMPLIES \QQEVAL{m}{x}{\TRUTH}}}{1, \RULENAME{Conseq}}
\end{NDERIVATION}
\begin{NDERIVATION}{3}
  \NLINE{\ASSERT{\FALSITY\SUBST{x}{a} \AND x\CONV}{x}{a}{\FALSITY}}{\RULENAME{Var}}
  \NLINE{\ASSERT{\FALSITY}{x}{a}{\TRUTH}}{\RULENAME{Conseq}, 3}
  \NLINE{\ASSERT{\TRUTH}{\QQ{x}}{u}{\FALSITY \IMPLIES \QQEVAL{u}{a}{\TRUTH}}}{\RULENAME{Quote}, 4}
  \NLINE{\ASSERT{\FALSITY \IMPLIES \TRUTH}{\QQ{x}}{u}{\TRUTH}}{\RULENAME{Conseq}, 5}
  \NLASTLINE{\ASSERT{\TRUTH}{\LETQQ{x}{\QQ{\Omega}}{\QQ{x}}}{u}{\TRUTH}}{\RULENAME{Unquote}, 2, 6}
\end{NDERIVATION}

\NI The examples below make use of the following convenient forms of the
recursion rule and [\RULENAME{Unquote}]. 
\[
	\ONEPREMISERULENAMEDRIGHT
	{
		\ASSERT{A^{\MINUS gn} \AND \forall 0 \leq i < n. B\SUBST{i}{n}\SUBST{g}{u}}{\lambda x.M}{u}{B^{\MINUS g}}
	}
	{
		\ASSERT{A}{\mu g.\lambda x.M}{u}{\forall n \geq 0.B}
	}{Rec'}
\]

\vspace{2mm}
\NI It is easily derived from [\RULENAME{Rec}] using
    [\RULENAME{Aux$_\forall$}].
\end{exa}

\begin{exa}%\PARAGRAPH{Example 6}
We now reason about $\LIFT{\INT}$ from Section~\ref{language}.  In the
proof we assume that $i, n$ range over non-negative integers.  Let
$A_{n}^{u} \DEFEQ \ONEEVAL{u}{n}{m}{m = \QQ{n}}$.  We now establish the
following assertion from Section~\ref{logic}:
$\ASSERT{\TRUTH}{\LIFT{\INT}}{u}{\forall n. A_n^u}$.  We set $C \DEFEQ
i \leq n \AND \forall j < n.A_j^g$, $D \DEFEQ i > 0 \AND \forall r.(0
\leq r < n \IMPLIES \ONEEVAL{g}{r}{m}{m = \QQ{r}})$ and $P \DEFEQ
\LETQQ{x}{g(i-1)}{\QQ{x+1}}$.  \raggedbottom
\begin{NDERIVATION}{1}
	\NLINE{\ASSERT{C}{i \leq 0}{b}{C \AND (b = \TRUE \equiv i \leq 0)}}{}
	\NLINE{\ASSERT{\TRUTH}{\QQ{0}}{m}{m = \QQ{0}}}{Like Ex.~1}
	\NLINE{\ASSERT{i = 0}{\QQ{0}}{m}{m = \QQ{i}}}{\RULENAME{Invar, Conseq}, 2}
	\NLINE{\ASSERT{(C \AND b = \TRUE \equiv i \leq 0)\SUBST{\TRUE}{b}}{\QQ{0}}{m}{m = \QQ{i}}}{\RULENAME{Conseq}, 3}
	\NLINE{\ASSERT{x = i-1}{\QQ{x + 1}}{m}{m = \QQ{i}}}{Like Ex.~3}
	\NLINE{\ASSERT{\TRUTH \IMPLIES x = i-1}{\QQ{x + 1}}{m}{m = \QQ{i}}}{\RULENAME{Conseq}, 5}
	\NLINE{\ASSERT{(C \AND b = \TRUE \equiv i \leq 0)\SUBST{\FALSE}{b}}{g}{s}{D}}{\RULENAME{Var}}
        \NLINE{\ASSERT{D}{i-1}{r}{\ONEEVAL{g}{r}{t}{t = \QQ{i-1}}}}{}
	\NLINE{\ASSERT{(C \AND b = \TRUE \equiv i \leq 0)\SUBST{\FALSE}{b}}{g( i - 1 )}{t}{t = \QQ{i-1}}}{\RULENAME{App}, 7, 8}
 	\NLINE{\ASSERT{(C \AND b = \TRUE \equiv i \leq 0)\SUBST{\FALSE}{b}}{P}{m}{m = \QQ{i}}}{\RULENAME{Unquote, Conseq}, 6, 9}
 	\NLINE{\ASSERT{C}{\IFTHENELSE{i \leq 0}{\QQ{0}}{P}}{m}{m = \QQ{i}}}{\RULENAME{If}, 4, 10}
	\NLINE{\ASSERT{\TRUTH}{\lambda i.\IFTHENELSE{i \leq 0}{\QQ{0}}{P}}{u}{\forall i.(C \IMPLIES  A^u_i)}}{\RULENAME{Abs}, 11}
	\NLINE{\ASSERT{\forall j < n.A_j^g}{\lambda i.\IFTHENELSE{i \leq 0}{\QQ{0}}{P}}{u}{\forall i \leq n.A^u_i}}{\RULENAME{Conseq} $\IMPLIES$-$\AND$, 12}
	\NLINE{\ASSERT{\TRUTH}{\LIFT{\INT}}{u}{\forall n. \forall i \leq n. A_n^u}}{\RULENAME{Rec'}, 13}
	\NLASTLINE{\ASSERT{\TRUTH}{\LIFT{\INT}}{u}{\forall n. A_n^u}}{\RULENAME{Conseq}, 14}
\end{NDERIVATION}
\end{exa}

\begin{exa}%\PARAGRAPH{Example 7}
We close this section by reasoning about the staged power function
from Section \ref{language}. Assuming that $i, j, k, n$ range
over non-negative integers, we define $B^u_n \DEFEQ
\ONEEVAL{u}{n}{m}{\QQEVAL{m}{y}{\forall j. y \bullet j = j^n}}$.  In
the derivation, we provide less detail than in previous proofs for
readability. 
\begin{NDERIVATION}{1}
	\NLINE{C \quad\DEFEQ\quad n \leq k \AND \forall i < k.B_i^p \qquad D \quad\DEFEQ\quad C \AND (b = \TRUE \AND n \leq 0)}{}
	\NLINE{P \quad\DEFEQ\quad \LETQQ{q}{p (n-1)}{\QQ{\lambda x. x \times (q\ x)}}}{}
 	\NLINE{\ASSERT{C}{n \leq 0}{b}{D}}{}
 	\NLINE{\ASSERT{D\SUBST{\TRUE}{b}}{\QQ{\lambda x. 1}}{m}{\QQEVAL{m}{y}{\forall j. y \bullet j = j^n}}}{Like prev.~examples}
 	\NLINE{\ASSERT{D\SUBST{\FALSE}{b}}{p (n-1)}{r}{\TRUTH \IMPLIES \QQEVAL{r}{q}{\forall j. q \bullet j = j^{n-1}}}}{Like Ex.~6}
 	\NLINE{\ASSERT{\TRUTH \IMPLIES \forall j. q \bullet j = j^{n-1}}{\QQ{\lambda x. x \times (q\ x)}}{m}{\QQEVAL{m}{y}{\forall j. y \bullet j = j^n}}}{Like Ex.~6}
 	\NLINE{\ASSERT{D\SUBST{\FALSE}{b}}{P}{m}{\QQEVAL{m}{y}{\forall j. y \bullet j = j^n}}}{\RULENAME{Unquote}, 5, 6}
 	\NLINE{\ASSERT{C}{\IFTHENELSE{n \leq 0}{\QQ{\lambda x. 1}}{P}}{m}{\QQEVAL{m}{y}{\forall j. y \bullet j = j^n}}}{\RULENAME{If}, 7}
 	\NLINE{\ASSERT{\TRUTH}{\lambda n.\IFTHENELSE{n \leq 0}{\QQ{\lambda x. 1}}{P}}{u}{\forall n \leq k.((\forall i < k.B_i^p) \IMPLIES B_n^u)}}{\RULENAME{Abs}, 8}
 	\NLINE{\ASSERT{\forall i < k.B_i^p}{\lambda n.\IFTHENELSE{n \leq 0}{\QQ{\lambda x. 1}}{P}}{u}{\forall n \leq k.B_n^u}}{\RULENAME{Conseq}, 9}
 	\NLINE{\ASSERT{\TRUTH}{\POWER}{u}{\forall k. \forall n \leq k.B_n^u}}{\RULENAME{Rec'}, 10}
 	\NLASTLINE{\ASSERT{\TRUTH}{\POWER}{u}{\forall n.B_n^u}}{\RULENAME{Conseq}, 11}
\end{NDERIVATION}
\end{exa}

\section{Completeness}\label{completeness2}

\NI This section poses, and then answers in the affirmative, three
important meta-logical questions about the logic introduced in
previous sections:
\begin{itemize}

\item Is the logic \emph{relatively complete} in the sense of
  Cook~\cite{CookSA:soucomaspv}?

\item Is the logic \emph{observationally
  complete}~\cite{HondaK:descriptive}?

\item Does the logic have \emph{characteristic
  formulae}~\cite{AcetoL:chaforfata}?

\end{itemize}

\NI The first question can be seen as a reversal of soundness: does
\[
  \models \ASSERT{A}{M}{m}{B}\ \text{imply}\ \vdash \ASSERT{A}{M}{m}{B}
\]
for all appropriate $A, B$?  Relative completeness means that in the
presence of an oracle for the ambient theory of arithmetic, e.g.~Peano
arithmetic or ZFC set-theory (used with [\RULENAME{Conseq}]), the
logic can syntactically derive all semantically true assertions, and
reasoning about programs does not need to concern itself with
models. Another way of saying this is that in relatively complete
program logics, the expressive power of the ambient theory of
arithmetic is the only source of incompleteness.\footnote{This does
  not violate Clarke's result \cite{ClarkeEN:prolancfwiiitoghas}
  because our logic has higher-order features (evaluation formulae and
  code evaluation predicates). See \cite{HondaK:obscomplfihofTECREP}
  for a more extensive discussion.}

The second question investigates if the program logic makes the same
distinctions as the observational congruence. In other words, is the
following characterisation true?
\[
     \ \qquad M \CONG N\ 
        \text{exactly when for all  $A, B$:}\  
     \ASSERT{A}{M}{m}{B}\ \text{iff}\  \ASSERT{A}{N}{m}{B}
\]
Observational completeness means that the operational semantics
(given by the contextual congruence) and the axiomatic semantics given
by logic cohere with each other. We believe that observational
completeness is a key property of program logics because it guarantees
that any operationally relevant program property can be
expressed.

If a logic is observationally complete, we may ask the third question
above about characteristic formulae: given a program $M$, can we find,
by induction on the syntax of $M$, a pair of formulae $A, B$ such that
\begin{itemize}
    
\item $\models  \ASSERT{A}{M}{m}{B}$ 
\item for all programs $N$: $M \LEQ N$ iff $\models \ASSERT{A}{N}{m}{B}$? 

\end{itemize}
Such formulae are called \emph{characteristic}.  If characteristic
formulae always exist, the semantics of each program can be expressed
succinctly in the logic, using just a pair of formulae, and we call
the logic \emph{descriptively complete} \cite{HondaK:descriptive}.
The reason we use the contextual precongruence $\LEQ$ from Section
\ref{language} in the definition of characteristic formulae above, and
not the congruence $\CONG$, is that our logic is for total
correctness, and cannot express program divergence. More precisely,
the following holds:
\[ 
   \left.
   \begin{array}{l}
     \models \ASSERT{A}{M}{m}{B}\\
      M \LEQ N
   \end{array}
   \right\}
     \text{implies}\ \models \ASSERT{A}{N}{m}{B}.
\]

\NI In other words, if $\ASSERT{A}{M}{m}{B}$ holds and we make some
parts of $M$ more defined (e.g.~by replacing a divergent with a
convergent subterm), obtaining $N$, then $\ASSERT{A}{N}{m}{B}$ holds,
too. Let's look at an example:
\[
   \models \ASSERT{\TRUTH}{\lambda x.\Omega}{m}{\TRUTH}.
\]
 If we replace $\Omega$ with $17$, we obtain $\lambda
x.17$, and clearly $\lambda x.\Omega \LEQ \lambda x.17$. But also:
\[
   \models \ASSERT{\TRUTH}{\lambda x.17}{m}{\TRUTH}.
\]
This indicates that in logics for total correctness, pairs $A, B$ talk
about upwards-closed sets of programs.  Upwards-closed sets with a
least member (up to $\CONG$) are especially nice, and for each such
set, its least element can be seen as representing the set.

\PARAGRAPH{Proof strategy} We prove the three completeness theorems 
promised at the beginning of this section following ideas developed in
\cite{BergerM:comlogfamltmp,HondaK:descriptive,GLOBAL,YHB07:local:full},
but adapted to the present logic. The proofs are broken down into
the following steps where we:
\begin{enumerate}

\item make precise the relevant notion of characteristic
  formula.

\item present an inference system for characteristic formulae.

\item prove that the inference system computes characteristic
  formulae.

\item show that the characteristic formulae are derivable
  using the rules and axioms of Section \ref{logic}.

\item use characteristic formulae to prove observational
  completeness.

\item employ characteristic formulae to prove relative
  completeness.

\end{enumerate}

\subsection{Formalising characteristic formulae}
We now precisely define we mean by characteristic formulae. Our
definition is split into three parts, one guaranteeing the soundness
of characteristic formulae, one to do with termination, and one that
is about divergence-related aspects of program behaviour.

\begin{defi}
A pair $(A, B)$ is a \emph{total characteristic assertion pair}, or
\emph{TCAP}, \emph{of $M$ at $u$}, if the following conditions hold
(in each clause we assume well-typedness).
\begin{itemize}

\item (soundness) $\models \ASSERT{A}{M}{u}{B}$.

\item (MTC, minimal terminating condition) For all appropriately typed
  models $\eta$, $M\eta \CONV$ if and only if $\eta \models A$.

\item (closure) \ If $\eta \models \ASSERT{E}{N}{u}{B}$ and $E\IMPLIES
  A$, then $\eta\models E$ implies $M\eta \LEQ N\eta$.

\end{itemize}
\end{defi}

\NI A TCAP of $M$ denotes a set of programs whose minimum element (up
to $\CONG$) is $M$, and in that sense characterises that behaviour
uniquely up to $\LEQ$. As mentioned above, characterisation up to
$\CONG$ is not possible in a logic for total correctness. Logics of
partial correctness suffer from a dual problem because they cannot
express convergence. To achieve logical characterisation up to $\CONG$
in a single pair of formulae, we need both total and partial
correctness, ideally combined into a logic of general correctness, see
e.g.~\cite{BergerM:comlogfamltmp}.

\begin{FIGURE}
\begin{RULES}
	\ONEPREMISERULENAMEDRIGHT
        {
		x\ \text{non-modal}
        }
        {
		\ASSERT{\TRUTH}{x}{m}{x = m}
        }{Var$^t$}
		\quad
	\ONEPREMISERULENAMEDRIGHT
        {
		x\ \text{modal}
        }
        {
		\ASSERT{x \CONV}{x}{m}{x = m}
        }{Var$_m^t$}
		\quad
	\ZEROPREMISERULENAMEDRIGHT
        {
		\ASSERT{\TRUTH}{\PROGRAM{c}}{m}{\LOGIC{c} = m}
        }{Const$^t$}
		\\\\
	\ONEPREMISERULENAMEDRIGHT
	{
		\ASSERT{A}{M}{m}{B}
	}
	{
		\ASSERT{\TRUTH}{\lambda x^{\alpha}.M}{u}{\forall x.(A \IMPLIES \ONEEVAL{u}{x}{m}{B}}
	}{Abs$^t$}
		\\\\
	\TWOPREMISERULENAMEDRIGHT
	{
		\ASSERT{A_i}{M}{m_i}{B_i}
	}
	{
		i = 1, ..., n
	}
	{
		\ASSERT
		{
			\displaystyle{\BIGAND}_i A_i
		}
		{
			\PROGRAM{op}(\VEC{M})
		}
		{u}
		{
			\exists \VEC{m}.(u = \LOGIC{op}(\VEC{m}) \AND \displaystyle{\BIGAND}_i B_i)
		}
	}{Op$^t$}
		\\\\
	\TWOPREMISERULENAMEDRIGHT
	{
		\ASSERT{A_1}{M}{m}{B_1}
	}
	{
		\ASSERT{A_2}{N}{n}{B_2}
	}
	{
		\begin{array}{c}
   	           \{ A_1 \AND A_2 \AND \forall mn.((B_1 \AND B_2) \IMPLIES \ONEEVAL{m}{n}{z}{\TRUTH}) \} \\
		      MN :_{u} \\
		   \{\exists mn.\ONEEVAL{m}{n}{z}{B_1 \AND B_2 \AND z = u}\}
                \end{array}
	}{App$^t$}
		\\\\
	\FOURPREMISERULENAMEDRIGHT
	{
		\ASSERT{A}{M}{m}{B}
	}
	{
		\ASSERT{A_i}{N}{u}{B_i}
	}
	{
		b_1 = \TRUE
        }
        {
                b_2 = \FALSE
	}
	{
		\ASSERT{A \AND \displaystyle{\BIGAND}_{i}(B\SUBST{b_i}{m} \IMPLIES A_i)}{\IFTHENELSE{M}{N_1}{N_2}}{u}{\displaystyle{\BIGOR}_i(B\SUBST{b_i}{m} \AND B_i)}
	}{If$^t$}
		\\\\
	\ONEPREMISERULENAMEDRIGHT
	{
		\ASSERT{\TRUTH}{\lambda x.M}{m}{A}
	}
	{
		\ASSERT{\TRUTH}{\mu g.\lambda x.M}{m}{A\SUBST{m}{g}}
	}{Rec$^t$}
		\quad
	\ONEPREMISERULENAMEDRIGHT
	{
		\ASSERT{A}{M}{m}{B}
	}
	{
		\ASSERT{\TRUTH}{\QQ{M}}{u}{A \IMPLIES \QQEVAL{u}{m}{B}}
	}{Quote$^t$}
		\\\\
                \hspace{-8mm}	
        \TWOPREMISERULENAMEDRIGHT
	{
		\ASSERT{A_1}{M}{m}{B_1}
	}
        {
		\ASSERT{A_2}{N}{u}{B_2}
        }
	{
		\begin{array}{c}
		\{
			A_1 \AND ((\forall x^{\square}.A_2) \OR \forall m.(B_1 \IMPLIES \QQEVAL{m}{x}{A_2}))
		\} \\
		   \LETQQ{x}{M}{N} :_{u} \\
		\{
			\exists mx^{\square}.((m = \QQ{\cdot} \IMPLIES m = \QQ{x}) \AND B_1 \AND B_2)
		\}
                \end{array}
        }{Unquote$^t$}
\end{RULES}
\caption{Inference system for TCAPs.}\label{figure:completeness:rules}
\end{FIGURE}

\PARAGRAPH{An inference system for TCAPs}
The definition of TCAPs is semantic. We now present an algorithm that
enables us to derive TCAPs for each \PCFDP-program by induction on the
typing derivation.  The rules are given in
Figure~\ref{figure:completeness:rules} and follow ideas
from~\cite{BergerM:comlogfamltmp,HondaK:descriptive,GLOBAL,YHB07:local:full}.
Rulenames are derived from those in Figure \ref{figure:logic:rules}
but with a superscript (e.g.~[\RULENAME{Var$^t$}] instead of
[\RULENAME{Var}]).  The assertion language is that of Section~\ref{logic}
with one extension: quantification over modal variables. That means
the assertions are now generated by the following extended grammar:

\begin{GRAMMAR}
	A
		&\ ::=\ &
	...
		\VERTICAL
	\MODALFORALL{x}{\alpha}.A
\end{GRAMMAR}

\NI We call $\MODALFORALL{x}{\alpha}.A$ \emph{modal universal
  quantification}, where the bound variable $x$ ranges over arbitrary
programs, not just values.  For $\MODALFORALL{x}{\alpha}.A$ to be
well-formed, $x$ must be modal and of type $\alpha$ in $A$, and modal
quantification is typed as follows:
\[
  \ONEPREMISERULE{
    \FORMULATYPES{\Gamma; \Delta, x : \alpha}{A} }{
    \FORMULATYPES{\Gamma; \Delta}{\MODALFORALL{x}{\alpha}. A} }
\]
The semantics of modal quantification is given by the following:
\[
   (\xi, \sigma) \models \MODALFORALL{x}{\alpha}.A\ \text{iff for all
   closed programs $M$ of type $\alpha$}:\ (\xi, \sigma \cdot x : M)
   \models A.
\]
Since the addition of modal quantification does not change our notions
of model and satisfaction relation, all proofs in Section \ref{logic}
stay valid.

The existential modal quantifier $\MODALEXISTS{x}{\alpha}.A$ is given
by de Morgan duality.  We often drop type annotations in modal
quantifiers, e.g.~writing $\MODALFORALL{x}{}.A$.  Axiomatising modal
quantification uses the standard axioms for first-order quantifiers
with the following addition:
\[
        (div_m) \qquad \neg \MODALFORALL{x}{}. x\CONV
\]
This axiom states that not all modal variables denote terminating programs,
which is immediately true from the model.

We write $\PROVESTCAP\ASSERT{A}{M}{u}{B}$ to indicate that the
assertion $\ASSERT{A}{M}{u}{B}$ is derivable using the rules of Figure
\ref{figure:completeness:rules} only (i.e.~without application of
rules from Figures \ref{figure:logic:rules} and
\ref{figure:logic:structuralRules}). As before, we assume that
assertions, programs and rules are well-typed, and newly introduced
variables are always fresh.

Before presenting proofs, we make a small observation: the pre-
and postcondition pairs in Figure~\ref{figure:completeness:rules}
constrain exactly the free variables of a program, together with the
anchor:
\begin{obs}\label{completeness:observation:1}
  Let $\PROVESTCAP \ASSERT{A}{M}{m}{B}$ then $\FV{A} = \FV{M}$ and
  $\FV{B} = \FV{M} \cup \{m\}$.
\end{obs}

\PARAGRAPH{Informal explanation of the rules} Except for
[\RULENAME{Unquote$^t$}] and [\RULENAME{Var$_m^t$}], all rules in Figure
\ref{figure:completeness:rules} are either unchanged from the corresponding rules in Figure
\ref{figure:logic:rules} or have already been used in some of
\cite{BergerM:comlogfamltmp,HondaK:descriptive,GLOBAL,YHB07:local:full}.
We now give an informal explanation of the rules not already in Figure~\ref{figure:logic:rules}.

[\RULENAME{Var$^t$}] says that the TCAP of a non-modal variable $x$ at $m$
is $(\TRUTH, x = m)$. The precondition is $\TRUTH$ because non-modal
variables always denote values and thus always terminate. The
postcondition $x = m$ says that whenever a program $M$ satisfies
$\ASSERT{\TRUTH}{M}{m}{x = m}$, then $M$ must be contextually equal to
$x$ in the ambient model.

[\RULENAME{Var$_m^t$}] for modal variables $x$ has a more elaborate
precondition than [\RULENAME{Var$^t$}], because modal variables can denote
non-terminating programs. The formula $x \CONV$ is true exactly when
the denotation of $x$ is terminating.  The postcondition is the same
as in the case of [\RULENAME{Var$^t$}].

[\RULENAME{Const$^t$}] says that the TCAP for constants $c$ at $m$ is
$(\TRUTH, \LOGIC{c} = m)$.  As with non-modal variables, the
precondition is $\TRUTH$ because constants are values.  The
postcondition $\LOGIC{c} = m$ says that whenever a program $M$
satisfies $\ASSERT{\TRUTH}{M}{m}{\LOGIC{c} = m}$, then $M$ must be
contextually equal to $\PROGRAM{c}$. For example, under the typing $x
: \INT; \epsilon$, the program $\IFTHENELSE{x}{5}{5}$ has this
property, and indeed $5 \CONG \IFTHENELSE{x}{5}{5}$ when $x$ is
non-modal.

[\RULENAME{Op$^t$}] computes all TCAPs for operands in the premise. As an
operation (e.g.~addition) terminates exactly when all operands
terminate, the precondition of the rule's conclusion is simply the
conjunction of all preconditions for operands.  The postcondition of
the rule conclusion states that the result of the computation is the
operation applied to some operands, and each operand is constrained by
the postconditions of the rule premises. Depending on the operations
used, additional constraints might be needed in the precondition: for
example division $M / N$ requires $N$ to evaluate to a non-zero value.

[\RULENAME{App$^t$}] works as follows. In a call-by-value language an
application $MN$ terminates if: the evaluations of both $M$ and $N$
terminate to $V$ and $W$, respectively; and, in addition, the
application $VW$ itself terminates. The first two requirements are
stated by putting $A_1 \AND A_2$ into the precondition of the
conclusion on the rule.  Here $A_i$ is obtained recursively by
computing the TCAPs of $M$ and $N$, so e.g.~$A_1$ holds exactly when
$M$ terminates. The additional assumption
\[
   \forall mn.((B_1 \AND B_2) \IMPLIES \ONEEVAL{m}{n}{z}{\TRUTH})
\]
says that no matter what $M$ and $N$ evaluate to, the program
terminates as long as $m$ is as constrained by $B_i$, $n$ is
constrained by $B_2$, and the application $m \bullet n$ terminates.
The postcondition of the conclusion says that the program $MN$
evaluates to the result of applying $M$ to $N$.

[\RULENAME{If$^t$}] makes the following assertion. A conditional
terminates exactly when the condition terminates and the branch chosen
by the conditional does, too. This is formalised by:
\[
   A \AND \displaystyle{\BIGAND}_{i}(B\SUBST{b_i}{m} \IMPLIES A_i)
\]
As exactly one of $B\SUBST{\TRUE}{m}$ and $B\SUBST{\FALSE}{m}$ is true and
exactly one is false, one implication is vacuously true, and the other
requires the corresponding $A_i$ to hold, giving the correct
termination condition. For the same reason exactly one of 
\[
   B\SUBST{b_i}{m} \AND B_i
\]
must be false, and one must hold exactly when the corresponding $B_i$
holds. Since these two formulae are connected by an outer disjunction,
the postcondition of the rule's conclusion give exactly the behaviour
of the program.

[\RULENAME{Unquote$^t$}] This rule is the main intellectual novelty of the
present section. Clearly, $\LETQQ{x}{M}{N}$ terminates exactly when:
\begin{itemize}

\item $M$ evaluates to some $\QQ{M'}$, and

\item $N\SUBST{M'}{x}$ terminates.

\end{itemize}
The former is reflected in the precondition of the conclusion of the
rule by adding $A_1$, which controls the termination of $M$. The
second condition is more complicated, because it is possible that $M$
evaluates to e.g.~$\QQ{\Omega}$, and yet $N\SUBST{\Omega}{x}$
terminates, for example in
\[
   \LETQQ{x}{\QQ{\Omega}}{\QQ{x}}
\]
This case is covered by the clause $\forall x^{\square}.A_2$.  We see
here the reason for using modal quantification. If the quantifier were
to range over values only, programs such as
\begin{align}\label{completeness:eq:9987}
   \LETQQ{x}{\QQ{\Omega}}{x},
\end{align}
which do not terminate, would cause trouble without modal
quantification, because only when $x$ is bound to a non-terminating
term would $N$ diverge. The TCAP for $x$ at $u$ is $(x\Downarrow, x =
u)$, making $\forall x.x\Downarrow$, unlike $\forall
x^{\square}. x\Downarrow$, trivially true, leading to the erroneous
precondition $\TRUTH$ for (\ref{completeness:eq:9987}).

One may also ask, why not use a simpler precondition like
\begin{align}\label{completeness:eq:9988}
  A_1 \AND  \forall m.(B_1 \IMPLIES \QQEVAL{m}{x}{A_2})
\end{align}
in the conclusion of [\RULENAME{Unquote$^t$}]? The answer is that this
would also be too weak for completeness. To see why, consider the
program:
\[
   \LETQQ{x}{\QQ{\Omega}}{8}.
\]
The TCAPs of $\QQ{\Omega}$ is
$\ASSERT{\TRUTH}{\QQ{\Omega}}{m}{\TRUTH}$, cf.~Example 16, and using
the rule [\RULENAME{Const$^t$}], we see that $\ASSERT{\TRUTH}{8}{u}{u =
  8}$ is the TCAP of 8.  That means (\ref{completeness:eq:9988}) gives
us a precondition
\[
   \ASSERT{\TRUTH \AND \forall m.\QQEVAL{m}{x}{\TRUTH}}{\LETQQ{x}{\QQ{\Omega}}{8}}{u}{...}
\]
which is equivalent to:
\begin{align}\label{completeness:eq:1182}
   \ASSERT{\FALSITY}{\LETQQ{x}{\QQ{\Omega}}{8}}{u}{...}
\end{align}
since $\forall m.\QQEVAL{m}{x}{\TRUTH}$ is equivalent to $\FALSITY$.
Now (\ref{completeness:eq:1182}) is clearly sound, but the
precondition too weak to capture the full meaning of the program
$\LETQQ{x}{\QQ{\Omega}}{8}$.

Next we look at the postcondition. It says that the result of
evaluating $\LETQQ{x}{M}{N}$ is, among other things, as described by
$B_2$, which is the postcondition of $N$ at $u$. However, by
Observation \ref{completeness:observation:1}, $B_2$ contains $x$ as
free variable. We hide it with a modal existential quantifier. But $x$
cannot be arbitrary, as it is the result of unquoting what $M$
evaluates to. Note that the postcondition $B_1$ speaks about $M$
named $m$. So $x$ is the unquoting of $m$. We cannot assert
$\QQEVAL{m}{x}{...}$, because that would stipulate that $M$ evaluates
to a term that, when unquoted, terminates, which cannot be guaranteed
(e.g.~if $M$ is $\QQ{\Omega}$). To deal with this issue, we explicitly require that  $x$ is the
unquoting of $m$, provided $m$ denotes a terminating meta-program:
\begin{align}\label{completeness:eq:123282}
   m = \QQ{\cdot} \IMPLIES m = \QQ{x}
\end{align}
which means, if $M$ converges to a quasi-quote $\QQ{M'}$ and $M'$
converges, say to $V$, then $x$ describes this value $V$. Finally, we
hide $m$ by an existential quantifier, and constrain $m$ by $B_1$.
Note that the conditional constraining of $x$ in
(\ref{completeness:eq:123282}) does not hold if $M$ diverges, or
converges to e.g.~$\QQ{\Omega}$. In the former case, the precondition
must be (equivalent to) $\FALSITY$, because the whole program
diverges. In the latter case, a logic for total correctness cannot
make an interesting assertion about the use of $x$ in $N$.

\begin{thm}\label{completeness:theorem:1}
\
\begin{enumerate}

\item\label{completeness:theorem:1:0} \emph{(descriptive completeness
  for total correctness)}\ \ Assume $\TYPES{\Gamma;
  \Delta}{M}{\alpha}$.  Then $\PROVESTCAP \ASSERT{A}{M}{u}{B}$ implies
  $(A, B)$ is a TCAP of $M$ at $u$.

\item\label{completeness:theorem:1:1} {\rm (observational
  completeness)} $M\CONG N$ if and only if, for each $A$ and $B$, we
  have $\models \ASSERT{A}{M}{u}{B}$ iff $\models
  \ASSERT{A}{N}{u}{B}$.

\item\label{completeness:theorem:1:2}{\rm (relative
  completeness)}\ Let $B$ be upward-closed at $u$.  Then $\models
  \ASSERT{A}{M}{u}{B}$ implies $\vdash\ASSERT{A}{M}{u}{B}$.

\end{enumerate}
\end{thm}

\NI Before giving a proof of Theorem \ref{completeness:theorem:1}
establish some helpful facts.

\begin{prop}\label{completeness:proposition:14}
\ 
\begin{enumerate}

\item\label{completeness:proposition:14:1} If $(A,B)$ is a TCAP of $M$
  at $u$ and if $\models \ASSERT{A}{N}{u}{B}$, then $M \LEQ N$.

\item\label{completeness:proposition:14:2} $(A,B)$ is a TCAP of $M$ at
  $u$ iff (soundness), (MTC) and the following condition hold:
  (\emph{closure-2}): if $(\xi, \sigma) \models A$ and for closed $V$ we have
  $(\xi \cdot u : V, \sigma) \models B$ then $M(\xi, \sigma) \LEQ V$.

\end{enumerate}
\end{prop}

\begin{proof}
We begin with (\ref{completeness:proposition:14:1}).  Assume that
$\eta = (\xi, \sigma) \models \ASSERT{A}{N}{u}{B}$. 
There are two cases.
\begin{itemize}

\item $\eta \models A$. In this case $M\eta \CONV V$ by soundness, and
  $(\xi \cdot u : V, \sigma) \models B$.  Now $M\eta \LEQ N\eta$ follows by
  (\emph{closure}).

\item $\eta \not \models A$. In this case, by (MTC) we have $M\eta
  \DIV$ and hence trivially $M\eta \LEQ N\eta$.

\end{itemize}
Now the result follows from Observation
\ref{completeness:lemma:15}.

For (\ref{completeness:proposition:14:2}) we begin with the (if)
direction.  Assume $\eta \models \ASSERT{E}{N}{u}{B}$ where $E
\IMPLIES A$ and $\eta \models E$.  Hence $N\eta \CONV V$ with $(\xi
\cdot u : V, \sigma) \models B$ by soundness.  From $E \IMPLIES A$ we
get $\eta \models A$, but then by (\emph{closure-2}) it must be the
case that $M\eta \LEQ V$ which in turn implies $M\eta \LEQ N\eta$
since $V \CONG N\eta$ by Proposition
\ref{lamguage:theorem:1}.\ref{lamguage:theorem:1:1}.

For the reverse direction, suppose $(A, B)$ is a TCAP for $M$ at
$u$. We must show that (\emph{closure-2}) holds. So let $\eta = (\xi,
\sigma) \models A$ and $(\xi \cdot u : V, \sigma) \models B$, with $V$
being closed and appropriately typed. Define:
\[
   E\ \DEFEQ\ A \AND \exists u.B
\]
Then clearly:
\begin{itemize}

\item $\eta \models E$,
\item $E \IMPLIES A$,
\item $\eta \models \ASSERT{E}{V}{u}{B}$.
\end{itemize}
Hence by (\emph{closure}) $M\eta \LEQ V\eta = V$
\end{proof}

\NI Proposition
\ref{completeness:proposition:14}.\ref{completeness:proposition:14:1}
shows that TCAPs of a program $M$ really represent a set of behaviours
whose minimal element is $M$.\medskip

%\PARAGRAPH{Proof of Theorem \ref{completeness:theorem:1}.\ref{completeness:theorem:1:0}}
\noindent\emph{Proof} \relax[\hspace{1 pt}of Theorem
  \ref{completeness:theorem:1}.\ref{completeness:theorem:1:0}].
The proof we are about to embark on is somewhat lengthy, and benefits
from having the following convenient proposition available.

\begin{prop}\label{completeness:proposition:tcapSoundness}
If $\PROVESTCAP \ASSERT{A}{M}{u}{B}$ then also $\vdash \ASSERT{A}{M}{u}{B}$.
\end{prop}

\begin{proof} We proceed by induction on the
derivation of $\PROVESTCAP \ASSERT{A}{M}{u}{B}$. The cases
[\RULENAME{Abs$^t$}, \RULENAME{Quote$^t$}] follow immediate from the (IH),
since these rules are identical in the rule systems of Figures
\ref{figure:logic:rules} and \ref{figure:completeness:rules}.

\begin{description}
\item[\rm\RULENAME{Var$^t$}] We proceed as follows.
\begin{NDERIVATION}{1}
  \NLINE{\ASSERT{x = m \SUBST{x}{m} \AND x \CONV}{x}{m}{x = m}}{\RULENAME{Var}}
  \NLINE{\ASSERT{x \CONV}{x}{m}{x = m}}{\RULENAME{Conseq}, 1}
  \NLASTLINE{\ASSERT{\TRUTH}{x}{m}{x = m}}{(term), \RULENAME{Conseq}, 2}
\end{NDERIVATION}

\item[\rm\RULENAME{Var$_m^t$}] This case is exactly like the previous, except
that the last line is omitted.

\item[\rm\RULENAME{Const$^t$}]
Similar to [\RULENAME{Var$^t$}].

\item[\rm\RULENAME{Op$^t$}]
We treat the special case of addition.
\begin{NDERIVATION}{1}
  \NLINE{\ASSERT{A_i}{M_i}{m_i}{B_i}}{(IH)}
  \NLINE{\ASSERT{A_1 \AND A_2}{M_1}{m_1}{B_1 \AND A_2}}{\RULENAME{Invar}, 1}
  \NLINE{\ASSERT{B_1 \AND A_2}{M_2}{m_2}{B_1 \AND B_2}}{\RULENAME{Invar}, 1}
  \NLINE{\ASSERT{B_1 \AND A_2}{M_2}{m_2}{m_1 + m_2 = m_1 + m_2 \AND B_1 \AND B_2}}{\RULENAME{Conseq}, 3}
  \NLINE{\ASSERT{B_1 \AND A_2}{M_2}{m_2}{(u = m_1 + m_2)\SUBST{m_1 + m_2}{u} \AND B_1 \AND B_2}}{4}
  \NLINE{\ASSERT{A_1 \AND A_2}{M_1 + M_2}{u}{u = m_1 + m_2 \AND B_1 \AND B_2}}{\RULENAME{Add}, 2, 5}
  \NLASTLINE{\ASSERT{A_1 \AND A_2}{M_1 + M_2}{u}{\exists m_1 m_2.(u = m_1 + m_2 \AND B_1 \AND B_2)}}{\RULENAME{Conseq}, 6}
\end{NDERIVATION}

\item[\rm\RULENAME{App$^t$}] The proof for this rule is the sole place in this
paper where the $(q_{\alpha})$ axiom is used. It is an open question
as to whether this axiom is strictly needed, but we have not yet
managed without it.
\begin{NDERIVATION}{1}
  \NLINE{\ASSERT{A_1}{M}{m}{B_1}}{(IH)}
  \NLINE{C \DEFEQ \forall mn.((B_1 \AND B_2) \IMPLIES \ONEEVAL{m}{n}{u}{\TRUTH})}{}
  \NLINE{\ASSERT{A_1 \AND A_2 \AND C}{M}{m}{A_2 \AND B_1 \AND C}}{\RULENAME{Invar}, 1}
  \NLINE{\ASSERT{A_2}{N}{n}{B_2}}{(IH)}
  \NLINE{\ASSERT{A_2 \AND B_1 \AND C}{N}{n}{B_1 \AND B_2 \AND C}}{\RULENAME{Invar}, 4}
  \NLINE{\ASSERT{A_2 \AND B_1 \AND C}{N}{n}{B_1 \AND B_2 \AND \ONEEVAL{m}{n}{u}{\TRUTH}}}{5}
  \NLINE{\ASSERT{A_2 \AND B_1 \AND C}{N}{n}{\ONEEVAL{m}{n}{u}{B_1 \AND B_2}}}{(q4), 6}
  \NLINE{\ASSERT{A_2 \AND B_1 \AND C}{N}{n}{\ONEEVAL{m}{n}{u}{\ONEEVAL{m}{n}{z}{B_1 \AND B_2 \AND u = z}}}}{($q_{\alpha}$), 7}
  \NLINE{\ASSERT{A_1 \AND A_2 \AND C}{MN}{u}{\ONEEVAL{m}{n}{z}{B_1 \AND B_2 \AND u = z}}}{\RULENAME{App}, 3, 8}
  \NLASTLINE{\ASSERT{A_1 \AND A_2 \AND C}{MN}{u}{\exists mn.(\ONEEVAL{m}{n}{z}{B_1 \AND B_2 \AND u = z)}}}{9}
\end{NDERIVATION}

\item[\rm\RULENAME{If$^t$}] In the derivation of this rule we make unusually heavy tacit use of the [\RULENAME{Conseq}]
rule.
\begin{NDERIVATION}{1}
  \NLINE{\ASSERT{A}{M}{m}{B}}{(IH)}
  \NLINE{C \DEFEQ \displaystyle{\BIGAND}_{i}(B\SUBST{b_i}{m} \IMPLIES A_i)}{}
  \NLINE{\ASSERT{A \AND C}{M}{m}{B \AND C}}{\RULENAME{Invar}, 1}
  \NLINE{B\SUBST{\TRUE}{m} \equiv \TRUTH \qquad B\SUBST{\FALSE}{m} \equiv \FALSITY}{Wlog.}
  \NLINE{\ASSERT{A_i}{N_i}{u}{B_i}}{(IH)}
  \NLINE{\ASSERT{B\SUBST{b_i}{m} \AND C}{N_i}{u}{B_i}}{5}
  \NLINE{\ASSERT{B\SUBST{b_i}{m} \AND C\SUBST{b_i}{m}}{N_i}{u}{B_i}}{$m \notin \FV{C}$, 6}
  \NLINE{\ASSERT{B\SUBST{b_i}{m} \AND B\SUBST{b_i}{m} \AND C\SUBST{b_i}{m}}{N_i}{u}{B\SUBST{b_i}{m} \AND B_i}}{\RULENAME{Invar}, 7}
  \NLINE{\ASSERT{B\SUBST{b_i}{m} \AND C\SUBST{b_i}{m}}{N_i}{u}{B\SUBST{b_i}{m} \AND B_i}}{8}
  \NLINE{\ASSERT{(B \AND C)\SUBST{b_i}{m}}{N_i}{u}{B\SUBST{b_i}{m} \AND B_i}}{9}
  \NLINE{D \DEFEQ \displaystyle{\BIGOR}_i(B\SUBST{b_i}{m} \AND B_i)}{}
  \NLINE{\ASSERT{(B \AND C)\SUBST{b_i}{m}}{N_i}{u}{D}}{10}
  \NLASTLINE{\ASSERT{A \AND C}{\IFTHENELSE{M}{N_1}{N_2}}{u}{D}}{3, 12}
\end{NDERIVATION}

\item[\rm\RULENAME{Unquote$^t$}] This is the last step in our proof. The
derivation uses [\RULENAME{Unquote}$^+$], the only use of that rule in
the paper. It is unclear if the simpler version of
[\RULENAME{Unquote}$^+$] presented in Section \ref{logic} is strong
enough to carry out this part of the proof.
\begin{NDERIVATION}{1}
  \NLINE{\ASSERT{A_1}{M}{m}{B_1}}{(IH)}
  \NLINE{C \DEFEQ (\forall x^{\square}.A_2) \OR \forall m.(B_1 \IMPLIES \QQEVAL{m}{x}{A_2}}{}
  \NLINE{\ASSERT{A_1 \AND C}{M}{m}{B_1 \AND C}}{\RULENAME{Invar}, 1}
  \NLINE{\ASSERT{A_1 \AND C}{M}{m}{B_1 \AND ((\forall x^{\square}.A_2) \OR \QQEVAL{m}{x}{A_2})}}{3}
  \NLINE{\ASSERT{A_1 \AND C}{M}{m}{B_1 \AND ((\neg\forall x^{\square}.A_2) \IMPLIES \QQEVAL{m}{x}{A_2})}}{4}
  \NLINE{D \DEFEQ m = \QQ{\cdot} \IMPLIES m = \QQ{x}}{}
  \NLINE{\ASSERT{A_2}{N}{u}{B_2}}{(IH)}
  \NLINE{\ASSERT{A_2 \AND D \AND B_1}{N}{u}{D \AND B_1 \AND B_2}}{\RULENAME{Invar}, 7}
  \NLINE{((\neg\forall x^{\square}.A_2) \IMPLIES A_2)\ \IMPLIES\ A_2}{\text{see below}}
  \NLINE{\ASSERT{B_1 \AND ((\neg\forall x^{\square}.A_2) \IMPLIES A_2) \AND D}{N}{u}{D \AND B_1 \AND B_2}}{\RULENAME{Conseq}, 8, 9}
  \NLINE{\ASSERT{A_1 \AND C}{\LETQQ{x}{M}{N}}{u}{D \AND B_1 \AND B_2}}{\RULENAME{Unquote}$^+$, 5, 10}
  \NLASTLINE{\ASSERT{A_1 \AND C}{\LETQQ{x}{M}{N}}{u}{\exists mx^{\square}.(D \AND B_1 \AND B_2)}}{\RULENAME{Conseq}, 11}
\end{NDERIVATION}

\NI It remains to justify Line 9. Rewriting 
\[
   ((\neg\forall x^{\square}.A_2) \IMPLIES A_2)\ \IMPLIES\ A_2
\]
in the equivalent form
\[
   ((\forall x^{\square}.A_2) \OR A_2)\ \IMPLIES\ A_2
\]
lets us see immediately that Line 9 is true. \qedhere
\end{description}
\end{proof}

\noindent Proposition \ref{completeness:proposition:tcapSoundness} together with
the soundness of the rules in Figure \ref{figure:logic:rules} immediately implies
the soundness of the TCAP rules. We record this fact:

\begin{cor}
The TCAP rules in Figure \ref{figure:completeness:rules} are sound.
\end{cor}

\NI Now we establish the first part of the theorem by induction on the
derivation of $\PROVESTCAP \ASSERT{A}{M}{u}{B}$, using
(\emph{closure-2}) from Proposition
\ref{completeness:proposition:14}.\ref{completeness:proposition:14:2}
instead of (\emph{closure}) for simplicity. We focus on the
interesting cases [\RULENAME{Var$_m^t$, Rec$^t$, Quote$^t$, Unquote$^t$}], leaving
the remaining ones to Appendix \ref{app:completenessProofs}.

We start the proof of Theorem
\ref{completeness:theorem:1}.\ref{completeness:theorem:1:0} with
    [\RULENAME{Var$_m^t$}]. For soundness, assume that $\eta \DEFEQ
    (\xi, \sigma \cdot x : M) \models x\CONV$. By definition of
    $x\CONV$ we can find a value $V$ of appropriate type such that
    $(\xi, y : V, \sigma \cdot x : M) \models x = y$ where $y$ is some
    fresh variable. Thus $M \CONG V$, hence $x\eta \CONV W$ for some
    value $W$ with $W \CONG V$ (Proposition
    \ref{lamguage:theorem:1}.\ref{lamguage:theorem:1:5}) and clearly
    $(\xi \cdot m : W, \sigma) \models x = m$.  (MTC) follows
    by the assumption in the precondition that $x \CONV$. For
    (\emph{closure-2}), we choose a model $\eta \DEFEQ (\xi, \sigma
    \cdot x : M)$ with $\eta \models x \CONV$ and $(\xi \cdot m : V,
    \sigma \cdot x : M) \models m = x$.  As above, $\eta \models
    x\CONV$ means that $M \CONV W$ for some appropriate closed value
    $W$. Hence $V \CONG M \CONV W$ which means in particular $M \LEQ
    V$ hence $x\eta \LEQ V$ as required.

Next is [\RULENAME{Quote$^t$}]. Let $\eta \DEFEQ (\xi \cdot u : \QQ{N},
\sigma)$.  Soundness has already been proven in Theorem
\ref{logic:theorem:1}.\ref{logic:theorem:1:1}, and (MTC) is trivial. So
suppose $\eta \models A \IMPLIES \QQEVAL{u}{m}{B}$. There are two
cases.
\begin{itemize}

\item $\eta \not\models A$. By (IH) we know that $A$ is an MTC for
  $M$, hence it must be the case that $M\eta \DIV$, thus trivially
  $M\eta \LEQ N$.  Since $\LEQ$ is a congruence by definition, we know
  that $\QQ{M\eta} \LEQ \QQ{N}$.  Now $\QQ{M}\eta = \QQ{M\eta}$ hence
      $\QQ{M}\eta \LEQ \QQ{N}$ as required.

\item $\eta \models A$ and $\eta \models \QQEVAL{u}{m}{B}$. Now we
  reason as given next.

\begin{NDERIVATION}{1}
  \NLINE{\eta \models A}{Assumption}
  \NLINE{(\xi, \sigma) \models A}{Prop.~\ref{app:soundness:9qw384}.\ref{app:soundness:9qw384:3}, $u \notin \FV{A}$, 1}
  \NLINE{\eta \models \QQEVAL{u}{m}{B}}{Assumption}
  \NLINE{N\CONV V\ \text{and}\ (\xi \cdot u : \QQ{N}, \sigma \cdot m : V) \models B}{3}
  \NLINE{\xi, \sigma \cdot m : V \models B}{Prop.~\ref{app:soundness:9qw384}.\ref{app:soundness:9qw384:3}, $u \notin \FV{B}$, 4}
  \NLINE{M(\xi, \sigma) \LEQ V}{by (IH), (\emph{closure-2}), 2, 5}
  \NLINE{\QQ{M}(\xi, \sigma) = \QQ{M(\xi, \sigma)} \LEQ \QQ{V}}{$\LEQ$ is a congruence, 6}
  \NLASTLINE{\QQ{M}(\xi, \sigma) \LEQ \QQ{N}}{Lem.~\ref{app:completenessProofs:lemma:1}.\ref{app:completenessProofs:lemma:1:1}, 4, 7 }
\end{NDERIVATION}

\end{itemize}

Next we deal with [\RULENAME{Unquote$^t$}], the most complicated case.  We
begin with soundness.  Let $A$ be the formula
\[
A_1 \AND (\forall x^{\square}.A_2) \OR \forall m.(B_1 \IMPLIES \QQEVAL{m}{x}{A_2})
\]
Assume that $(\xi, \sigma) \models A$. 
\begin{NDERIVATION}{1}
  \NLINE{(\xi, \sigma) \models A}{Assumption}
  \NLINE{(\xi, \sigma) \models A_1}{}
  \NLASTLINE{M(\xi, \sigma) \CONV \QQ{M'} \quad\text{and}\quad (\xi \cdot m : \QQ{M'}, \sigma) \models B_1}{(IH), 2}
\end{NDERIVATION}
Now we have two cases, here is the first.
\begin{NDERIVATION}{4}
  \NLINE{(\xi, \sigma) \models \forall x^{\square}.A_2}{First case}
  \NLINE{\text{For all appropriate programs}\  U: (\xi, \sigma \cdot x : U) \models A_2}{4}
  \NLINE{(\xi, \sigma \cdot x : M') \models A_2}{Specialisation of 5}
  \NLASTLINE{N(\xi, \sigma \cdot x : M') \CONV V\quad\text{and}\quad (\xi \cdot u : V, \sigma \cdot x : M') \models B_2}{(IH), 6}
\end{NDERIVATION}

\NI We now consider reductions where we set $\eta \DEFEQ (\xi,
\sigma)$.
\[
\begin{array}{lclcl}
  (\LETQQ{x}{M}{N})\eta
  &\ \quad=\quad\ &
  \LETQQ{x}{M\eta}{N\eta}
  \\
  &\NRED&
  \LETQQ{x}{\QQ{M'}}{N\eta}
  & \ \qquad\ &
  by\ 3
  \\
  &\NRED&
  N\eta\SUBST{M'}{x}
  \\
  &=&
  N(\xi, \sigma \cdot x : M')
  && Prop.~\ref{app:soundness:9qw384}.\ref{app:soundness:9qw384:4}
  \\
  &\CONV&
  V
  && by\ 7
\end{array}
\]

\NI Using this fact, we continue to reason as follows. Define
\[
   \eta'' \DEFEQ (\xi \cdot u : V \cdot m : \QQ{M'}, \sigma \cdot x : M')
\]
We need to show that
\begin{align}\label{completeness:eq:asd89}
   \eta'' \models (m = \QQ{\cdot} \IMPLIES m = \QQ{x}) \AND B_1 \AND B_2.
\end{align}
Since by (3) we have $(\xi \cdot m : \QQ{M'}, \sigma) \models B_1$ and
$x, u \notin \FV{B_1}$ by Observation
\ref{completeness:observation:1}, we can apply Proposition
\ref{app:soundness:9qw384}.\ref{app:soundness:9qw384:2} to get $\eta''
\models B_1$. By similar reasoning we get $\eta'' \models B_2$ from
(8). Hence $\eta'' \models B_1 \AND B_2$. That leaves the implication.
Assume $\eta'' \models m = \QQ{\cdot}$. By definition that means there
is a value $W$ such that $M' \CONV W$ and
\[
   (\xi \cdot u : V \cdot m : \QQ{M'}, \sigma \cdot x : W) \models m = \QQ{x}
\]
By Proposition \ref{lamguage:theorem:1}.\ref{lamguage:theorem:1:1} then
$M' \CONG W$, hence we can apply Proposition
\ref{app:soundness:9qw384}.\ref{app:soundness:9qw384:3}, giving us the required 
(\ref{completeness:eq:asd89}), which in turn implies
\[
   (\xi \cdot u : V, \sigma) \models \exists mx^{\square}.((m = \QQ{\cdot} \IMPLIES m = \QQ{x}) \AND B_1 \AND B_2)
\]
which finishes the soundness proof for this case.

We now consider the second case. 
\begin{NDERIVATION}{4}
  \NLINE{(\xi, \sigma) \models \forall m.(B_1 \IMPLIES \QQEVAL{m}{x}{A_2})}{Second case}
  \NLINE{\text{For all appropriate programs}\ L. (\xi \cdot m : L, \sigma) \models B_1 \IMPLIES \QQEVAL{m}{x}{A_2}}{4}
  \NLINE{(\xi \cdot m : \QQ{M'}, \sigma) \models B_1 \IMPLIES \QQEVAL{m}{x}{A_2}}{Specialisation of 5}
  \NLINE{(\xi \cdot m : \QQ{M'}, \sigma) \models \QQEVAL{m}{x}{A_2}}{3, 6}
  \NLINE{M' \CONV W \quad\text{and}\quad  (\xi \cdot m : \QQ{M'}, \sigma \cdot x : W)\models A_2}{(IH), 7}
  \NLINE{(\xi, \sigma \cdot x : W)\models A_2}{Obs.~\ref{completeness:observation:1}, $m \notin \FV{A_2}$, Prop.~\ref{app:soundness:9qw384}.\ref{app:soundness:9qw384:3}, 8}
  \NLASTLINE{N(\xi, \sigma \cdot x : W) \CONV V\quad\text{and}\quad (\xi \cdot u : V, \sigma \cdot x : W) \models B_2}{(IH), 9}
\end{NDERIVATION}

\NI Now we consider reductions where we set $\eta \DEFEQ (\xi,
\sigma)$.
\[
\begin{array}{lclcl}
  (\LETQQ{x}{M}{N})\eta
  &\ \quad=\quad\ &
  \LETQQ{x}{M\eta}{N\eta}
  \\
  &\NRED&
  \LETQQ{x}{\QQ{M'}}{N\eta}
  & \ \qquad\ &
  by\ 3
  \\
  &\NRED&
  N\eta\SUBST{M'}{x}
  \\
  &=&
  N(\xi, \sigma \cdot x : M')
  && Prop.~\ref{app:soundness:9qw384}.\ref{app:soundness:9qw384:4}
  \\
  &\CONV&
  V
  &&by\ 10
\end{array}
\]
The rest of this case is essentially identical to the corresponding
reasoning for the first case, and omitted.

Now we establish (MTC).  Choose a model $(\xi, \sigma)$ and assume that 
\[
	(\LETQQ{x}{M}{N})(\xi, \sigma) \CONV.
\]
We will show that $(\xi, \sigma) \models A$. The reverse implication is
part of soundness. Notice that this assumption implies the existence of
a reduction sequence as follows.
\begin{eqnarray}
	(\LETQQ{x}{M}{N})(\xi, \sigma) 
		&\ \quad=\quad\ &
	\LETQQ{x}{M(\xi, \sigma)}{N(\xi, \sigma)} 
                \notag\\
		&\NRED&
	\LETQQ{x}{\QQ{M'}}{N(\xi, \sigma)} 
	       \label{app:completenessProof:eq:0}
		\\
		&\RED&
	N(\xi, \sigma)\SUBST{M'}{x}
	       \notag
	       \\
	       &=&
	N(\xi, \sigma \cdot x : M')
	       \\
	       &\CONV&
	V
		\label{app:completenessProof:eq:2}
\end{eqnarray}
by (\ref{app:completenessProof:eq:0}) we know that 
\[
	M(\xi, \sigma) \CONV \QQ{M'}
\]
Since by (IH) $A_1$ is an MTC for $M$, we know that 
\begin{eqnarray}
	(\xi, \sigma) \models A_1
	      \quad\text{and hence}\quad
	(\xi \cdot m : \QQ{M'}, \sigma) \models B_1
		\label{app:completenessProof:eq:22}
\end{eqnarray}
We have two cases. First assume that $M' \DIV$. By
(\ref{app:completenessProof:eq:2}) and Lemma
\ref{app:completenessProofs:lemma:1}.\ref{app:completenessProofs:lemma:1:3}
we know that the following holds.
\[
	\text{For all appropriately typed and closed programs}\ L:\ N(\xi, \sigma \cdot x : L) \CONV.
\]
By (IH) $A_2$ is an MTC for $N$ at $u$, so we can reason as follows.
\begin{NDERIVATION}{1}
  \NLINE{\text{for all}\ L. (\xi, \sigma \cdot x : L) \models A_2}{}
  \NLINE{(\xi, \sigma) \models \MODALFORALL{x}{}.A_2}{1}
  \NLASTLINE{(\xi, \sigma) \models A_1 \AND ((\forall x^{\square}. A_2) \AND \forall m.(B_1 \IMPLIES \QQEVAL{m}{x}{A_1}))}{2, (\ref{app:completenessProof:eq:22})}
\end{NDERIVATION}

\NI The second case is that $M' \CONV$. We proceed as follows.

\begin{NDERIVATION}{1}
  \NLINE{(\xi \cdot m : L, \sigma) \models B_1}{Fresh assumption, $L = \QQ{L'}$ arbitrary value}
  \NLINE{(\xi, \sigma) \models A_1}{\ref{app:completenessProof:eq:22}}
  \NLINE{M(\xi, \sigma) \LEQ \QQ{L'}}{By (IH) (closure-2) holds for $A_1, B_1$, 1, 2}
  \NLINE{M(\xi, \sigma) \CONV \QQ{M'}}{\ref{app:completenessProof:eq:0}}
  \NLINE{\QQ{M'}\LEQ \QQ{L'}}{Lem.~\ref{app:completenessProofs:lemma:1}.\ref{app:completenessProofs:lemma:1:4}, 3, 4}
  \NLINE{M' \LEQ L}{Lem.~\ref{app:completenessProofs:lemma:1}.\ref{app:completenessProofs:lemma:1:2}, 5}
  \NLINE{N(\xi, \sigma \cdot x : M') \CONV}{\ref{app:completenessProof:eq:2}}
  \NLINE{N(\xi, \sigma \cdot x : L') \CONV}{Lem.~\ref{app:completenessProofs:lemma:1}.\ref{app:completenessProofs:lemma:1:4}, 6, 7}
  \NLINE{(\xi, \sigma \cdot x : L') \models A_2}{By (IH) $A_2$ is (MTC) for $N$, 8}
  \NLINE
  {(\xi \cdot m : L, \sigma \cdot x : L') \models A_2}
  {$m \notin \FV{A_2}$, Prop.~\ref{app:soundness:9qw384}.\ref{app:soundness:9qw384:2}, 9}
  \NLINE{(\xi \cdot m : L, \sigma) \models \QQEVAL{m}{x}{A_2}}{10}
  \NLINE{(\xi \cdot m : L, \sigma) \models B_1 \IMPLIES \QQEVAL{m}{x}{A_2}}{1, 11}
  \NLINE{(\xi, \sigma) \models \forall m.(B_1 \IMPLIES \QQEVAL{m}{x}{A_2})}{$L$ was arbitrary, 12}
  \NLASTLINE{(\xi, \sigma) \models A_1 \AND ((\forall x^{\square}. A_2) \AND \forall m.(B_1 \IMPLIES \QQEVAL{m}{x}{A_1}))}{2, 13}
\end{NDERIVATION}

\NI This establishes (MTC).

We conclude this case by proving (\emph{closure-2}).  Let $\eta \DEFEQ
(\xi, \sigma)$ be an appropriately typed model such that:
\begin{itemize}

\item $\eta \models A_1 \AND ((\forall x^{\square}.A_2) \OR \forall
  m.(B_1 \IMPLIES \QQEVAL{m}{x}{A_2}))$.

\item $(\xi \cdot u : V, \sigma) \models \exists mx^{\square}.((m =
  \QQ{\cdot} \IMPLIES m = \QQ{x}) \AND B_1 \AND B_2)$.

\end{itemize}
This means in particular that there are $M_m$ and $M_x$ such that
\begin{align}\label{completeness:eq:hhh}
   (\xi \cdot u : V \cdot m : \QQ{M_m}, \sigma \cdot x : M_x) \models (m = \QQ{\cdot} \IMPLIES m = \QQ{x}) \AND B_1 \AND B_2.
\end{align}
We first note that since
\[ 
   (\xi \cdot u : V \cdot m : \QQ{M_m}, \sigma \cdot x : M_x) \models (m = \QQ{\cdot} \IMPLIES m = \QQ{x})
\]
it must be the case that:
\[
  M_m \CONV \ \text{implies}\  M_m \CONG M_x.
\]
This is an immediate consequence of the definition of the satisfaction
relation for $m = \QQ{\cdot}$ and $m = \QQ{x}$. At the same time,
trivially:
\[
  M_m\DIV  \ \text{implies}\  M_m \LEQ M_x.
\]
Taking those two facts together, we see that the following holds.
\begin{align}\label{completeness:eq:termination}
  M_m \LEQ M_x.
\end{align}

\NI Since $u, x \notin \FV{B_1}$, we can use (\ref{completeness:eq:hhh})
and Proposition \ref{app:soundness:9qw384}.\ref{app:soundness:9qw384:3} to
conclude that
\begin{align}\label{completeness:eq:simpllle}
   \eta \models A_1
      \qquad\qquad
   (\xi \cdot m : \QQ{M_m}, \sigma) \models B_1
\end{align}
which, in turn, enables us to use the (IH), so by (\emph{closure-2})
we know that
\begin{align}\label{completeness:eq:leqM}
   M\eta \LEQ \QQ{M_m}.
\end{align}

\NI In a similar way we establish that
\begin{align}\label{completeness:eq:hallo}
   (\xi \cdot u : V, \sigma \cdot x : M_x) \models B_2
\end{align}

\NI Now we have to distinguish the following two cases.
\begin{itemize}

\item $\eta \models \MODALFORALL{x}{}.A_2$.
\item $\eta \models \forall m.(B_1 \IMPLIES \QQEVAL{m}{x}{A_2})$.

\end{itemize}

\NI In the first case clearly
\[
   \eta' \DEFEQ (\xi, \sigma \cdot x : M_x) \models A_2
\]
which together with (\ref{completeness:eq:hallo}) means we can use the (IH) on 
$\PROVESTCAP \ASSERT{A_2}{N}{u}{B_2}$, where (\emph{closure-2}) means that
\begin{align}\label{completeness:eq:termination2}
   N\eta' \LEQ V.
\end{align}
This together with (\ref {completeness:eq:leqM}) means
\[
\begin{array}{lclcl}
  (\LETQQ{x}{M}{N})\eta
     &=&
  \LETQQ{x}{M\eta}{N\eta}
     \\
     &\LEQ&
  \LETQQ{x}{\QQ{M_m}}{N\eta} && by\ (\ref{completeness:eq:leqM})
     \\
     &\RED&
  N\eta\SUBST{M_m}{x}
     \\
     &\LEQ&
  N\eta\SUBST{M_x}{x} &\ \quad\ & by\ (\ref{completeness:eq:termination})
     \\
     &=&
  N\eta' && Prop.~\ref{app:soundness:9qw384}.\ref{app:soundness:9qw384:4}
     \\
     &\LEQ&
  V && by\ \ref{completeness:eq:termination2}
\end{array}
\]
Since $\RED\ \subseteq\ \CONG\ \subseteq\ \LEQ$ (Theorem \ref{lamguage:theorem:1}.\ref{lamguage:theorem:1:1}), and
$\LEQ$ is transitive, we can thus conclude to:
\[
   (\LETQQ{x}{M}{N})\eta\ \LEQ\ V
\]
as required.

Now we consider the second case $\eta \models \forall m.(B_1 \IMPLIES \QQEVAL{m}{x}{A_2})$.
\begin{NDERIVATION}{1}
  \NLINE{\eta \models \forall m.(B_1 \IMPLIES \QQEVAL{m}{x}{A_2})}{}
  \NLINE{(\xi \cdot m : \QQ{M_m}, \sigma) \models B_1 \IMPLIES \QQEVAL{m}{x}{A_2}}{1}
  \NLASTLINE{(\xi \cdot m : \QQ{M_m}, \sigma) \models \QQEVAL{m}{x}{A_2}}{(\ref{completeness:eq:simpllle}), 2}
\end{NDERIVATION}

\begin{NDERIVATION}{4}
  \NLINE{\QQ{M_m} \CONV W, \quad (\xi \cdot m : \QQ{M_m}, \sigma \cdot x : W) \models A_2}{3}
  \NLASTLINE{\eta' \DEFEQ (\xi, \sigma \cdot x : W) \models A_2}{Prop.~\ref{app:soundness:9qw384}.\ref{app:soundness:9qw384:3}, $m \notin \FV{A_2}$, 4}
\end{NDERIVATION}

\NI The rest of this case is handled exactly like the previous case,
concluding the proof of (\emph{closure-2}).

\subsection{Proofs of Theorems \ref{completeness:theorem:1}.\ref{completeness:theorem:1:1} 
and \ref{completeness:theorem:1}.\ref{completeness:theorem:1:2}}
We conclude this section by proving observational and relative completeness.

%\PARAGRAPH{Proof of Theorem \ref{completeness:theorem:1}.\ref{completeness:theorem:1:1}}
\noindent\emph{Proof} \relax[\hspace{1 pt}of Theorem \ref{completeness:theorem:1}.\ref{completeness:theorem:1:1}].
Assume that $M \CONG N$. Now let $\eta \DEFEQ (\xi, \sigma)$, $\eta
\models A$, $M\eta \CONV V$ and $(\xi \cdot u : V, \sigma) \models B$.
Since $M \CONG N$ we know that $M\eta \CONG N\eta$ by Observation
\ref{completeness:lemma:15}, $N\eta \CONV W$ and $V \CONG W$. Hence we
can apply Proposition
\ref{app:soundness:9qw384}.\ref{app:soundness:9qw384:3} to obtain
$(\xi \cdot u : W, \sigma) \models B$ as required.  The remaining
case, $\eta \not\models A$, is immediate.

For the reverse direction, let $\PROVESTCAP \ASSERT{A}{M}{u}{B}$.
Then $(A, B)$ is a TCAP of $M$ at $u$, hence by soundness of TCAPs 
and Theorem
\ref{completeness:theorem:1}.\ref{completeness:theorem:1:0} we know
that $\models \ASSERT{A}{M}{u}{B}$. Then by assumption also $\models
\ASSERT{A}{N}{u}{B}$. Since $(A, B)$ is a TCAP for $M$ at $u$ we apply
Proposition
\ref{completeness:proposition:14}.\ref{completeness:proposition:14:1}
to obtain $M \LEQ N$.  Similarly we derive $N \LEQ M$, which together
implies $M \CONG N$. This establishes observational completeness.\qed

%\PARAGRAPH{Proof of Theorem \ref{completeness:theorem:1}.\ref{completeness:theorem:1:2}}
\noindent\emph{Proof} \relax[\hspace{1 pt}of Theorem
  \ref{completeness:theorem:1}.\ref{completeness:theorem:1:2}]. 
Relative completeness is equally easy to justify.  We start from the following
assumption.
\[
  \models \ASSERT{A}{M}{u}{B}
\]
Using the rules in Figure
\ref{figure:completeness:rules}, we obtain a TCAP $(A', B')$ for $M$ at
$u$ such that
\[
  \PROVESTCAP \ASSERT{A'}{M}{u}{B'} 
\]
holds. With these assumptions, the proof of Theorem
\ref{completeness:theorem:1}.\ref{completeness:theorem:1:2} has the
following form:
\[
   \TWOPREMISERULENAMEDRIGHT
   {
     \TWOPREMISERULENAMEDRIGHT
     {
       \models \ASSERT{A}{M}{u}{B}
     }
     {
       \PROVESTCAP \ASSERT{A'}{M}{u}{B'}
     }
     {
       A \IMPLIES (A' \AND (B' \IMPLIES B))
     }{(*)}
   }
   {
     \ONEPREMISERULENAMEDRIGHT
     {
       \PROVESTCAP \ASSERT{A'}{M}{u}{B'}
     }
     {
       \vdash \ASSERT{A'}{M}{u}{B'}
     }{Prop.~\ref{completeness:proposition:tcapSoundness}}
   }
   {
     \vdash \ASSERT{A}{M}{u}{B}
   }{Conseq-Kl}
\]
It remains to establish step (*). For this purpose, let $\eta \DEFEQ
(\xi \cdot u : W, \sigma)$ be a model and assume $ \eta \models
A$. 

We first establish that $\eta \models A'$.  Since $u \notin \FV{A}$,
we know from Proposition
\ref{app:soundness:9qw384}.\ref{app:soundness:9qw384:3} that with
$\eta' \DEFEQ (\xi, \sigma)$ also $ \eta' \models A.  $ This fact
together with the assumption $\models \ASSERT{A}{M}{u}{B}$ means that
$ M\eta' \CONV V $ for some closed value $V$. As $(A', B')$ is a TCAP
for $M$ at $u$, (MTC) holds so it must also be the case that $ \eta'
\models A'$. Applying $u \notin \FV{A}$ with Proposition
\ref{app:soundness:9qw384}.\ref{app:soundness:9qw384:3} again we now
obtain $ \eta \models A' $ as required. This shows that $A \IMPLIES
A'$.

To prove that $(A \AND B') \IMPLIES B$, assume $\eta \models A \AND
B'$.  From $\eta \models B'$ and Proposition
\ref{completeness:proposition:14}.\ref{completeness:proposition:14:2}
we obtain $M\eta' \LEQ W$.  We showed above that $M\eta' \CONV V$, so
in fact $M\eta' \CONG V$ by Proposition
\ref{lamguage:theorem:1}.\ref{lamguage:theorem:1:1}, hence clearly
\[
   V \LEQ W.
\]
From $\eta \models A$, $M\eta' \CONV V$ (see above) and the assumption
that $\models \ASSERT{A}{M}{u}{B}$ we obtain
\[
   (\xi \cdot u : V, \sigma) \models B.
\]
Now we use the upwards-closure of $B$ to conclude that:
\[
   \eta = (\xi \cdot u : W, \sigma) \models B.
\]

\section{Examples of characteristic formulae}

\NI In this section we look at some example inferences for TCAPs. To
make the derivations more readable, we will make simplifications such
as writing $\TRUTH$ for $\TRUTH \AND \TRUTH$. Note that these
simplifications are not admissible using the inference system of
Figure \ref{figure:completeness:rules} only.\qed

\begin{exa}%\PARAGRAPH{Example 8}
We begin with a simple program $2+3$.
\begin{NDERIVATION}{1}
  \NLINE{\ASSERT{\TRUTH}{2}{a}{a = 2}}{\RULENAME{Const$^t$}}
  \NLINE{\ASSERT{\TRUTH}{3}{b}{b = 3}}{\RULENAME{Const$^t$}}
  \NLASTLINE{\ASSERT{\TRUTH \AND \TRUTH}{2 + 3}{c}{\exists ab.(c = a+b \AND a = 2 \AND b = 3)}}{\RULENAME{Plus$^t$}, 1, 2}
\end{NDERIVATION}
Clearly the conclusion in Line (3) is logically equivalent to 
\[
   \ASSERT{\TRUTH}{2+3}{c}{c = 5}
\]
as expected.
\end{exa}

\begin{exa}%\PARAGRAPH{Example 9}
We continue with a variant of Example 8, using a non-modal variable
$x$. This example is preparation, of sorts, for more involved examples.
\begin{NDERIVATION}{1}
  \NLINE{\ASSERT{\TRUTH}{x}{a}{a = x}}{\RULENAME{Var$^t$}}
  \NLINE{\ASSERT{\TRUTH}{1}{b}{b = 1}}{\RULENAME{Const$^t$}}
  \NLASTLINE{\ASSERT{\TRUTH \AND \TRUTH}{x + 1}{c}{\exists ab.(c = a+b \AND a = x\AND b = 1)}}{\RULENAME{Plus$^t$}, 1, 2}
\end{NDERIVATION}
As expected,  the conclusion in Line (3) is logically equivalent to 
\[
   \ASSERT{\TRUTH}{x+1}{c}{c = x+1}
\]
\end{exa}

\begin{exa}%\PARAGRAPH{Example 10}
We use the previous example to derive the TCAP for a
abstraction $\lambda x.x+1$.
\begin{NDERIVATION}{1}
  \NLINE{\ASSERT{\TRUTH}{x+1}{c}{c = x+1}}{Ex.~9}
  \NLASTLINE{\ASSERT{\TRUTH}{\lambda x.x + 1}{u}{\forall x.(\TRUTH \IMPLIES \SIMPLEEVAL{u}{x}{x+1})}}{\RULENAME{Abs$^t$}, 1}
\end{NDERIVATION}
As before, the derived TCAP is easily seen to be logically 
equivalent to
\[
   \ASSERT{\TRUTH}{\lambda x.x + 1}{u}{\forall x.\SIMPLEEVAL{u}{x}{x+1}}
\]
\end{exa}

\begin{exa}%\PARAGRAPH{Example 11}
This example shows how the rule for application works.
\begin{NDERIVATION}{1}
  \NLINE{\ASSERT{\TRUTH}{\lambda x.x + 1}{m}{\forall x.\SIMPLEEVAL{m}{x}{x+1}}}{Ex.~10}
  \NLINE{\ASSERT{\TRUTH}{2}{n}{n = 2}}{\RULENAME{Const$^t$}}
  \NLASTLINE{
    \begin{array}[t]{c}
      \{\TRUTH \AND \TRUTH \AND \forall mn.((\forall x.\SIMPLEEVAL{m}{x}{x+1} \AND n = 2) \IMPLIES \EVALCONV{m}{n})\} \\
      (\lambda x.x+1)2\ 
      :_u\\
      \{\exists mn.(\ONEEVAL{m}{n}{z}{\forall x.\SIMPLEEVAL{m}{x}{x+1} \AND n = 2 \AND z = u})\}
    \end{array}}{\RULENAME{App$^t$}, 1, 2}
\end{NDERIVATION}
It is easy to see that 
\[
   \TRUTH \AND \TRUTH \AND \forall mn.((\forall x.\SIMPLEEVAL{m}{x}{x+1} \AND n = 2) \IMPLIES \EVALCONV{m}{n})
\]
simplifies to $\TRUTH$ via $\forall mn.(\EVALCONV{m}{n} \ \IMPLIES\ \EVALCONV{m}{n})$.
The postcondition can be simplified to $u = 3$ as follows:
\begin{NDERIVATION}{1}
  \NLINE{\exists mn.(\ONEEVAL{m}{n}{z}{\forall x.\SIMPLEEVAL{m}{x}{x+1} \AND n = 2 \AND z = u})}{}
  \NLINE{\exists mn.(\forall x.\SIMPLEEVAL{m}{x}{x+1} \AND n = 2 \AND \ONEEVAL{m}{n}{z}{z = u})}{1}
  \NLINE{\exists mn.(\SIMPLEEVAL{m}{2}{3} \AND \ONEEVAL{m}{2}{z}{z = u})}{2}
  \NLINE{\exists mn.(\ONEEVAL{m}{2}{z}{z = 3} \AND \ONEEVAL{m}{2}{z}{z = u})}{\text{Unwinding of shorthand}, 3}
  \NLINE{\exists mn.(\ONEEVAL{m}{2}{z}{z = 3 \AND z = u})}{Axiom (e1) from Fig.~\ref{figure:logic:PCFaxioms}, 4}
  \NLINE{\exists mn.(\ONEEVAL{m}{2}{z}{u = 3})}{5}
  \NLINE{\exists mn.(u = 3 \AND \ONEEVAL{m}{2}{z}{\TRUTH})}{Axiom (e4) from Fig.~\ref{figure:logic:PCFaxioms}, 6}
  \NLASTLINE{u = 3}{7}
\end{NDERIVATION}
\end{exa}

\begin{exa}%\PARAGRAPH{Example 12}
The TCAP in the previous example turned out to be logically equivalent
to a very simple assertion, albeit only after simplification starting
with rather large formulae. This simplification was possible because
both parts of the application were concrete terms. In an application
like $gx$ this is not the case as we show now, even when neither $g$
nor $x$ are modal.

\begin{NDERIVATION}{1}
  \NLINE{\ASSERT{\TRUTH}{g}{a}{a = g}}{\RULENAME{Var$^t$}}
  \NLINE{\ASSERT{\TRUTH}{x}{b}{b = x}}{\RULENAME{Var$^t$}}
  \NLINE{
    \begin{array}[t]{c}
      \{\TRUTH \AND \TRUTH \AND \forall ab. ((a = g \AND b = x ) \IMPLIES \EVALCONV{a}{b})\}\\
      g x\ 
      :_m\\
      \{\exists ab.(\ONEEVAL{a}{b}{z}{a = g \AND b = x \AND z = m})\}
    \end{array}}{\RULENAME{App$^t$}, 1, 2}
  \NLASTLINE{\ASSERT{\EVALCONV{g}{x}}{g x}{m}{\SIMPLEEVAL{g}{x}{m}}}{3}
\end{NDERIVATION}

\NI Here the (simplified) TCAP explicitly assumes that the application
converges, and states that the result of the program is simply the
result of the application.
\end{exa}

\begin{exa}%\PARAGRAPH{Example 13}
We use the previous example to derive the TCAP for $\omega$.  Our
preceding discussion indicated that
$\ASSERT{\TRUTH}{\omega}{u}{\TRUTH}$ is the strongest assertion we can
make about a program such as $\omega$ in a logic for total correctness.
This is borne out by the derivation to follow. 

\begin{NDERIVATION}{1}
  \NLINE{\ASSERT{\EVALCONV{g}{x}}{g x}{m}{\SIMPLEEVAL{g}{x}{m}}}{Ex.~11}
  \NLINE{\ASSERT{\TRUTH}{\lambda x.g x}{u}{\forall x.(\EVALCONV{g}{x}\ \IMPLIES\ \ONEEVAL{u}{x}{m}{\SIMPLEEVAL{g}{x}{m}})}}{\RULENAME{Abs$^t$}, 1}
  \NLINE{\ASSERT{\TRUTH}{\omega}{u}{\forall x.(\EVALCONV{u}{x}\ \IMPLIES\ \ONEEVAL{u}{x}{m}{\SIMPLEEVAL{u}{x}{m}})}}{\RULENAME{Rec$^t$}, 2}
  \NLINE{\ASSERT{\TRUTH}{\omega}{u}{\forall x.(\EVALCONV{u}{x}\ \IMPLIES\ \VERYSIMPLEEVAL{u}{x} = \VERYSIMPLEEVAL{u}{x})}}{3}
  \NLASTLINE{\ASSERT{\TRUTH}{\omega}{u}{\TRUTH}}{4}
\end{NDERIVATION}
\end{exa}

\begin{exa}%\PARAGRAPH{Example 14}
We build on Example 13 to derive the TCAP for $\Omega$. As $\Omega$
diverges, the precondition of the TCAP must be falsity.

\begin{NDERIVATION}{1}
    \NLINE{\ASSERT{\TRUTH}{\omega}{m}{\TRUTH}}{Ex.~12}
    \NLINE{\ASSERT{\TRUTH}{()}{n}{n = ()}}{\RULENAME{Const$^t$}}
    \NLASTLINE{\ASSERT{\forall m. \EVALCONV{m}{()}}{\Omega}{u}{\exists m. \SIMPLEEVAL{m}{()}{u}}}{\RULENAME{App$^t$}, 3}
\end{NDERIVATION}

\NI Clearly $\forall m. \EVALCONV{m}{()}$ is false, because not every
function is terminating. This is intuitively obvious, and follows
formally from the axiom $(div)$ in Figure
\ref{figure:logic:PCFaxioms}. Consequently, the TCAP for $\Omega$ is
logically equivalent to
\[
   \ASSERT{\FALSITY}{\Omega}{u}{\TRUTH}
\]
as expected.
\end{exa}

\begin{exa}%\PARAGRAPH{Example 15}
We will now look at examples involving MP.
\begin{NDERIVATION}{1}
    \NLINE{\ASSERT{\TRUTH}{3}{m}{m=3}}{\RULENAME{Const$^t$}}
    \NLINE{\ASSERT{\TRUTH}{\QQ{3}}{u}{\TRUTH \IMPLIES \QQEVAL{u}{m}{m=3}}}{\RULENAME{Quote$^t$}, 1}
    \NLASTLINE{\ASSERT{\TRUTH}{\QQ{3}}{u}{u = \QQ{3}}}{2}
\end{NDERIVATION}

\NI The result is not surprising because 
    [\RULENAME{Quote$^t$}] is unchanged from Figure
    \ref{figure:logic:rules}.
\end{exa}

\begin{exa}%\PARAGRAPH{Example 16}
Next we tackle an example that uses [\RULENAME{Unquote$^t$}]. Clearly
the program $\LETQQ{x}{\QQ{3}}{x}$ evaluates to 3. Note that
$x$ is modal.

\begin{NDERIVATION}{1}
  \NLINE{\ASSERT{\TRUTH}{\QQ{3}}{m}{m = \QQ{3}}}{Ex.~14}
  \NLINE{\ASSERT{x\CONV}{x}{n}{n = x}}{\RULENAME{Var$_m^t$}}
  \NLINE{
    \begin{array}[t]{c}
      \{\TRUTH \AND ((\forall x^{\square}.x\CONV) \OR \forall m.(m=\QQ{3} \IMPLIES \QQEVAL{m}{x}{x\CONV}))\} \\
      \LETQQ{x}{\QQ{3}}{x}\ 
      :_n \\
      \{\exists mx^{\square}.((m = \QQ{\cdot} \IMPLIES m = \QQ{x}) \AND m = \QQ{3} \AND n = x)\}
    \end{array}}
    {\RULENAME{Unquote$^t$}, 1, 2}
  \NLINE{
    \begin{array}[t]{c}
      \{\TRUTH\} \\
      \LETQQ{x}{\QQ{3}}{x}\ 
      :_{n}\\
      \{\exists mx^{\square}.((m = \QQ{\cdot} \IMPLIES m = \QQ{x}) \AND m = \QQ{3} \AND n = x)\}
    \end{array}}
        {3}
        \NLASTLINE{\ASSERT{\TRUTH}{\LETQQ{x}{\QQ{3}}{x}}{n}{n = 3}}{4}
\end{NDERIVATION}

\NI We now explain the last two simplification steps.  First
$(\MODALFORALL{x}{}.x\CONV)$ must be equivalent to $\FALSITY$ because
not all denotations of the modal variable $x$ terminate. This is
formalised by Axiom $(div_m)$ from Section \ref{completeness2}.  The formula
$\forall m.(m=\QQ{3} \IMPLIES \QQEVAL{m}{x}{x\CONV}$ is equivalent to
$\TRUTH$ because $m=\QQ{3}$ is a shorthand for $\QQEVAL{m}{x}{x = 3}$,
and clearly $x = 3$ implies $x\CONV$. This justifies Line 4.
Regarding the last line, clearly $m = \QQ{3}$ implies $m = \QQ{\cdot}$, so
$\exists mx^{\square}.(m = \QQ{x} \AND m = \QQ{3} \AND n = x)$
holds. Using Axioms $(q1)$ and $(q4)$ allows us to obtain $\exists
mx^{\square}.(x = 3 \AND n = x)$, which in turn simplifies to $n = 3$
as required.
\end{exa}

\begin{exa}%\PARAGRAPH{Example 17}
We now derive a simple result from Example 14 that is used later.
\begin{NDERIVATION}{1}
    \NLINE{\ASSERT{\FALSITY}{\Omega}{x}{\TRUTH}}{Ex.~13}
    \NLINE{\ASSERT{\TRUTH}{\QQ{\Omega}}{m}{\FALSITY \IMPLIES \QQEVAL{m}{x}{\TRUTH}}}{\RULENAME{Quote$^t$}, 1}
    \NLASTLINE{\ASSERT{\TRUTH}{\QQ{\Omega}}{m}{\TRUTH}}{2}
\end{NDERIVATION}
\end{exa}

\begin{exa}%\PARAGRAPH{Example 18}
We continue with an example where quasi-quotes divergent code gets
unquoted, and then re-quoted without further use.
\begin{NDERIVATION}{1}
    \NLINE{\ASSERT{\TRUTH}{\QQ{\Omega}}{m}{\TRUTH}}{Ex.~16}
    \NLINE{\ASSERT{x\CONV}{x}{b}{x = b}}{\RULENAME{Var$_m^t$}}
    \NLINE{\ASSERT{\TRUTH}{\QQ{x}}{u}{x\CONV\ \IMPLIES\ \QQEVAL{u}{b}{x = b}}}{\RULENAME{Quote$^t$}, 2}
    \NLINE{
    \begin{array}[t]{c}
      \{\TRUTH \AND ((\forall x^{\square}.\TRUTH) \OR \forall m.(\TRUTH \IMPLIES \QQEVAL{m}{x}{\TRUTH}))\}\\
      \LETQQ{x}{\QQ{\Omega}}{\QQ{x}}\ 
      :_{u}\\
      \{\exists mx^{\square}.((m = \QQ{\cdot} \IMPLIES m = \QQ{x}) \AND \TRUTH \AND (x\CONV \IMPLIES \QQEVAL{u}{b}{x = b}))\}
    \end{array}
    }{\RULENAME{Unquote$^t$}, 1, 3}
    \NLINE{
    \begin{array}[t]{c}
      \{\TRUTH\}\\
      \LETQQ{x}{\QQ{\Omega}}{\QQ{x}}\ 
      :_{u}\\
      \{\exists mx^{\square}.((m = \QQ{\cdot} \IMPLIES m = \QQ{x}) \AND \TRUTH \AND (x\CONV \IMPLIES \QQEVAL{u}{b}{x = b}))\}
    \end{array}
    }{4}
    \NLASTLINE{\ASSERT{\TRUTH}{\LETQQ{x}{\QQ{\Omega}}{\QQ{x}}}{u}{\TRUTH}}{5}
\end{NDERIVATION}

\NI Line 6 follows because clearly $\forall x^{\square}.\TRUTH$ is
equivalent to $\TRUTH$. Finally, the simplification in Line 6 is
immediate, because everything implies $\TRUTH$.
\end{exa}
\begin{exa}%\PARAGRAPH{Example 19}
We continue by determining the TCAP for $\LETQQ{x}{\QQ{\Omega}}{x}$,
which is a divergent program, which  we expect to be equivalent to
\[
   \ASSERT{\FALSITY}{\LETQQ{x}{\QQ{\Omega}}{x}}{u}{\TRUTH}.
\]
We infer the TCAP as follows:
\begin{NDERIVATION}{1}
  \NLINE{\ASSERT{\TRUTH}{\QQ{\Omega}}{x}{\TRUTH}}{Ex.~16}
  \NLINE{\ASSERT{x\CONV}{x}{u}{u = x}}{\RULENAME{Var$_m^t$}}
  \NLINE{
    \begin{array}[t]{c}
      \{\TRUTH \AND ((\forall x^{\square}.x\CONV) \OR \forall m.(\TRUTH \IMPLIES \QQEVAL{m}{x}{x\CONV}))\} \\
      \LETQQ{x}{\QQ{\Omega}}{x}\ 
      :_{u}\\
      \{\exists mx^{\square}.((m = \QQ{\cdot} \IMPLIES m = \QQ{x}) \AND \TRUTH \AND u = x)\}
    \end{array}}{\RULENAME{Unquote$^t$}, 1, 2}
  \NLASTLINE{\ASSERT{\FALSITY}{\LETQQ{x}{\QQ{\Omega}}{x}}{u}{\TRUTH}}{3}
\end{NDERIVATION}

\NI We now explain the simplifications leading to Line 4.  By the
axiom $(div_m)$ which says that not all modal variables denote a
terminating program, we know that $\forall x^{\square}.x\CONV$ is
false. Likewise $\forall m.(\TRUTH \IMPLIES \QQEVAL{m}{x}{x\CONV})$ is
equivalent to $\forall m.\QQEVAL{m}{x}{x\CONV}$ which claims that any
quasi-quote must contain a terminating program, which is false by the
axiom $(div)$ in Figure \ref{figure:logic:axioms}. Consequently, the
precondition in Line 3 is equivalent to $\FALSITY$. Simplification of
the postcondition to $\TRUTH$ is immediate.
\end{exa}

\section{Conclusion}\label{conclusion}

\NI We have proposed a program logic for an HGRTMP language, and
established key metalogical properties like completeness and the
correspondence between axiomatic and operational semantics. We are not
aware of previous work on program logics for
meta-programming. So far, only typing systems for
statically enforcing program properties have been investigated; the
two most expressive are $\Omega$mega~\cite{SheardT:proome} and
Concoqtion~\cite{FogarthyS:concoqtionitn}. Both use indexed typed to
achieve expressivity.  $\Omega$mega is a call-by-value variant of
Haskell with generalised algebraic datatypes (GADTs) and an extensible
kind system.  In $\Omega$mega, GADTs can express easily datatypes
representing object-programs, whose meta-level types encode the
object-level types of the programs represented. Tagless interpreters
can directly be expressed and typed for these object
programs. $\Omega$mega is expressive enough to encode the MetaML
typing system together with a MetaML interpreter in a type-safe
manner.  Concoqtion is an extension of MetaOCaml and uses the term
language of the theorem prover Coq to define index types, specify
index operations, represent the properties of indexes and construct
proofs.  Basing indices on Coq terms opens all mathematical insight
available as Coq libraries to use in typing meta-programs.  Types in
both languages are not as expressive with respect to properties of
\emph{meta-programs themselves} as our logics, which capture exactly
the observable properties.  Nevertheless, program logic and
type-theory are not mutually exclusive; on the contrary, reconciling
both in the context of meta-programming is an important open problem.

\PCFDP\ lacks the ability, vital for realistic
meta-programming, to manipulate \emph{open} code, i.e.~code
with free variables.  
 The
$\lambda^{\circ}$-calculus~\cite{DaviesR:temlogatbta} is a small language for
HGRTMP where code
with free variables can be manipulated. As with \PCFDP\ (without
recursion), there is a Curry-Howard correspondence: $\lambda^{\circ}$
is a proof calculus for a temporal logic. Due to its simplicity,
$\lambda^{\circ}$ is an ideal object of study to see how the logic
presented here can be generalised to open code. The  simplicity of $\lambda^{\circ}$ has
a price: the calculus cannot be directly extended with a
construct expressing the evaluation of generated code.  A more
ambitious target that allows the manipulation of terms with free
variables, but also the evaluation of generated code, is Taha's and
Nielsen's system of environment classifiers~\cite{TahaW:envClas}, which
also forms the basis of MetaOCaml, the most widely studied
meta-programming language in the MetaML tradition.
Moreover,~\cite{TsukadaT:logfoufecLONG} presents a Curry-Howard
correspondence between a typing system closely related to that of~\cite{TahaW:envClas}
and a modal logic.  We believe that a logical
account of meta-programming with open code is a key challenge in
bringing program logics to realistic meta-programming languages.  

A different challenge is to add state to \PCFDP\ and extend the
corresponding logics. We expect the logical treatment of state given
in~\cite{ALIASfull,YHB07:local:full} to extend smoothly to a
meta-programming setting. The main  question is what typing
system to use to type stateful meta-programming. The system used in
MetaOCaml, based on~\cite{TahaW:envClas}, is unsound in the presence
of state due to a form of scope extrusion. Recent versions of
MetaOCaml add a dynamic check to detect this behaviour.  As an
alternative to dynamic typing, the Java-like meta-programming language
Mint~\cite{WestbrookE:minjmspuws} simply prohibits the \emph{sharing}
of state between different meta-programming stages, resulting in a
statically sound typing system.  Yet another approach is given in~\cite{KameyamaY:shistaswdc}
where a two-level HGRTMP language is
introduced with delimited control operators and a restriction of side
effects during code generation to the scope of generated binders. That
guarantees well-typedness.  We believe that these approaches can all
be made to coexist with modern logics for higher-order
state~\cite{ALIASfull,YHB07:local:full}.

Relatedly, \cite{DaviesR:modanaosc} presents a unstaging translation
from \PCFDP\ to \PCF. The key idea is that a quasi-quote $\QQ{M}$ is
turned into a thunk $\lambda ().M'$ where $M'$ is the translation of
$M$. Consequently the type $\QQ{\alpha}$ is translated to $\UNIT \FS
\alpha'$ where $\alpha'$ is the translation to $\alpha$.  What are the
properties of this translation? Is it fully abstract?  The translation
can be extended to translating \PCFDP\ assertions and proofs into the
logic for \PCF. Would this latter translation be logically fully
abstract in the sense of~\cite{PlotkinG:logfula}?  If unstaging is
fully abstract, then it should be possible to recover the logic
presented here from the logic for \PCF\ and the translation.
 
Reasoning about HGRTMP using unstaging translations looks promising.
In \cite{ChoiW:staanaomspvut} a complex HGRTMP language that allows
the manipulation of open code and the capture of free variables is
unstaged. However, logical reasoning about meta-programs in the target
language of an unstaging translation incurs a cost: it leads to larger
formulae and proofs in comparison with reasoning about the
meta-programs directly using the source language. Moreover, this cost is paid in
every reasoning process.  In contrast, the cost of developing a logic
for the meta-programming language is paid only once.  An additional
question is whether unstaging translations are fully abstract for more
complicated HGRTMP languages.

A technical issue we left open is to do with the size of
characteristic formulae.  The inference system in Section
\ref{completeness2} may lead to an exponential blow up of TCAPs
vis-a-vis the programs they are derived from. We believe that it is
possible to give an alternative inference system for TCAPs such that
the size of the TCAP is linear, i.e.~$O(n)$, in the size of the
program. In~\cite{ChargueraudA:provertcf,HondaK:descriptive} this is
achieved for logics of partial correctness for \PCF-like languages,
and in~\cite{ChargueraudA:chaforftvoip} for a simple imperative
language.

Finally we have a question about modal quantification: 'normal'
reasoning about \PCFDP-programs using the rules and axioms of Section
\ref{logic} appears to be possible entirely without modal
quantification.  Can we abolish modal quantification altogether? If
not, why is the lack of modal quantification no issue in practise?

\PARAGRAPH{Acknowledgements} 
We thank Dana Xu for careful comments on the short version of this
article, Arthur Chargu\'eraud for discussions about characteristic
formulae and completeness, and Jacques Carette, Billiejoe Charlton,
Rowan Davies, Oleg Kiselyov, Chung-chieh Shan, and Walid Taha
for answering questions about meta-programming. We also thank the
anonymous reviewers for their insightful comments.

\bibliographystyle{abbrv}
\bibliography{../../../mala}

\appendix

\section{Axioms for \texorpdfstring{\PCF}{Pcf}\ that are also valid
  for \texorpdfstring{\PCFDP}{Pcf sub dp}}\label{app:axioms}

\NI Section \ref{logic} presented the axioms of our logic that involve
meta-programming features. Other axioms are listed in Figure
\ref{figure:logic:PCFaxioms}. The axioms are standard, and
explanation, as well as soundness proofs can be found in
\cite{ALIASfull,HondaK:obscomplfihofTECREP,YHB07:local:full}.
Moreover, soundness proofs for the axioms given in Figure
\ref{figure:logic:PCFaxioms} are straightforward adaptations of the
proofs for the axioms in Figure \ref{figure:logic:axioms}.
 The presentation uses the
following abbreviations:
\begin{center}

$\EXT{xy}$ stands for $\forall az.(\ONEEVAL{x}{z}{w}{w = a} \equiv
  \ONEEVAL{y}{z}{w}{w = a})$. 

\end{center}
 
\NI Note that it is vital for $x$ and $y$ to be non-modal. The direction
 $\EXT{xy} \IMPLIES x = y$ is unsound otherwise, because $\EXT{xy}$
 cannot distinguish between e.g.~appropriately typed $\Omega$ and
 $\lambda x.\Omega$.

\begin{FIGURE}
\begin{RULES}
\begin{array}{lrclll}
	(e1) & \ONEEVAL{x}{y}{z}{A} \AND \ONEEVAL{x}{y}{z}{B} 
	     &\equiv&
	\ONEEVAL{x}{y}{z}{A \AND B}
  			    \\
	(e2) & \ONEEVAL{x}{y}{z}{\neg A} 
	     &\IMPLIES& 
	\neg\ONEEVAL{x}{y}{z}{A}
  			    \\
	(e3) & \ONEEVAL{x}{y}{z}{A} \AND \neg \ONEEVAL{x}{y}{z}{B} 
	     &\equiv&
	\ONEEVAL{x}{y}{z}{A \AND \neg B}
  			    \\
	(e4) & \ONEEVAL{x}{y}{z}{A \AND B} 
	     &\equiv& 
	A \AND \ONEEVAL{x}{y}{z}{B} 
          	   &
	z \notin \FV{A}
  			    \\
	(e5) & \ONEEVAL{x}{y}{z}{\forall a^{\alpha}. A} 
	     &\equiv& 
	\forall a^{\alpha}.\ONEEVAL{x}{y}{z}{A}
        &a \neq x, y, z
  			    \\
	(e6) & (A \IMPLIES B) \AND \ONEEVAL{x}{y}{z}{A}
	     &\IMPLIES& 
	\ONEEVAL{x}{y}{z}{B}
			&
	z \notin \FV{A, B} \\
        (div) & \neg \forall x^{\alpha}.\EVALCONV{m}{x}

			\\
	(ext) & x = y
	      &\equiv&
	\EXT{xy}
           & 
        x, y\ \text{of function type}\\
        &&&&\text{both non-modal} \\
        (e_{\alpha}) &\ONEEVAL{x}{y}{z}{A}
                        &\equiv&
                      \ONEEVAL{x}{y}{a}{\ONEEVAL{x}{y}{z}{A \AND a = z}} \hspace{-15mm} \\
                      &&&& \hspace{-23mm} a \notin \{x, y\}, a \in \FV{A}\ \text{implies}\ a = z 
\end{array}
\end{RULES}
\caption{\PCFDP\ axioms not involving meta-programming
  constructs. Except were noted otherwise, all free variables can be modal or non-modal.}\label{figure:logic:PCFaxioms}
\end{FIGURE}

\section{Omitted proofs for Section \ref{completeness2} (Completeness)}\label{app:completenessProofs}

\NI It remains to establish Theorem
\ref{completeness:theorem:1}.\ref{completeness:theorem:1:1} for the
rules [\RULENAME{Var$^t$}, \RULENAME{Const$^t$}, \RULENAME{Abs$^t$},
  \RULENAME{App$^t$}, \RULENAME{Op$^t$}, \RULENAME{If$^t$}, \RULENAME{Rec$^t$}]. All
proofs here are variants of the proofs in the
unpublished long version of~\cite{HondaK:descriptive}. 
\begin{description}
\item[\rm\RULENAME{Var$^t$}] The MTC is trivially true. For (\emph{closure-2}),
assume that $(\xi \cdot x : V, \sigma) \models \TRUTH$ and $(\xi \cdot x : V, m
: W, \sigma)\models x = m$. Then immediately $V \CONG W$, hence $V
\LEQ W$ as required.

\item[\rm\RULENAME{Const$^t$}] Similar to [\RULENAME{Var$^t$}] and omitted.

\item[\rm\RULENAME{Abs$^t$}] Since abstractions are values, (MTC) is trivially
true. For (\emph{closure-2}), assume that $(\xi, \sigma)$ is a model
and $(\xi \cdot u : V, \sigma) \models \forall x.(A \IMPLIES
\ONEEVAL{u}{x}{m}{B})$.

\begin{NDERIVATION}{1}
  \NLINE{(\xi \cdot u : V, \sigma) \models \forall x.(A \IMPLIES \ONEEVAL{u}{x}{m}{B})}{Assumption}
  \NLINE{(\xi \cdot u : V \cdot x : W, \sigma) \models A \IMPLIES \ONEEVAL{u}{x}{m}{B}}{W arbitrary, 1}
  \NLINE{(\xi \cdot u : V \cdot x : W, \sigma) \models A }{Assumption}
  \NLINE{(\xi \cdot u : V \cdot x : W, \sigma) \models \ONEEVAL{u}{x}{m}{B}}{2, 3}
  \NLINE{VW \CONV U \qquad (\xi \cdot u : V \cdot x : W \cdot m : U, \sigma) \models B}{4}
  \NLINE{\eta \DEFEQ (\xi \cdot x : W \cdot m : U, \sigma) \models B}{$u \notin \FV{B}$, Prop.~\ref{app:soundness:9qw384}.\ref{app:soundness:9qw384:3}, 5}
  \NLINE{(\xi \cdot x : W, \sigma) \models A }{$u \notin \FV{A}$, Prop.~\ref{app:soundness:9qw384}.\ref{app:soundness:9qw384:3}, 3}
  \NLINE{M\eta \LEQ U}{(IH), (closure-2), 6, 7}
  \NLINE{VW \CONG U}{Prop.~\ref{lamguage:theorem:1}.\ref{lamguage:theorem:1:1}, 5}
  \NLINE{M\eta \LEQ VW}{8, 9}
  \NLINE{\eta' \DEFEQ (\xi \cdot m : U, \sigma) }{}
  \NLASTLINE{(\lambda x.M\eta')W \RED M\eta'\SUBST{W}{x} = M\eta}{Prop.~\ref{app:soundness:9qw384}.\ref{app:soundness:9qw384:4}}
\end{NDERIVATION}

\begin{NDERIVATION}{13}
  \NLINE{(\lambda x.M\eta')W \LEQ M\eta}{Prop.~\ref{lamguage:theorem:1}.\ref{lamguage:theorem:1:1}, 12}
  \NLINE{(\lambda x.M\eta')W = (\lambda x.M)\eta'W \LEQ M\eta \LEQ U \LEQ VW}{9, 10, 13}
  \NLINE{\text{for all $W$}\ (\lambda x.M)\eta'W \LEQ VW}{W arbitrary, 14}
  \NLASTLINE{ (\lambda x.M)\eta' \LEQ V}{Lem.~\ref{app:completenessProofs:lemma:1}\ref{app:completenessProofs:lemma:1:2}, 15}
\end{NDERIVATION}

\item[\rm\RULENAME{App$^t$}] We begin with (MTC). Let $\eta$ be an appropriately
typed model such that
\[
   (MN)\eta \CONV V
\]
Then in particular $M\eta \CONV W$, $N\eta \CONV U$ and $WU \CONV C$. By (IH) the
first two mean that
\[
   \eta \models A_1
      \qquad
   \eta \models A_2
\]

\item[\rm\RULENAME{Op$^t$}] We treat the special case of addition. The MTC
follows directly from the (IH), noting that $(M+N)\eta \CONV$ holds
exactly when $M\eta\CONV$ and $N\eta\CONV$. For (\emph{closure-2}) we
reason as follows.
\begin{NDERIVATION}{1}
  \NLINE{\eta \DEFEQ (\xi, \sigma) \models A_1 \AND A_2}{Assumption}
  \NLINE{(\xi \cdot u : V, \sigma) \models \exists m_1 m_2.(u = m_1 + m_2 \AND B_1 \AND B_2)}{Assumption}
  \NLINE{(\xi \cdot u : V \cdot m_1 : W_1 \cdot m_2 : W_2, \sigma) \models u = m_1 + m_2 \AND B_1 \AND B_2}{2}
  \NLINE{(\xi \cdot m_i : W_i, \sigma) \models B_i}{i = 1, 2, $u, m_{3-i} \notin \FV{B_i}$, Prop.~\ref{app:soundness:9qw384}.\ref{app:soundness:9qw384:3}, 3}
  \NLINE{M_i\eta \LEQ W_i}{(IH), 1, 4}
  \NLINE{(M_i + M_2)\eta \LEQ W_1 + W_2}{5}
  \NLINE{W_1 + W_2 \CONG V}{3}
  \NLASTLINE{(M_i + M_2)\eta \LEQ V}{6, 7}
\end{NDERIVATION}

\item[\rm\RULENAME{If$^t$}] For (MTC), with $\eta \DEFEQ (\xi, \sigma)$, $b_1 =
\TRUE$, $b_2 = \FALSE$, assume wlog that:
\begin{eqnarray}
  (\IFTHENELSE{M}{N_1}{N_2})\eta 
     &\ \NRED\ &
  \IFTHENELSE{\TRUE}{N_1}{N_2} \label{app:completenessProofs:eq:if1}
     \\
     &\RED& 
  N_1 \notag
     \\
     &\CONV& 
  W_1 \label{app:completenessProofs:eq:if2}
\end{eqnarray}
Now we reason as follows.
\begin{NDERIVATION}{1}
  \NLINE{M\eta \CONV \TRUE}{(\ref{app:completenessProofs:eq:if1})}
  \NLINE{\eta \models A \quad (\xi \cdot m : \TRUE, \sigma) \models B}{(IH), (MTC), 1}
  \NLINE{\eta \models B\SUBST{\TRUE}{m}}{Prop.~\ref{app:soundness:9qw384}.\ref{app:soundness:9qw384:4}, 2}
  \NLINE{N\eta \CONV W_1}{(\ref{app:completenessProofs:eq:if2})}
  \NLINE{\eta \models A_1 \quad (\xi \cdot m : \TRUE, \sigma) \models B_1}{(IH), (MTC), 4}
  \NLINE{\eta \models B\SUBST{\TRUE}{m} \IMPLIES A_1}{3, 5}
  \NLASTLINE{\eta \models B\SUBST{\FALSE}{m}}{Assumption towards a contradiction}
\end{NDERIVATION}

\begin{NDERIVATION}{8}
  \NLINE{\eta \models A \quad (\xi \cdot m : \FALSE, \sigma) \models B}{2, 7}
  \NLINE{M\eta \LEQ \FALSE}{(IH), (closure-2), 8}
  \NLINE{\eta \not \models B\SUBST{\FALSE}{m}}{19 contradicts 1}
  \NLINE{\eta \models B\SUBST{\FALSE}{m} \IMPLIES A_2}{10}
  \NLASTLINE{\eta \models A \AND \BIGAND_i (B\SUBST{b_i}{m} \IMPLIES A_i)}{6, 11}
\end{NDERIVATION}

\NI The reverse direction follows from soundness.

For (\emph{closure-2}) the following derivation gets us towards the
result.
\begin{NDERIVATION}{1}
  \NLINE{\eta \DEFEQ (\xi, \sigma) \models A \AND \displaystyle{\BIGAND}_{i}(B\SUBST{b_i}{m} \IMPLIES A_i)}{Assumption}
  \NLINE{\eta' \DEFEQ (\xi \cdot V, \sigma) \models B\SUBST{\TRUE}{m} \AND B_1}{Assumption wlog}
  \NLINE{\eta \models B\SUBST{\TRUE}{m}}{$u \notin \FV{B}$, Prop.~\ref{app:soundness:9qw384}.\ref{app:soundness:9qw384:3}, 2}
  \NLINE{\eta \models A_1}{1, 3}
  \NLINE{N_1\eta \LEQ V }{(IH), closure-2, 2, 4}
  \NLINE{\eta \models A}{1}
  \NLINE{M\eta \CONV }{(MTC), 6}
  \NLINE{(\xi \cdot m : \TRUE, \sigma) \models B}{3}
  \NLINE{M\eta \LEQ \TRUE}{(IH), closure-2, 6, 8}
  \NLASTLINE{M\eta \CONV \TRUE}{7, 9}
\end{NDERIVATION}
We use these facts to derive:
\[
\begin{array}{lclcl}
  (\IFTHENELSE{M}{N_1}{N_2})\eta 
     &\ \NRED\ &
  \IFTHENELSE{\TRUE}{N_1\eta}{N_2\eta}
     &\ \qquad\ &
  (10)
     \\
     &\RED&
  N_1\eta
     \\
     &\LEQ&
  V
     &&
  (5)
\end{array}
\]
Using Proposition \ref{lamguage:theorem:1}.\ref{lamguage:theorem:1:1},
this implies the required
\[
   (\IFTHENELSE{M}{N_1}{N_2})\eta \LEQ V.
\]

\item[\rm\RULENAME{Rec$^t$}] In this case too, (MTC) is trivial. For
(\emph{closure-2}) let $\eta \DEFEQ (\xi, \sigma)$ and assume that
\[
   (\xi \cdot m : V, \sigma) \models A\SUBST{m}{g}
\]
which, by Proposition \ref{app:soundness:9qw384}.\ref{app:soundness:9qw384:4} is equivalent to
\[
   (\xi \cdot m : V \cdot g : V, \sigma) \models A.
\]
We now show by nested induction on $n$ that for all $n \geq 0$ it is
the case that
\[
   W_n\eta \ \LEQ \ V
\]
where the $W_n$ are defined as follows (cf.~Proposition
\ref{lamguage:theorem:1}.\ref{lamguage:theorem:1:6}).
\[
   W_0 \DEFEQ \Omega
      \qquad\qquad
   W_{n+1} \DEFEQ \lambda x.M\SUBST{W_n}{g}.
\]
The base case $n = 0$ is trivial. For the inductive step of the inner
induction, let $W_n\eta \LEQ V$.
\begin{eqnarray}
   W_{n_1} 
      & \ = \ &
   (\lambda x.M\SUBST{W_n}{g})\eta
      \notag
      \\
      &=&
   \lambda x.M\SUBST{W_n\eta}{g}
      \notag
      \\
      &\LEQ&
   \lambda x.M\SUBST{V}{g}
      \label{app:completenessProofs:eq:rec1}
      \\
      &=&
   \lambda x.M(\xi \cdot g : V, \sigma)
      \notag      \\
      &\LEQ&
   V
      \label{app:completenessProofs:eq:rec2}   
\end{eqnarray}
Here (\ref{app:completenessProofs:eq:rec1}) follows from the inner
(IH) together with $\SUBST{\cdot}{g}$'s being monotonic w.r.t.~to
$\LEQ$ (Lemma
\ref{app:completenessProofs:lemma:1}.\ref{app:completenessProofs:lemma:1:5}).
On the other hand, (\ref{app:completenessProofs:eq:rec2}) is directly
by the outer (IH) and (\emph{closure-2}).

Hence we have $W_n\eta \LEQ V$ for all $n$. Since $\mu g.\lambda
x.M\eta \not \LEQ V$, then $W_n \not \LEQ$ for some $n$. Using
Proposition \ref{lamguage:theorem:1}.\ref{lamguage:theorem:1:6} we
conclude that $\mu g.\lambda x.M\eta \LEQ V$.
\end{description}
\end{document}